\documentclass[manuscript]{aastex}
\usepackage{epstopdf}

\slugcomment{Submitted to the ApJ Supplement Series 2016-03-29; accepted 2016-09-12}

\shorttitle{NPOI Bright Star Survey}
\shortauthors{Hutter et al.}

\begin{document}

\title{Surveying the Bright Stars by Optical Interferometry I: A Search for
Multiplicity Among Stars of Spectral Types F$-$K}

\author{D. J. Hutter,\altaffilmark{1} R. T. Zavala,\altaffilmark{1}
C. Tycner,\altaffilmark{2} J. A. Benson,\altaffilmark{1} C. A.
Hummel,\altaffilmark{3} J. Sanborn,\altaffilmark{4} O. G.
Franz,\altaffilmark{4} and K. J. Johnston\altaffilmark{5,6}}

\altaffiltext{1}{U.S. Naval Observatory, Flagstaff Station, 10391 West Naval
Observatory Road, Flagstaff, AZ 86005-8521; djh@nofs.navy.mil}
\altaffiltext{2}{Central Michigan University, Department of Physics, Mount
Pleasant, MI 48859}
\altaffiltext{3}{European Southern Observatory, Karl-Schwarzschild-Str. 2,
Garching, 85748, Germany}
\altaffiltext{4}{Lowell Observatory, 1400 West Mars Hill Road, Flagstaff, AZ
86001-4499}
\altaffiltext{5}{U.S. Naval Observatory, 3450 Massachusetts Avenue, NW,
Washington, DC 20392-5420}
\altaffiltext{6}{Retired}

\begin{abstract}

We present the first results from an ongoing survey for multiplicity among the
bright stars using the Navy Precision Optical Interferometer (NPOI).  We first
present a summary of NPOI observations of known multiple systems, including the
first detection of the companion of $\beta$ Scuti with precise relative astrometry,
to illustrate the instrument's detection sensitivity for binaries at magnitude
differences $\Delta$$m$ $\lessapprox$ 3 over the range of angular separation 3 -
860 milliarcseconds (mas).  A limiting $\Delta$$m_{700}$ $\sim$ 3.5 is likely for
binaries where the component spectral types differ by less than two.  Model fits to
these data show good agreement with published orbits, and we additionally present a
new orbit solution for one of these stars, $\sigma$ Her.  We then discuss early
results of the survey of bright stars at $\delta$ $\geq$ -20$\arcdeg$.  This survey,
which complements previous surveys of the bright stars by speckle interferometry,
initially emphasizes bright stars of spectral types F0 through K2.  We report
observations of 41 stars of apparent visual magnitude $m_V$ $\leq$ 4.30, all having
been observed on multiple nights.  Analysis of these data produces fitted angular
separations, position angles, and component magnitude differences for six previously
known visual binaries.  Three additional systems were examined as possible binaries,
but no conclusive detection could be made.  No evidence of close stellar companions
within our detection limit of $\Delta$$m$ $\approx$ 3 was found for the remaining 32
stars observed; however, uniform-disk angular diameters are reported for 11 of the
resolved stars in this last group.

\end{abstract}


\section{Introduction} \label{intro}

Knowledge of the frequency of multiplicity among stars is fundamental to furthering
our understanding of many areas of astrophysics and has direct impact on the design
of future experiments to detect and image extrasolar planets.  However,
multiplicity surveys of even the brightest stars using modern techniques were
surprisingly incomplete until the 1990s \citep{har00}.  Early surveys of the bright
stars by speckle interferometry \citep{mca87,mca93}, although themselves
incomplete, revealed that the frequency of visual binaries in the range of angular
separations accessible to that technique was several times that previously known,
including substantial numbers of wider binaries missed by classical visual
observers.  These imaging surveys, combined with radial velocity observations and
other techniques, contributed to bright star multiplicity catalogs.  \citet{etv08}
present a catalog of bright, multiple (two or more components) star systems and
discuss the implications of multiplicity upon several topics in astrophysics.
Their catalog is derived from a number of input catalogs and observational
techniques and consists of more than 4,500 stars.  A catalog consisting of bright
systems with three or more stars, the $\mathit{Multiple\ Star\ Catalog}$
\citep[MSC,][]{tok97}, also draws its sources from different observational
techniques.  The MSC is available online \footnote{Catalog J/A+AS/124/75 in the
VizieR catalog access tool \citep{och00}.} and was updated in 2010.

As noted by \citet{rag10}, continued multiplicity survey efforts using speckle
interferometry \citep[e.g.,][]{mas98,hor08} have now resulted in nearly complete
coverage, at least for bright, solar-type stars, down to the diffraction limit of
large telescopes ($\sim$ 30 mas).  These surveys, along with systematic,
higher-precision radial velocity observations have also largely closed the
historical gap in orbital period coverage between these techniques \citep{rag12}.
\citet{mca92} points out, however, that only long-baseline optical interferometry,
using multiple-telescope arrays with baseline lengths of hundreds of meters, offers
a single technique for multiplicity detection throughout the
angular-separation/period range from classical visual doubles to interacting
binaries \citep{hrs09,zav10} and contact binaries.  Optical interferometry not only
provides the data products of speckle interferometry with improved precision, but
at the narrower angular separations not accessible to speckle where most
spectroscopic binaries reside, offers sensitivity for binary detection in a range
of orbital inclinations complementary to spectroscopy.  The combination of visual
orbits from interferometry with spectroscopic orbits for SB2 systems can yield
stellar masses and orbital parallax and, if one or more of the components are
resolved, stellar angular and potentially linear diameters \citep{hum94}.  High
precision mass determinations are potentially possible for SB1 systems as well
should GAIA parallaxes become available for such bright stars.  Improving our
knowledge of stellar multiplicity for both physical and ${\it optical}$ systems of
small angular separation also has important implications for precision navigation,
where the presence of stellar companions and their relative motions can affect the
determination of the ``center of light'' by relatively low-resolution star trackers
\citep{har00}.

We report here the first results of what is anticipated to be an ongoing survey of
the brighter stars using the Navy Precision Optical Interferometer (NPOI).  We
first describe the capabilities of the NPOI for such a survey in \S~\ref{npoi}.
Based on these capabilities (${\it c.}$ 2004), we next discuss the selection of the
targets for the initial survey in \S~\ref{targsel}.  The standard observing
procedures and data reduction are described in \S~\ref{seqred}, including the
selection and observation of calibration stars, and use of the resulting data to
produce the accurately calibrated fringe visibility data for the program stars upon
which all subsequent source modeling depends.  The remainder of \S~\ref{obs},
followed by \S~\ref{modeling}, present the results of models in three areas; first,
the systematic examination of the calibrated data for evidence of binary systems
among the program stars (\S~\ref{bindet}) and subsequent detailed astrometric and
photometric modeling of the detected systems (\S~\ref{binBSS},~\ref{mardet});
second, the observation and modeling of previously known binary systems
(\S~\ref{obslog},~\ref{binknown}), plus detected binaries among the program stars
(\S~\ref{bindet},~\ref{binBSS}), to determine the maximum detected magnitude
difference ($\Delta$$m$) of binary star pairs in our survey as a function of their
angular separation (\S~\ref{limbindet}); and, third, the subsidiary result of
accurate angular diameters for the resolved, single stars among the program sample
(\S~\ref{diamBSS}).  Plans for future stages of this survey are also discussed
(\S~\ref{resdisc}).

\section{The NPOI} \label{npoi}

The NPOI \citep{arm98} located on Anderson Mesa, AZ, is a joint project of the U.
S. Naval Observatory and the Naval Research Laboratory in cooperation with Lowell
Observatory.  A brief description of the instrument and the specific configuration
used in the observations reported in this paper are as follows.

\subsection{Instrumentation} \label{npoides}

\subsubsection{Siderostat Arrays} \label{arrays}

The NPOI includes arrays for both imaging and astrometry.  The four stations of the
astrometric array (AC, AE, AN, and AW) are fixed and feature a laser metrology
system for monitoring the siderostat pivot \citep{hae02}.  Six additional imaging
siderostats are operational at the E03, E06, E07, N03, W04, and W07 stations.  The
resulting baselines range from 9.8 m (AC-W04) to 97.6 m (E07-W07).  Recently
constructed shelters at the E10, N06, N07, and W10 piers will allow the
commissioning of additional baselines of up to 432 m (E10-W10) in the near future
through reconfiguration of the six imaging siderostats.  The unvignetted aperture
is the same for all siderostats (35 cm), but is stopped down to a 12.5 cm diameter
by the feed-system optics.  All stations are equipped with wave-front tip-tilt
correctors.  The light feed system to the beam combining lab is evacuated and
contains remotely actuated mirrors that allow the configuration of light paths from
each of up to six siderostats to a corresponding delay line.  The NPOI has six fast
vacuum delay lines (FDLs) that can add up to 35 m of optical path.  These delay
lines are used to track the atmospheric and sidereal motion of the fringes.  Long
delay lines \citep[LDLs,][]{jhc98}, which use pop-up mirrors to switch in and out
additional path, are being integrated into the feed paths and will enable
observations on all baselines in the original NPOI design, to a maximum baseline of
437 m (E10-N10).

\subsubsection{Beam Combination and Fringe Detection} \label{bcfd}

\begin{figure}
\epsscale{0.75}
\plotone{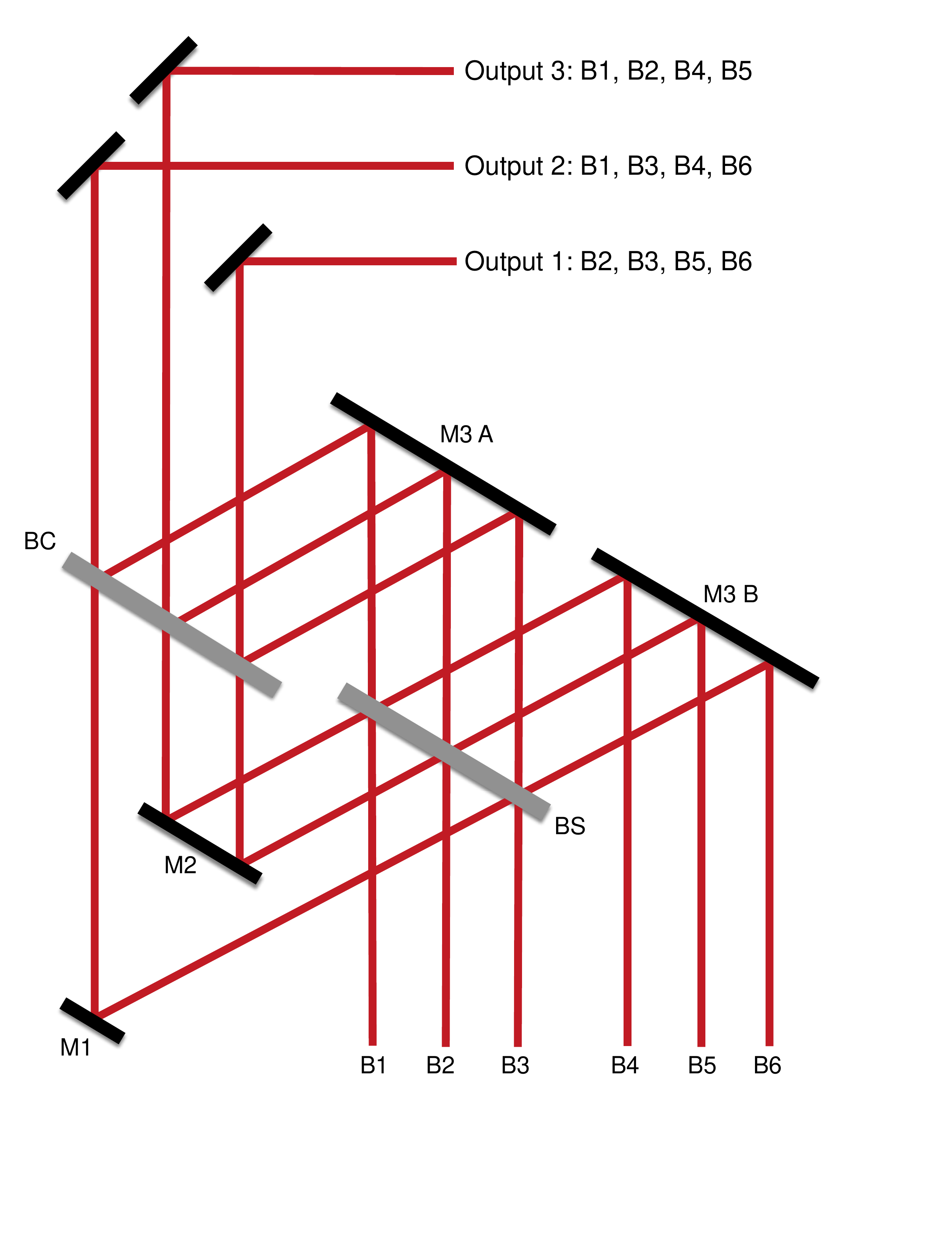}
\caption{A schematic of the NPOI beam combiner.  B1 through B6 are the incoming
beams, BS and BC are beam splitters.  Outputs 1 and 3, and the corresponding
spectrometers, were used for all of the reported observations.  The three
complementary output beams from the other side of BC, which are discarded, are not
shown.}
\label{beamcombiner}
\end{figure}

The beam combiner used for all the observations reported here is a pupil-plane,
free-space, bulk-optics system (Figure~\ref{beamcombiner}).  The original
three-beam combiner used for all observations prior to 2002 was subsequently
expanded into a six-beam combiner simply by adding a mirror, M3B, which injects the
next three beams at the back of the first beam splitter (BS) to combine with the
original three beams.  After interferometric combination at BC, the three combined
beams each contain light from up to four stations.  (The three complimentary output
beams from the other side of BC are discarded.)  Thus, our beam combiner can be
considered a \textit{hybrid} design intermediate between an all-in-one combiner,
which places all beams onto a single detector, and a pairwise combiner, which puts
single pairs of beams on separate detectors \citep{moz94,moz00}.

The combined beams, after spatial filtering using pinholes, pass through prism
spectrometers and are then collimated onto lenslet arrays and detected by three
banks of photon-counting avalanche photodiodes (APDs).  The spectral range covered
is from 450 to 860 nm in 32 channels, equally-spaced in wave number.  In addition
to tracking the fringe motions, each FDL also imposes on its beam a 500 Hz
triangle-wave modulation.  The resulting delay modulation on a baseline is the
difference between two FDL modulations.  The modulation sweeps the fringe pattern
of that baseline across the detectors, causing an intensity that varies
sinusoidally with time.  Changing the amplitude of the modulations (also called
\textit{strokes}) changes the frequencies at which the fringes pass over the
detectors.  Since the three output beams of our hybrid six-beam combiner contain
contributions from up to four input beams, there are a maximum of six baselines
present on each.  There are many stroke amplitude solutions that will place the
baselines at separate frequencies.  As long as no two baselines on the same
detector have the same fringe frequency, the fringes can be cleanly separated
\citep{moz94}.  The NPOI fringe detection algorithms grew out of the work pioneered
by the Mark III stellar interferometer \citep{sha88}.  The path length modulation
and stroke pattern used to separate the fringes were initially laid out in
\citet{sha77}.

Custom electronics and software, described in greater detail elsewhere
\citep{ben98, ben03, hum03}, are required to bin the detected photons in synchrony
with the delay modulation and compute real-time fringe tracking error signals.  A
binner board generates timing signals for 64 bins for up to 32 different wavelength
channels.  Custom APD processor boards use the timing signals to accumulate the
incoming photons into the bins.  Digital signal processors (DSPs) on the APD
boards calculate the real and imaginary components of the Fourier transforms along
the bin direction at eight frequencies.  In addition, the real and imaginary
components of the Fourier transform along the wavelength direction at each of the
eight frequencies are also computed to produce the group delay, which in turn is
used to calculate the fringe tracking error signals.  Our implementation of
group-delay fringe tracking is described in \citet{ben98}.

Observations taken before 2002 used the original pairwise beam combiner.  For
observations after 2002 presented here, only the binner board and one 32-channel
APD processor board were integrated into our fringe engine.  In order to detect and
process six beams with this less-than-complete hardware implementation, we feed the
output from two spectrometers to our single APD processor board.  Sixteen
wavelength channels (550 - 860 nm) from each spectrometer are sent to separate
channels on the 32-channel APD board.  This arrangement enables us to observe 11 of
the 15 baselines that are available with six stations.  One of the 11 baselines
appears on both spectrometers.  The beam combinations entering any two
spectrometers are sufficient to phase the array.  The sixteen detected wavelength
channels are centered near 700 nm, approximating the original Johnson \textit{R}
band \citep{jon66, hin11}.

\subsubsection{Array Phasing and Control} \label{phasing}

Until recently, our array-phasing algorithm was very simple.  A reference FDL
station (AC for most of the observations reported here) and the five baselines
connecting the reference station to the other five FDL stations are designated as
\textit{tracking baselines}.  The fringe tracker then looks only at the frequencies
of the designated tracking baselines to calculate and apply its fringe tracking
error signal.  The error signal is always applied to the non-reference FDL.  The
beam-combiner design, the thermal stability of the beam-combining room, and the
judicious choice of tracking baselines ensure that once all five tracking baselines
are locked on, the array is truly phased up, that is, fringes on all 15 baselines
provided by the six beams are present and constrained by the tracking baselines.

The NPOI control system includes a high degree of operational automation.  The
observer-level control system is based on a graphical user interface that provides
control of the various subsystems, such as the tip-tilt star tracking system, the
FDLs, and the fringe tracking system.  After system set up and alignment, selection
of a target causes the control system to acquire the star simultaneously with all
specified stations.  Once this is accomplished, fringe search begins on all
tracking baselines.  The observer can specify a threshold corresponding to the
minimum fringe amplitude signal-to-noise ratio required before the control system
switches from the fringe-search mode to the fringe-tracking mode.  After all
required fringe data are acquired, a sequence of 2 ms fringe frames is sent to a
data recorder.  The NPOI currently averages $\approx$ 120 multi-baseline
observations (\S~\ref{seqred}) per night to a limiting visual magnitude $m_V$
$\approx$ 5.5 under typical observing conditions and to $m_V$ $\gtrsim$ 6.0 during
periods of excellent seeing.  The NPOI was the world's first long-baseline optical
interferometer to simultaneously co-phase six elements \citep{hum03}.  The wide
detection bandwidth and rapid observing duty cycle of the NPOI make rapid surveys
(\S~\ref{obs}) and snapshot imaging \citep{zav07} practical.

\subsubsection{Wavelength Calibration} \label{channels}

Both binary star models (\S~\ref{binknown},~\ref{binBSS}) and single-star angular
diameter models (\S~\ref{diamBSS}) are fit to calibrated squared visibility
amplitude ($V^2$) data as a function of ${\it u}$ and ${\it v}$, or
${\it uv}$-radius for circularly-symmetric diameters.  Since ${\it u}$ and
${\it v}$ are expressed as a spatial frequency, which is the baseline length
divided by the wavelength, uncertainties in our knowledge of the wavelength scale
of our observations translate linearly into errors in the fitted angular
separations of binary star components, the scale (e.g., semimajor axis) of binary
orbits, and errors in fitted stellar diameters.  The NPOI has had the capability to
operate in a Fourier transform spectrometer (FTS) mode since 2005 March that allows
very accurate, simultaneous measurement of the central wavelengths of all
spectrometer channels.  FTS observations are currently made on a regular basis,
mostly on cloudy nights or after the APD lenslet arrays are translated with respect
to the output of the spectrometers for specialized, spectral line observations
(e.g., H$\alpha$).

Repeated FTS measurements, made during a single night, have shown that the lenslet
arrays can each be positioned to a desired translation with a 1$\sigma$ accuracy of
$\approx$ 0.7 nm (0.1$\%$ at 656 nm).  Repeated FTS measurements made when the
lenslet arrays have not been actively translated for long periods (intervals of
weeks to months, spanning 2012 - 2016) show slow drifts in the wavelength scale on
individual spectrometers of $\approx$ 0.6 nm RMS, again $\approx$ 0.1$\%$ at 656
nm.  This latter result is of particular significance because the great majority of
the observations reported here were made prior to the start of regular FTS
measurements.  Prior to 2005 May an assumed wavelength scale \citep{moz05},
calculated using the as-designed geometry of the spectrometers and the index of
refraction of the glass (BK7) used in the manufacture of the spectrometer prisms,
was used in the modelling of NPOI data.  Measurements of the wavelength scale made
in 1998 (prior to the upgrade of the beam combiner to 6-beam operation) using a
prototype FTS system, along with those made in 2005 March using the current system,
both agreed with the older, calculated scale to within 2 - 3 nm ($\approx$ 0.4$\%$
at 700 nm).  Since there were no significant changes made to any of the
spectrometer optics or mechanics between the epochs of the observations reported
here and the present, we assume that the FTS measurements made after early 2005 are
characteristic of the level of wavelength scale stability at prior times, and that
the 1998 and 2005 measurements are representative of our knowledge of the
wavelength scale pre-2005 as well.  Therefore, we assume that the systematic errors
in our wavelength scale are likely of order $\pm$ 0.5$\%$.

\subsubsection{Interferometric Field of View} \label{ifov}

As discussed in \S~\ref{bcfd}, the beam combiner used for the observations
reported here is a free-space, bulk-optics combiner.  Thus, singe-mode optical
fibers are not used in the beam combination process.  Single-mode fibers are
likewise not used for spatial filtering.  [The 50 $\mu$m diameter pinholes used as
spatial filters in the spectrometers (\S~\ref{bcfd}) serve only to restrict the
\textit{photometric} field of view to a radius of $\approx$ 1.5 arcseconds
\citep{moz05}.]  Therefore, our beam combiner does not suffer from the various
effects \citep{guy03,abs11} that effectively limit the interferometric field of
view of combiners employing single-mode fibers to typically $\leq$ 50 - 100 mas.
One of the most important of these effects derives from the coupling of the input
telescope pupils with the fibers \citep{dac99}.

\begin{figure}
\epsscale{0.6}
\plotone{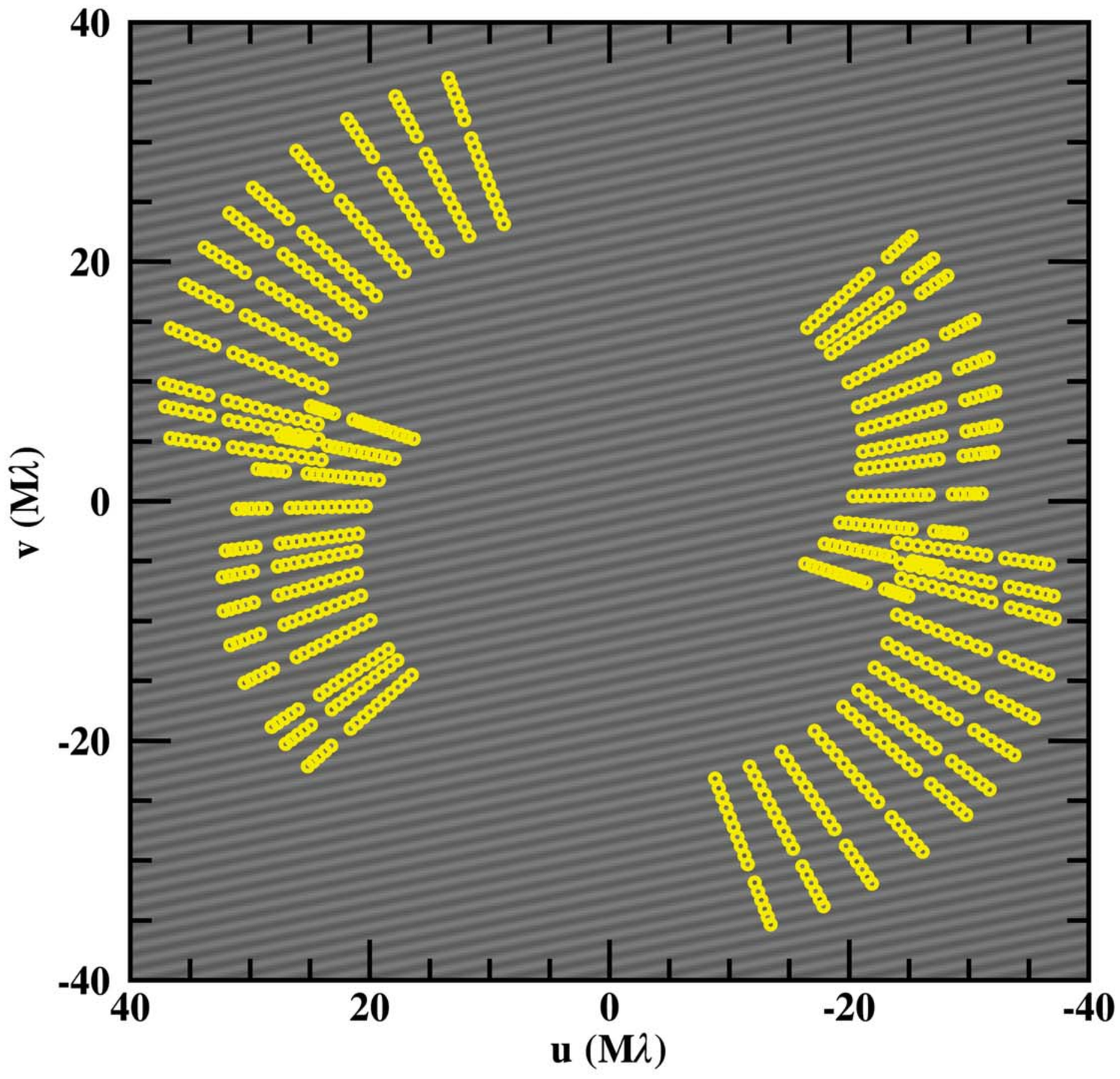}
\epsscale{0.6}
\plotone{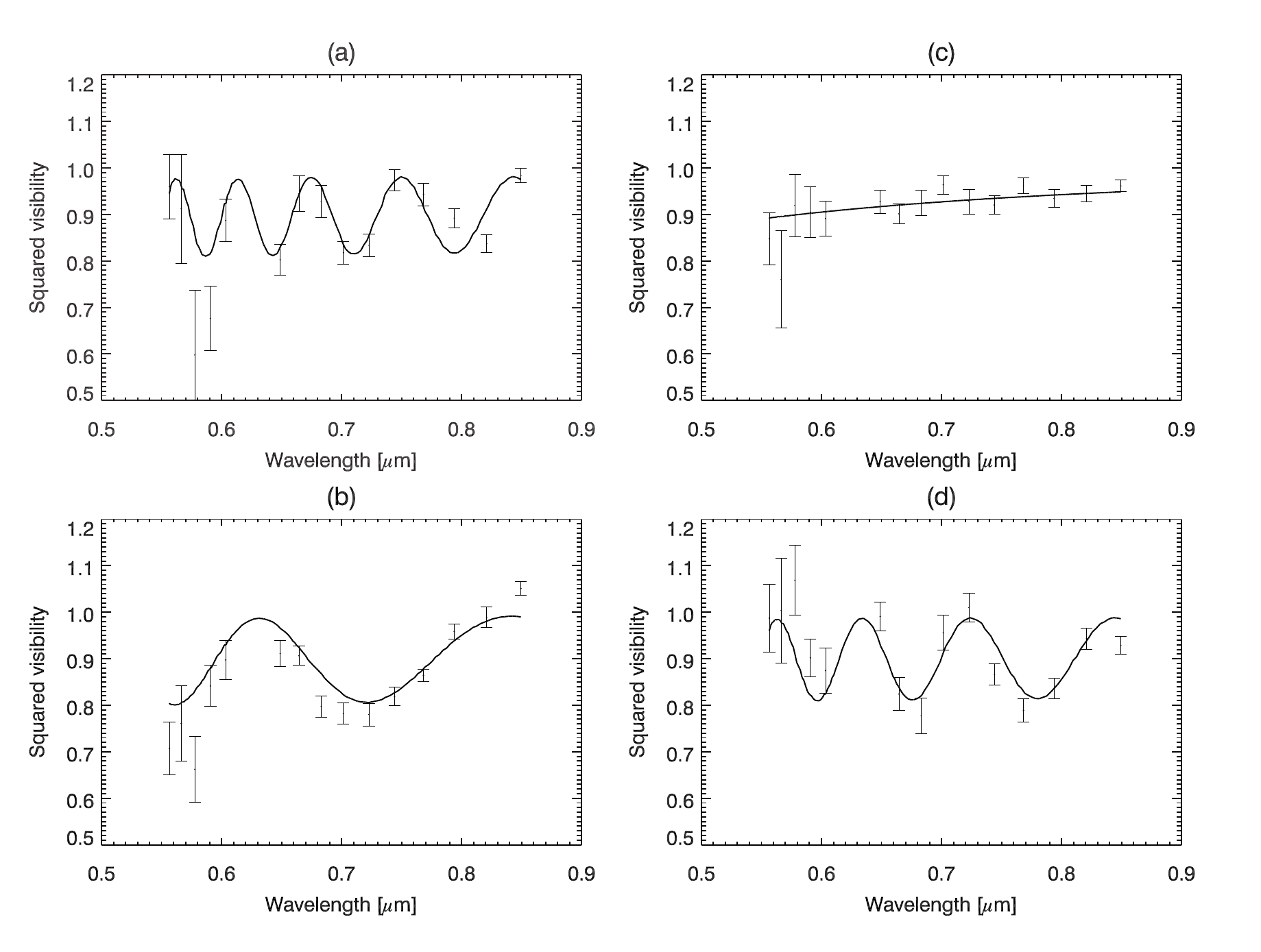}
\caption{Changes in interferometer response with baseline projection: ${\it Upper\
Panel:}$ ${\it uv}$ plot for the 12 observations (yellow circles) of 59 Cyg on 2004
July 30 UT on the AC-AE (lower-left, upper right) and AC-AW (upper left, lower
right) baselines, superimposed on the Fourier transform of the best-fit binary
model for that date (Tables~\ref{knownfit} \&~\ref{knowndelmag}).  The time
sequence of the observations on each baseline proceeds in clockwise order.  ${\it
Lower\ Panel:}$ Plots ${\it a}$ through ${\it d}$ display, for the 2nd, 5th, 7th,
and 11th observation on the AC-AE baseline, respectively, the calibrated $V^2$
${\it vs}$ wavelength with the best fit model for that date ($\rho$ = 173.65 mas,
$\theta$ = 11$\fdg$62, $\Delta$$m_{700}$ = 2.83) overplotted at the time of each
observation (solid lines).}
\label{59cyg-uv-vis-fit}
\end{figure}

\begin{figure}
\epsscale{.80}
\plotone{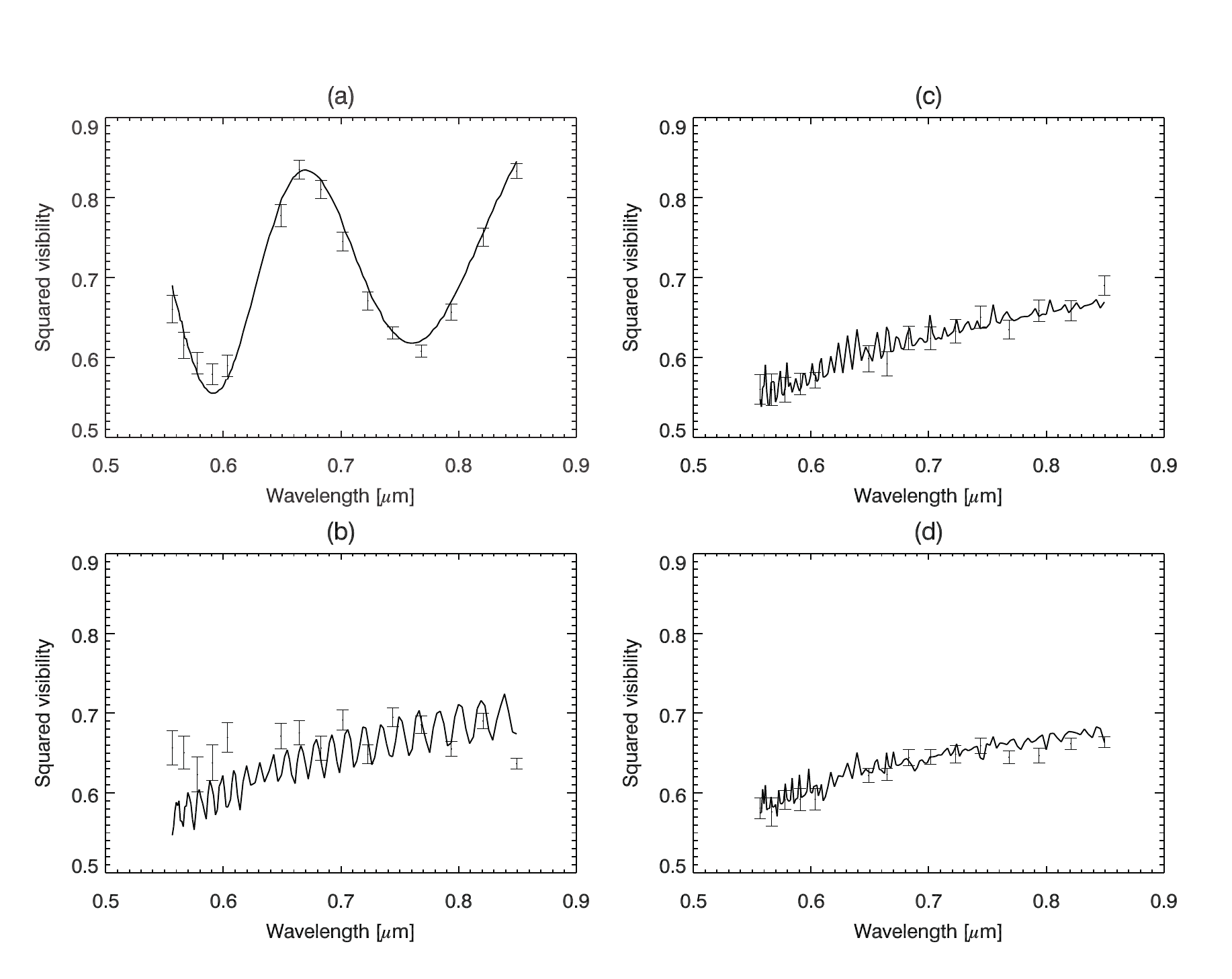}
\caption{Observations and modeling of the very wide binary system $\zeta$ Her:
Using data obtained on 2004 May 13 UT, plots ${\it a}$ through ${\it d}$ display,
for the 1st, 2nd, 4th, and 6th observation on the AC-AE baseline, respectively,
the calibrated $V^2$ ${\it vs}$ wavelength with the best fit model for that date
($\rho$ = 859.06 mas, $\theta$ = 235$\fdg$98, $\Delta$$m_{700}$ = 1.52;
Tables~\ref{programfit} \&~\ref{programdelmag}) overplotted (solid lines).  The
differences in the oscillation period with wavelength between the plots are
analogous to those seen in Figure~\ref{59cyg-uv-vis-fit}, but are more extreme due
to the much wider component separation of $\zeta$ Her.  The differences in the
amplitude of the $V^2$ oscillations between the plots are primarily due to the
varying effects of bandwidth smearing as a function of baseline projection.}
\label{V2-zeta-Her}
\end{figure}

However, the finite bandwidth of our beam combiner's 16 spectral channels reduces
visibility contrast and, as is well known from radio interferometry, will produce
radial smearing over a large interferometric field of view \citep{bas99,tms01}.
In the context of our survey, this is seen as a reduction in the $V^2$ of stellar
companions at large angular separations from the primary star.  The magnitude of
this effect can be estimated, for example, from Eq. 6.76 of \citet{tms01}.  Using
typical values of $\lambda_{0}$ = 700 nm (channel wavelength), $\Delta$$\lambda$ =
21 nm (channel width), and D = 18.9 m (baseline length, \S~\ref{config}), the
reduction in the $V^2$ of a secondary star at 30 mas separation would be $\approx$
1$\%$, but grows to $\approx$ 10$\%$ at 100 mas, and to $\approx$ 90$\%$ at 860
mas, the extreme limit of detection with the NPOI (\S~\ref{zether},
Figure~\ref{V2-zeta-Her}).  The effects of bandwidth smearing are explicitly
accounted for in the modelling software used to search for stellar companions in
our survey data (GRIDFIT, \S~\ref{binknown}) and for detailed modelling of detected
binary systems (OYSTER, \S~\ref{binknown}).  A monochromatic formula for a binary
observed with an optical interferometer was presented in Eq. 17 of \citet{hbd67b}.
Formulae for model fitting to visibility data for relative astrometry and magnitude
differences of binary stars observed with rectangular bandpasses are provided in
Eqs. 5, 6, and 7 of \citet{pan92}.  In practice, our modelling software computes
complex visibilities on a fine grid of wavelengths and then averages these over
each spectral channel (Figures~\ref{59cyg-uv-vis-fit},~\ref{V2-zeta-Her}).

\subsection{System Configuration} \label{config}

In anticipation of the very large number of observations and subsequent data
reduction effort required in the initial (\S~\ref{obslog}) and subsequent
installments of this survey, we chose one of the simplest configurations of the
NPOI, having the advantages of maximum sky coverage, simple and robust operation
of the instrument (ensuring the maximum number of observations each night), and
relatively easy fringe visibility calibration.

\begin{figure}
\epsscale{1.0}
\plotone{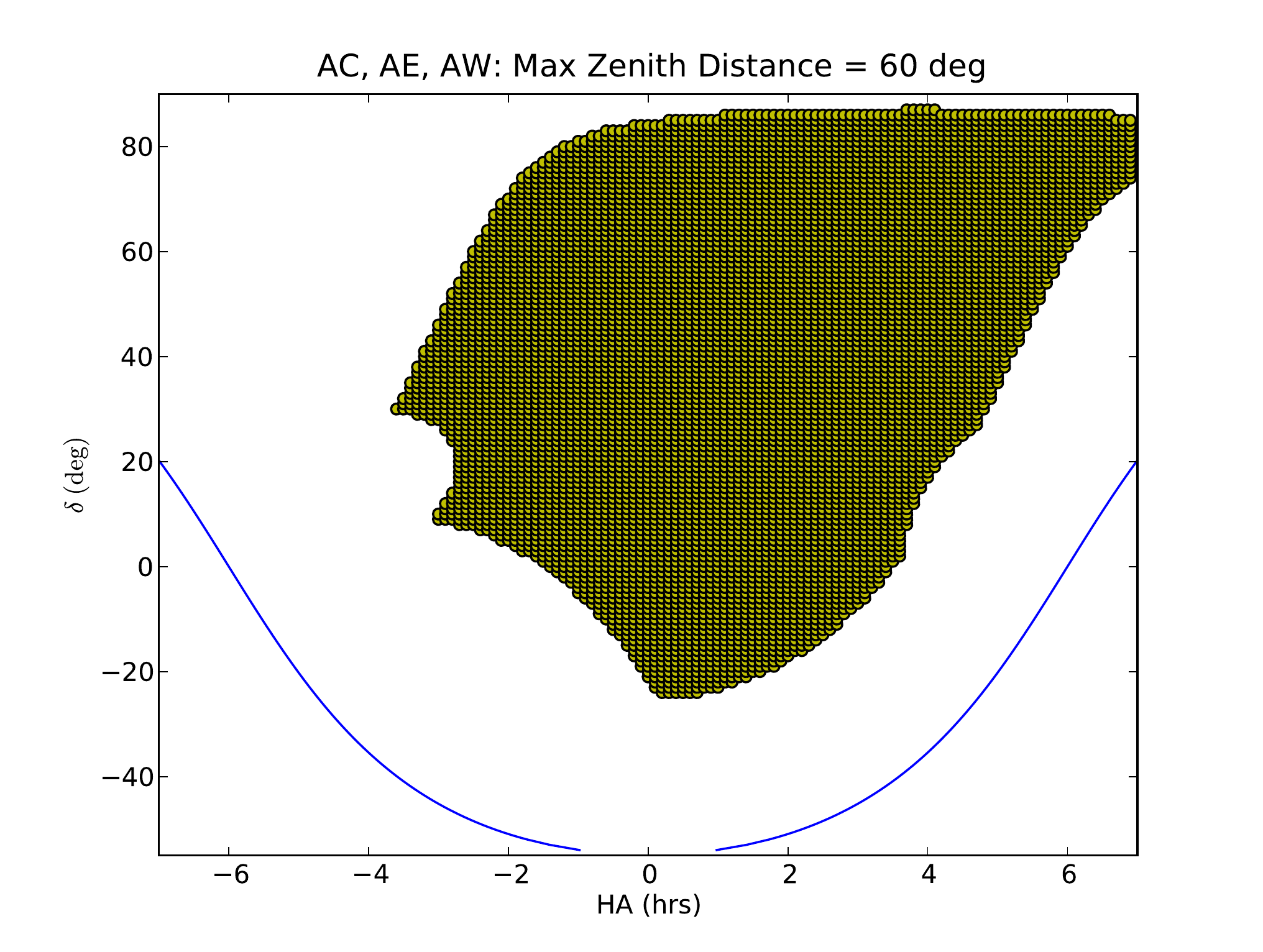}
\caption{Sky coverage plot for the NPOI bright star survey observations utilizing
the Center (AC), East (AE), and West (AW) stations of the astrometric array.  The
blue, solid curve represents the horizon, and the crosshatched area represents the
observable sky.}
\label{sky}
\end{figure}

\begin{figure}
\epsscale{.30}
\plotone{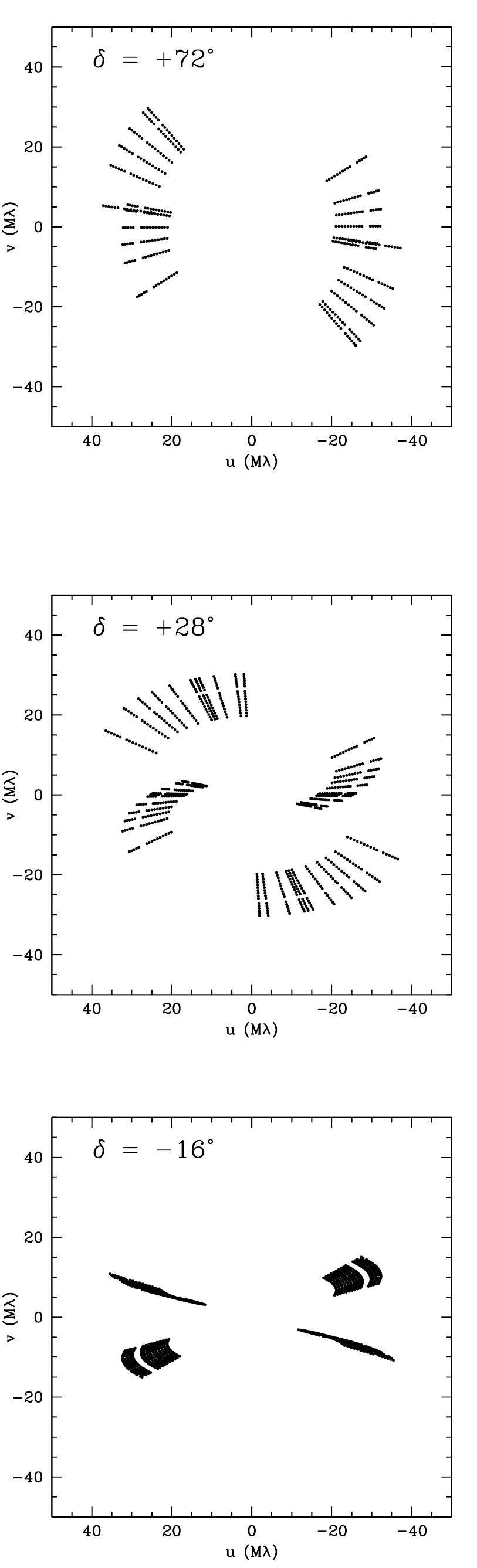}
\caption{Example ${\it uv}$ coverage plots for the NPOI bright star survey
observations on the AC-AE and AC-AW baselines at three different declinations.  The
declinations shown are for the northernmost ($\chi$ Dra), a midrange ($\beta$ Gem),
and southernmost ($\eta$ Crv) stars from Table~\ref{targetlist}.  The ${\it uv}$
points represent actual interferometric observations of these stars.  The ${\it u}$
and ${\it v}$ spatial frequency axes are in mega-wavelengths.  Each baseline
produces data from 16 spectral/spatial channels and thus results in a radial ray in
the ${\it uv}$-diagram for each observation.}
\label{uv}
\end{figure}

First, while as noted above (\S~\ref{bcfd}), the NPOI is capable of coherently
combining light from as many as six stations, we chose to use a subset of the
available stations for our initial survey.  Unless otherwise noted
(\S~\ref{obslog}), all observations were made utilizing the Center (AC), East (AE),
and West (AW) stations of the astrometric array corresponding to baselines of
lengths 18.9 m (AC-AE), and 22.2 m (AC-AW), with azimuths of 113$\arcdeg$ and
244$\arcdeg$, respectively, for the East and West stations relative to the Center
station.  Figure~\ref{sky} shows the sky coverage for this array, while plots of
the typical ${\it uv}$-plane coverage for a night's observation of program stars
(\S~\ref{targsel}) are shown in Figure~\ref{uv} for the AC-AE and AC-AW baselines.

Second, unless otherwise noted (\S~\ref{obslog}), the transport optics of the
starlight feed system were configured to place each of the beams from the AC, AE,
and AW stations at a separate input to the beam combiner (\S~\ref{bcfd}) such that
the interferometric combination of the light for each of the three baselines
appeared uniquely and simultaneously at each of the three beam combiner outputs.
Due to the limited fringe processing electronics available at the epoch of most of
the observations reported in this paper (2004, \S~\ref{bcfd}), only two of the
outputs, corresponding to the AC-AE and AC-AW baselines, respectively, were
utilized for the great majority of the observations.  This arrangement
significantly simplifies the calibration of the raw fringe data relative to other
configurations that produce multiple superimposed fringe patterns at one or more
beam combiner outputs.  (The third beam combiner output, typically containing the
AE-AW baseline, was utilized for observations in 1997 through 2001
\S~\ref{sigher}.)

\section{Program Star Selection} \label{targsel}

For the purposes of our initial survey, the target list (``Program Stars,'' see
Table~\ref{targetlist}) was limited to a small subset of the stars in the
$\mathit{Hipparcos\ Catalogue}$ \citep{per97}.  Specifically, a spreadsheet
chartered by the Terrestrial Planet Finder Interferometer Science Working Group
\citep[TPF-I SWG;][]{law07,tur04}, was used as a basis for our source selection.
This list includes all ${\it Hipparcos}$ stars within 30 pc [2,350 sources, per
the original ${\it Hipparcos}$ reduction \citep{esa97}], augmented with data on
age indicators, kinematics, and spectral type, plus data from the
$\mathit{Washington\ Double\ Star}$ \citep[WDS;][]{mas01}, $\mathit{2MASS}$
\citep{skr06}, and $\mathit{IRAS}$ \citep{neg84} catalogs.  This list was
systematically culled using four criteria designed to down-select a modest list
of sources that could be practically observed given the capabilities of the NPOI
in 2004:

1) The list was first culled to eliminate stars that were too faint for practical
observation with the NPOI in 2004.  The stars selected (148 sources) were required
to have an apparent visual magnitude in the Johnson system [550 nm effective
wavelength \citep{jon66}] $m_V$ $\leq$ 4.30 (Table~\ref{targetlist}, Col. 5).

2) Next, the list was culled to include only stars with ${\bv}$ $\geq$ 0.30
(Table~\ref{targetlist}, Col. 6; 98 sources).  This eliminated all stars with
spectral types earlier than approximately F0.  This criterion was originally
adopted in consistency with the TPF-I SWG, but also eliminated overlap with other
current NPOI multiplicity studies of A and B stars \citep{pat12,rdr14}.

3) The third cut was made to include only stars with declinations $\delta$ $\geq$
-20$\arcdeg$ (61 sources).  This declination cutoff limits the sample to only those
stars for which NPOI observations over a reasonable range of hour angles can be
obtained during a single night ($\approx$ 2 hours, Figure~\ref{sky}) using the
array configuration described in \S~\ref{config}.

4) Lastly, the two reddest stars in the remaining list [HR 5340, $\alpha$
Bo\"{o}tis; $\bv$ = 1.24 \citep{moz03}, and HR 1457, $\alpha$ Tauri; $\bv$ = 1.54
\citep{waf87}] were dropped since their angular diameters ($\approx$ 20 mas) are
too ${\it large}$ for practical observation with the baselines available in 2004,
due to very low interferometric fringe contrast.  This last cull has the effect of
setting a red cutoff to the list at $\bv$ = 1.20 (mid-K spectral type).

\begin{figure}
\epsscale{1.0}
\plotone{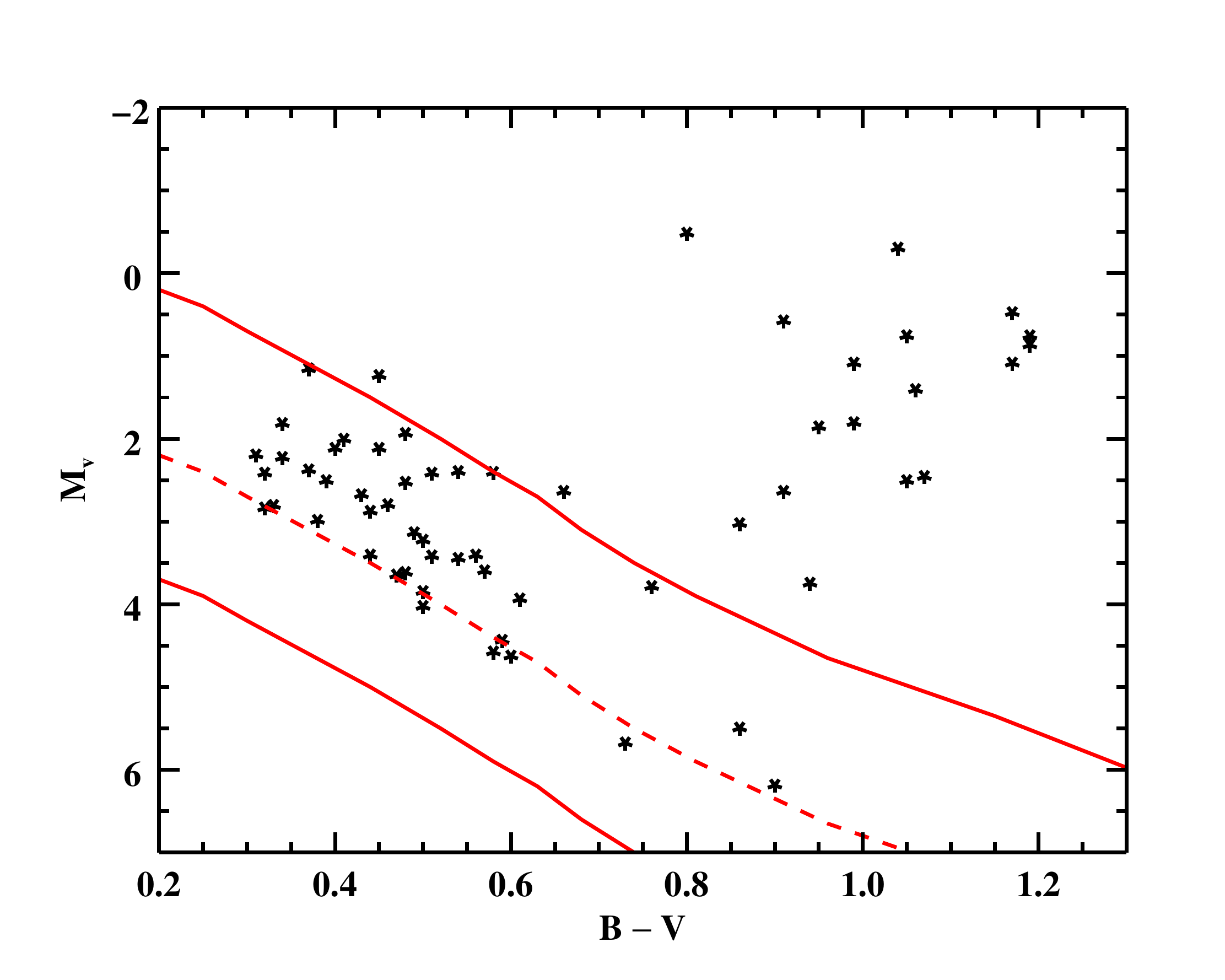}
\caption{Color-Magnitude diagram of the stars in the program list
(\S~\ref{targsel}).  The dashed red line is the Main Sequence \citep{sck82}, while
the solid red lines at -2 and +1.5 magnitudes with respect to this curve represent
nominal limits for Main Sequence stars for the purposes of the adopted program star
list.}
\label{color-mag-bss-with-Main-Seq}
\end{figure}

Thus, the final program star list (Table~\ref{targetlist}) consists of 59 stars.
HR \citep{hof91} and FK5 \citep{frk88} designations are included in this and
subsequent tables for consistency with the NPOI observation planning and data
archiving software.  A color-magnitude diagram for these sources is given in
Figure~\ref{color-mag-bss-with-Main-Seq}.  Per \citet{rag10}, we adopt absolute
magnitude limits of -2 and +1.5 with respect to the \citet{sck82} Main Sequence as
the nominal range of Main Sequence stars.  We will next discuss some statistical
properties of various subsets of our sample, referencing
Figure~\ref{color-mag-bss-with-Main-Seq}.

First, assuming that the line at -2 mag above the nominal Main Sequence in
Figure~\ref{color-mag-bss-with-Main-Seq} is the threshold above which stars are
significantly evolved, there are 19 giants and sub-giants in our sample.  By
comparison, if one were to apply the same color and declination selection
criteria as used in selecting our sample but increased the apparent magnitude
cutoff to $m_V$ = 7.3, the limit to which the $\mathit{Hipparcos\ Catalogue}$ is
complete for all spectral types and galactic latitudes \citep{rag10}, the resulting
list would contain 24 stars above the Main Sequence limit.  Of these, 13 are within
22 pc, the distance at which $m_V$ = 7.3 corresponds to the upper demarcation of
the Main Sequence at $\bv$ = 1.20 ($M_V$ = 5.6;
Figure~\ref{color-mag-bss-with-Main-Seq}).  Of the 19 giants/sub-giants in our
sample, 12 are within the same distance cutoff.  Therefore, even with the modest
apparent magnitude limit of our initial program star list, we are capturing $\sim$
90$\%$ (12/13) of all the (same) evolved stars in a complete, volume limited sample
within 22 pc.

Next, we considered the stars in our sample within the Main Sequence bounds of
Figure~\ref{color-mag-bss-with-Main-Seq}.  Given the modest magnitude limit of
our initial sample, it is not surprising that a similar analysis shows we sample
a much lower percentage of the Main Sequence stars.  This analysis proceeds in
two parts.

First, examination of Figure~\ref{color-mag-bss-with-Main-Seq} shows that a
substantial fraction of our sample (35 stars) lie within our adopted Main Sequence
boundaries at ${\bv}$ $\leq$ 0.60.  Of these stars, 34 lie within 25 pc, the
distance to which \citet{rag10} demonstrated the completeness of the
$\mathit{Hipparcos\ Catalogue}$ for Main Sequence stars in an overlapping color
range.  There are 122 stars in the $\mathit{Hipparcos\ Catalogue}$ within the same
color, Main Sequence, declination, and distance bounds, and within the completeness
limits in apparent magnitude ($m_V$ = 7.3).  Therefore, we are likely sampling only
34/122 $\approx$ 28$\%$ of the complete, volume-limited ${\it Hipparcos}$ sample in
this part of the Main Sequence.

Second, our sample contains only five stars within our Main Sequence bounds at
$\bv$ $>$ 0.6, and none redder than $\bv$ = 0.9.  The corresponding number of stars
in the $\mathit{Hipparcos\ Catalogue}$ to $m_V$ = 7.3, within the same color, Main
Sequence, declination, and distance range is 153, or 18 to the distance (10.5 pc)
of the most distant of our five stars.  Therefore, we are sampling less than a
third of the ${\it Hipparcos}$ sources to 10.5 pc and only $\approx$ 3$\%$ of the
complete, volume-limited ${\it Hipparcos}$ sample of Main Sequence stars in this
color range.

\section{Observations and Data Reduction} \label{obs}

Here we describe the observing sequence and aspects of the data reduction
(\S~\ref{seqred}), and summarize the log of observations (\S~\ref{obslog}).

\subsection{Observation Sequence and Reduction} \label{seqred}

\begin{figure}
\epsscale{1.0}
\plotone{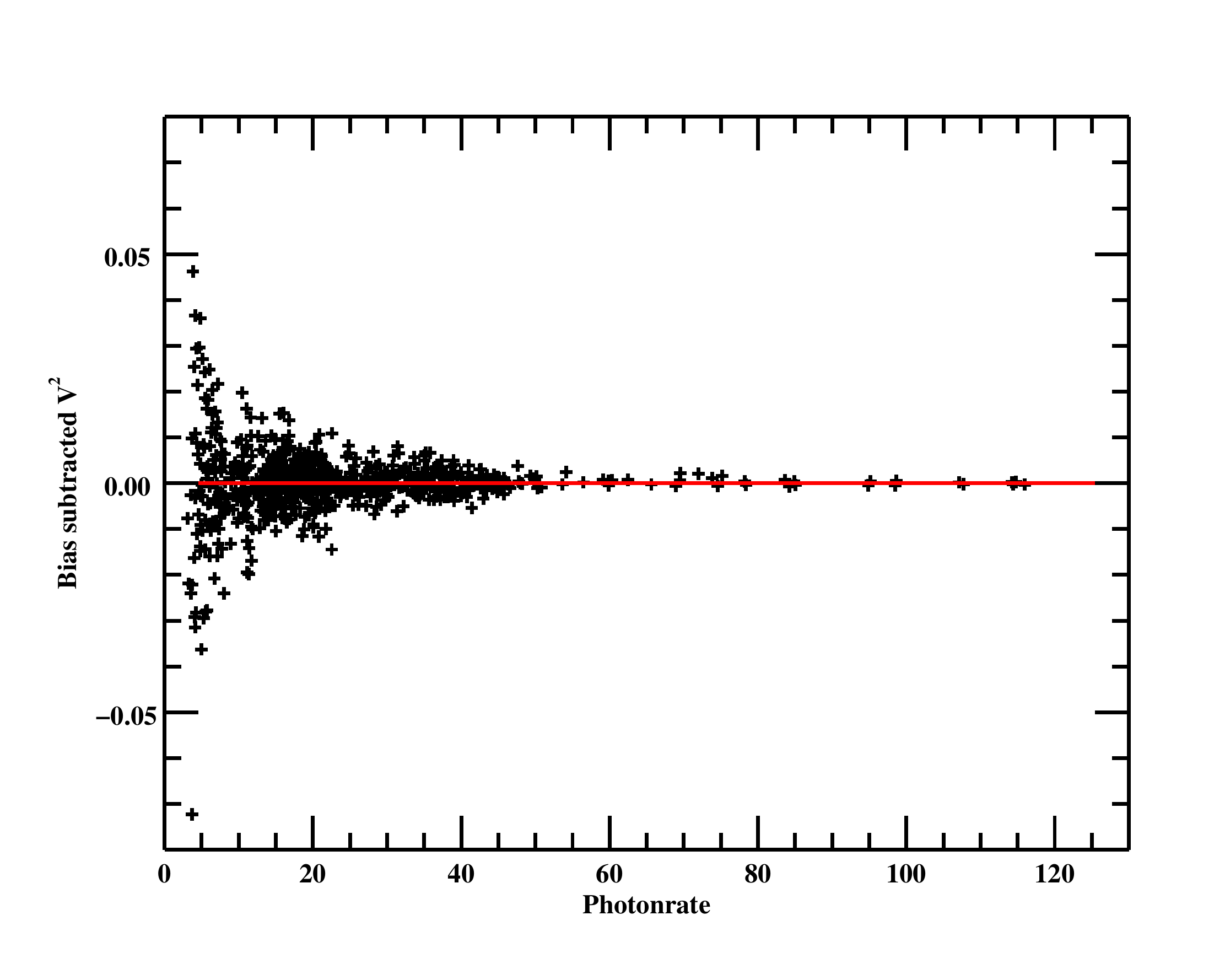}
\caption{The incoherent $V^2$ ${\it vs}$ scan averaged photon rate as measured
after the bias subtraction has been performed for the NPOI spectrometer 3, for
observations made on 2004 May 10 UT.}
\label{incoherentAfterBias}
\end{figure}

Each observation of a program star consisted of 30 seconds of continuous (or nearly
continuous) coherent fringe measurements, sampled at 500 Hz (hereafter, an
${\it observation}$).  Each coherent observation of a star was immediately followed
by an ${\it incoherent}$ observation where the delay lines were deliberately moved
so as to collect the total flux on each baseline outside the fringe packet.
${\it Background}$ observations were also obtained while pointing the siderostats
off-target.  The incoherent and background observations are both utilized in the
data reduction process to perform the photon noise bias correction in the
calculation of the squared visibility amplitude $V^2$ \citep[e.g., ][]{hum03}.
After subtraction of the bias using the incoherent observations, we would expect
the incoherent $V^2$ to have a mean of zero.  Examination of the incoherent
observations after the bias correction confirms this expectation
(Figure~\ref{incoherentAfterBias}).

Because the raw visibility amplitudes are degraded by atmospheric turbulence and
instrumental effects, observations of calibrator stars, weakly resolved or
unresolved stars, were interleaved with those of the program stars.  Since the
theoretical response of an interferometer to an unresolved source is known, the
observed visibilities of the calibrator stars can be used to normalize the
visibilities of the program stars.  Once normalized, the program star visibilities
can be reliably compared to single or multiple source models.  Basic data on each
of the calibrator stars are listed in Table~\ref{calibs}.  All calibrators were
chosen to be within 15$\arcdeg$ of their respective program stars and similar in
magnitude, where  possible.  Frequently, several program stars shared a common
calibrator.  In most cases, the adopted limb darkened angular diameters
($\theta_{LD}$) for the calibrator stars (Table~\ref{calibs}, Col. 8) were
estimated from the photometric properties of the stars ($m_V$ and ${\it V - K}$;
Cols. 5 and 6, respectively) using Eqs. 5 and 7 of \citet{moz03}, while for five
stars (Table~\ref{calibs}, note ``b'') diameters estimated from the infrared flux
method \citep{blk91} were used.  For one star (HIP 26311, $\epsilon$ Orionis;
Table~\ref{calibs}, note ``a''), a measured uniform disk diameter
\citep[$\theta_{UD}$;] [] {moz91}, corrected to a $\theta_{LD}$ via Eq. 5 of
\citet{hbd74}, was used.  The methodology for the determination of the
limb-darkening coefficients is discussed below.

To estimate the uncertainty of our estimated calibrator diameters, we compared the
values in Table~\ref{calibs} with those found in the literature from the
$\mathit{JMMC\ Stellar\ Diameters\ Catalogue}$
\citep[JSDC;][26 stars in common]{laf10} and with the
$\mathit{PTI\ Calibrator\ Catalog}$ \citep[14 stars in common]{gvb08}.  In the
former case, estimated $\theta_{LD}$ were also computed from $m_V$ and
${\it V - K}$ using Eqs. 9 and 10 of \citet{bon06}, while in the latter case,
spectral energy distribution (SED) fitting was used.

In the case of the comparison with the JSDC angular diameters (average quoted
errors of 7\% for the 26 in-common stars), our values are systematically
${\it larger}$ with the exception of two stars, the average difference being 0.067
$\pm$ 0.049 mas (approximately 11\% at the mean JSDC diameter of 0.606 mas).
However, this discrepancy can be accounted for by the difference in the surface
brightness ${\it vs}$ ${\it V - K}$ relations of \citet{moz03} and \citet{bon06}.
If, for example, one calculates the predicted angular diameters over the range of
${\it V - K}$ of the 26 stars common with the JSDC (approximately -0.7 to 1.2),
assuming their average $m_V$ = 3.83, using both the \citet{moz03} and \citet{bon06}
equations, one finds that \citet{bon06} predicts progressively smaller angular
diameters (as ${\it V - K}$ is decreased) with respect to \citet{moz03} for
${\it V - K}$ $\lesssim$ 0.5, the range of ${\it V - K}$ that includes 23 of the 26
stars common with the JSDC.

In the case of the comparison with the PTI angular diameters (average quoted error
of 9\% for the 14 in-common stars), our values are again systematically larger, but
the average difference is only 0.037 $\pm$ 0.081 mas (approximately 6\% at the mean
PTI diameter of 0.593 mas).  Given the different methodology used in the estimation
of the PTI angular diameters, the somewhat better agreement with our estimates is
reassuring.

Based on these comparisons, we conservatively estimate our calibrator angular
diameters to be good to $\pm$ 10\% for purposes of estimating the effects of
visibility calibration uncertainty in our later analysis (e.g., \S~\ref{diamBSS}).

The calibration of NPOI visibility data utilizes the calibrator stars as follows:
The spectral types from the $\mathit{The\ Bright\ Star\ Catalogue}$
\citep[][Table~\ref{calibs}, Col. 7]{hof91} are utilized to obtain
$\mathrm{T_{eff}}$ and log ${\it g}$ for each calibrator from the appropriate
tables of \citet{sck82} and \citet{str92}, respectively.  These quantities are then
used with the tables of monochromatic limb-darkening coefficients of \citet{vhm93},
via a quadratic interpolation in the range 450 - 860 nm, to determine the value of
the linear limb-darkening coefficient at the center wavelength of each NPOI
spectral channel.  The expected value for the squared visibility amplitude of the
calibrator at the time of each observation is then calculated via Eq. 4 of
\citet{hbd74}.  Comparison of the expected and observed squared visibility
amplitudes for the calibrator were used to generate the multiplicative correction
for the instrumental and atmospheric reduction of $V^2$ for the program star.

\begin{figure}
\epsscale{1.0}
\plotone{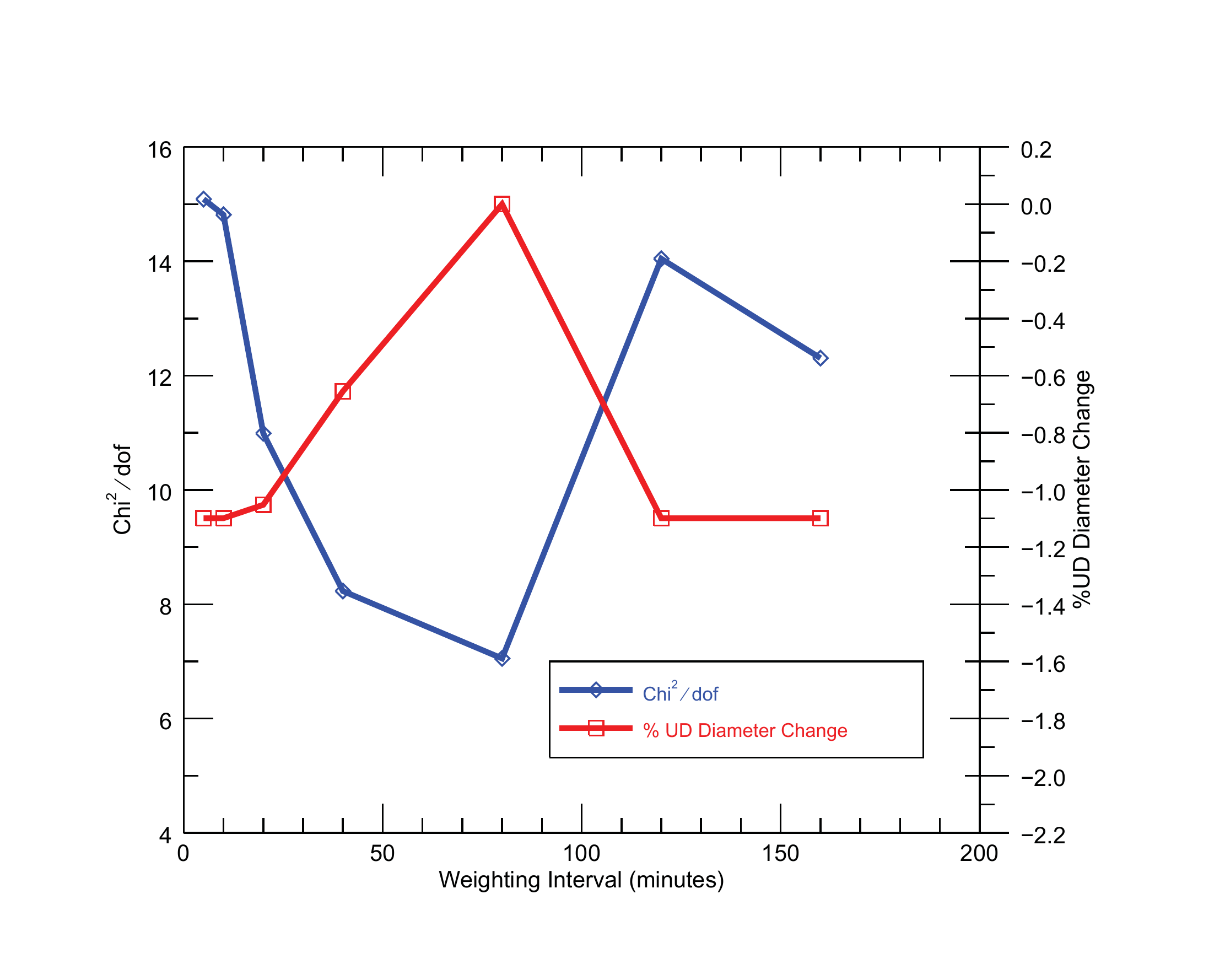}
\caption{The goodness of fit of the diameter model (blue curve) and the resulting
percentage change in the fitted angular diameter (red curve) are both functions of
the time weighting of the calibrator star observations in the calibration of the
program star interferometric visibilities.  Data plotted are for $\beta$ Ophiuchi
on 2004 May 03 UT.}
\label{windowwidth}
\end{figure}

Additional details of the reduction of the visibility data are described in
\citet{hum98}.  As noted there, the seeing conditions and the squared visibility
amplitudes of the calibrators vary significantly during a night, often on the time
scale of an hour or less.  Standard reduction procedure of NPOI data includes
smoothing the systematic variations in the squared visibilities with time in the
calibrator for each program star with a Gaussian weighting function, whose FWHM,
the ``weighting interval,'' is expressed in minutes.  However, the exact value of
the weighting interval to use is not obvious and the choice can significantly
affect the results of subsequent modeling of the calibrated squared visibilities
(e.g., \S~\ref{diamBSS}).  Extensive empirical experimentation was performed to
determine the best overall value for the weighting interval: The data for each of
eight resolved program stars of various angular diameters were repeatedly
calibrated for a number of values of the weighting function on each of several
nights.  The calibrated data for each star, on each night, were then fit with a
uniform angular disk model.  The values of the fitted diameters and the goodness of
fit ($\chi^{2}$) were seen to change significantly with choice of the weighting
interval (e.g., Figure~\ref{windowwidth}).  However, comparison of such results for
the various stars and nights showed a weighting interval of approximately 80
minutes to be an optimal overall value in terms of producing the smallest variation
in the fitted diameters relative to those individually optimized with respect to
weighting interval on a nightly basis.  Adopting a standard value for the
calibration weighting interval significantly reduced the labor involved in
visibility calibration for all the nights at the expense of introducing a
systematic error of perhaps $\pm$ 1\% in the angular diameter fits for resolved
stars (\S~\ref{diamBSS}).

\subsection{Observation Logs} \label{obslog}

\begin{figure}
\epsscale{1.0}
\plotone{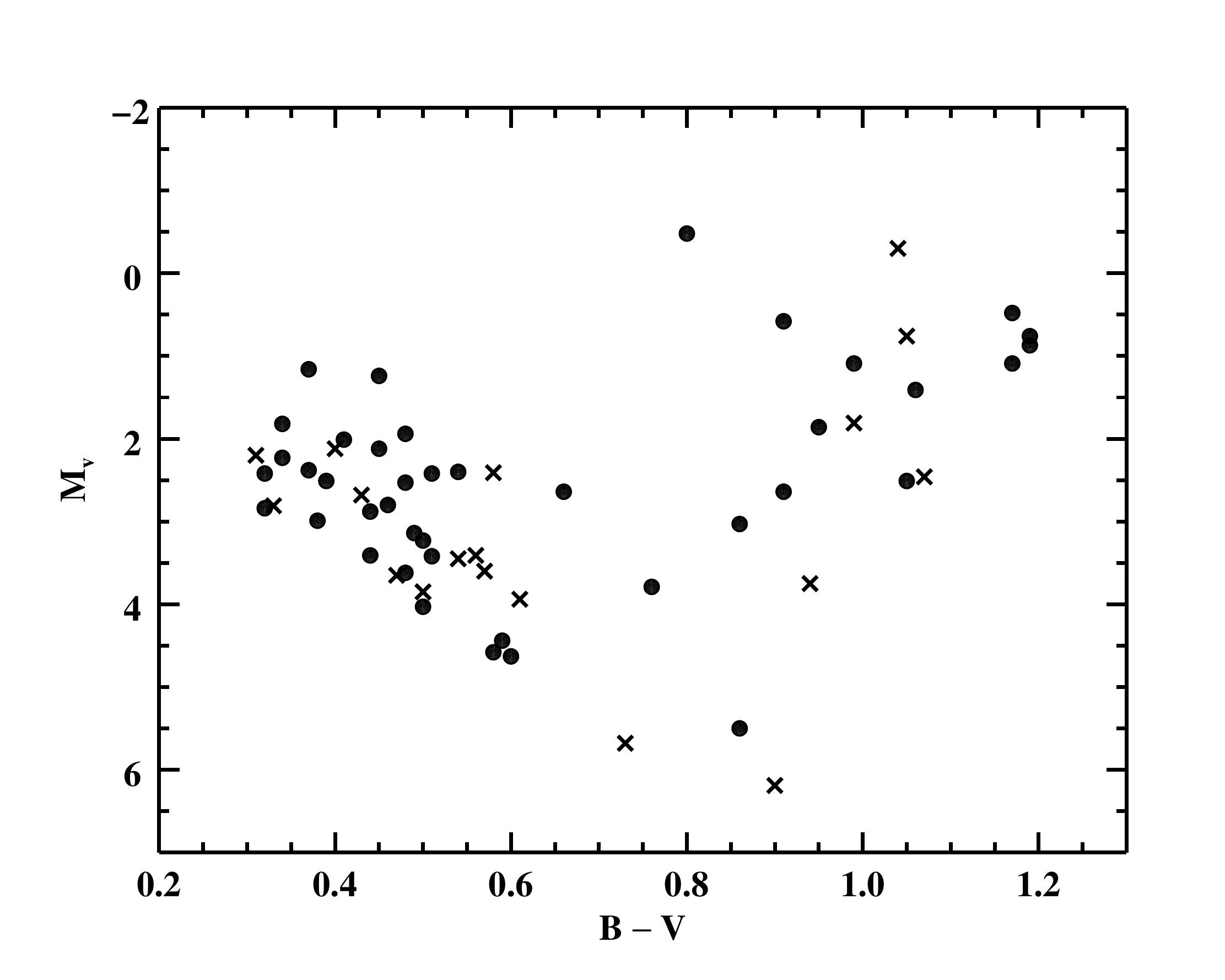}
\caption{Color-Magnitude diagram of the stars in the program star list showing
sources observed to date (filled circles) and not yet observed (crosses).}
\label{color-mag-observed-and-not}
\end{figure}

The observations of the program stars discussed here were obtained at the NPOI over
46 nights from 2004 March 10 UT through 2004 October 08 UT
(Table~\ref{progstarsobs}, with stars listed in the same order as in
Table~\ref{targetlist}).  A total of 1,389 coherent, multi-baseline observations
were obtained for 41 of the 59 program stars
(Figure~\ref{color-mag-observed-and-not}).  Program stars were typically each
observed on five nights, with all the stars reported here having at least two
nights of observations.  Table~\ref{progstarsobs} lists the UT dates of the
observations, the number of multi-baseline observations on each night, the total
number of $V^2$ measurements on each night, and the calibrator star observed.

Additionally, known, ${\it non-program}$ binary systems (\S~\ref{binknown})
covering a wide range of component $\Delta$$m$ and angular separation were
observed on most nights.  These systems were observed as ``test'' binaries in the
sense that their observations were to be reduced and analyzed in the same manner as
the program stars to determine the maximum $\Delta$$m$ detectible by NPOI
observations at various component angular separations.  A total of 705
multi-baseline observations of 15 such binaries was obtained (including
observations obtained on additional nights during 1997 May - 2001 May, 2004 June -
August, 2009 November - December, and 2013 March - May).  Basic data for these
systems are presented in Table~\ref{known}, and a summary of the nights of
observation and observation totals are presented in
Tables~\ref{binobs} \&~\ref{sigHerobs}.

\section{Data Modeling} \label{modeling}

After the data reduction procedures were completed, we systematically examined all
of the calibrated visibility data for both our program stars (\S~\ref{targsel},
Table~\ref{progstarsobs}) and the $\Delta$$m$ test binaries that were not part of
the program sample (\S~\ref{obslog}, Table~\ref{known}) for the possible sinusoidal
signature of a binary star \citep[][\S~\ref{binknown},~\ref{bindet}]{brn74}.  For
those systems where a statistically significant binary detection was obtained, we
proceeded to detailed modeling of the separation ($\rho$), position angle
($\theta$), and magnitude difference ($\Delta$$m$) of the binary components
(\S~\ref{binknown},~\S~\ref{binBSS}).  Lastly, we also performed uniform-disk,
angular diameter model fits to those resolved program stars that proved to have no
detectable stellar companions (\S~\ref{diamBSS}).

\subsection{Modeling of $\Delta$$m$ Test Binaries} \label{binknown}

\begin{figure}
\epsscale{.80}
\plotone{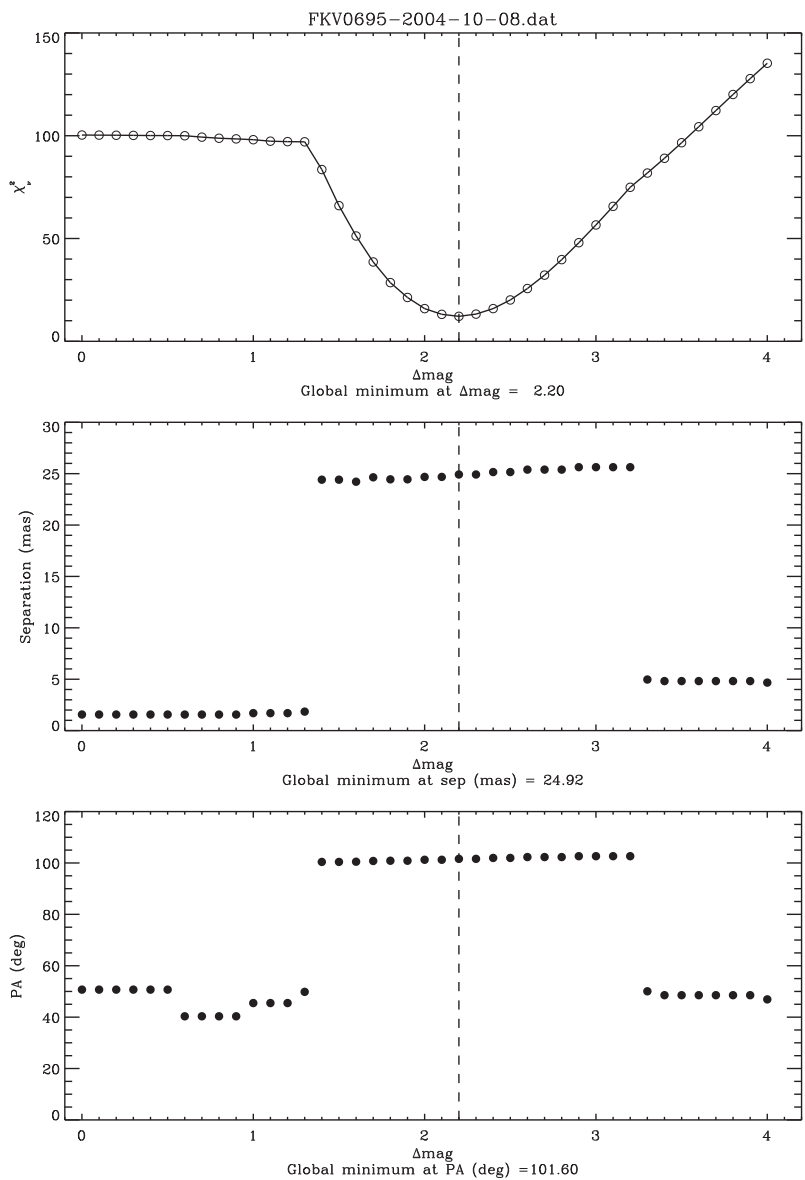}
\caption{Output of the GRIDFIT program for observations of $\chi$ Dra on 2004
October 8 UT.  The top panel shows the minimum value of the reduced $\chi$$^2$ for
grid searches over separation and position angle at successive, fixed component
magnitude differences ($\Delta$$m$), in 0.1 magnitude increments, over the range 0
to 4 magnitudes.  The second and third panels display the corresponding values of
component separation and position angle for the best-fit model at each $\Delta$$m$
value.  The dashed vertical lines correspond to the global $\chi$$^2$ minimum.}
\label{grid-fit-chi-dra}
\end{figure}

When embarking on a survey of stellar multiplicity one naturally would want to know
the expected detection limits of that survey in angular separation and $\Delta$$m$.
An empirical assessment of the NPOI's capabilities was performed by observing a
number of known binary systems, including program stars, with estimated $\Delta$$m$
values $\lesssim$ 4 in $m_V$ (hereafter $\Delta$$m_V$) over a range of angular
separations between 3 mas and 860 mas (\S~\ref{obslog}).  Once the reduction of the
observations to calibrated $V^2$ data was completed (\S~\ref{seqred}), we next
proceeded to systematically examine the calibrated visibility data for all of the
sources observed to date (Tables~\ref{progstarsobs} \&~\ref{binobs}) for
statistically significant evidence of a stellar companion using the software
GRIDFIT, written by R. T. Zavala.  This procedure performs a gridded search of the
$\Delta$$m$, $\rho$, $\theta$ space for evidence of a stellar companion by locating
the values of these parameters for which a binary star model, fit to all the
calibrated $V^2$ data for a source on a given night, produces a minimum in the
reduced $\chi$$^2$ ($\chi_{\nu}$$^2$).  Beginning at a fixed $\Delta$$m$ = 0,
GRIDFIT calculates the value of $\chi_{\nu}$$^2$ for a binary model fit to the
$V^2$ data at each point of a semicircular grid of radius 500 mas in increments of
0.1 mas in both RA and Dec.  The lowest resulting value of $\chi_{\nu}$$^2$ was
then saved, the $\Delta$$m$ value incremented by 0.1 mag, and the process repeated
up through $\Delta$$m$ = 4.0.  Figure~\ref{grid-fit-chi-dra} illustrates the
results of this process for one star ($\chi$ Dra) on a single night: The top panel
shows the minimum value of $\chi_{\nu}$$^2$ over the position-grid search at each
$\Delta$$m$ value, as a function of $\Delta$$m$.  The second and third panels
display the corresponding values of the component separation and position angle,
respectively, for the best-fit model at each $\Delta$$m$ value.

The plots analogous to Figure~\ref{grid-fit-chi-dra} for each star, on each night,
were examined for evidence of a ${\it significant}$ $\chi_{\nu}$$^2$ minimum
[$\chi_{\nu}$$^2$(min)], here defined as a minimum with a corresponding 99\%
confidence interval \citep[$\chi_{\nu}$$^2$(min) + 11.3 for a three-parameter
$\Delta$$m$, $\rho$, $\theta$ fit;][]{avn76,wal96} with a half-width of $\pm$ 1 or
less in $\Delta$$m$ (i.e., relatively small with respect to the whole $\Delta$$m$
search range).

Thus, the GRIDFIT search is a brute force method that searches all possible
locations and $\Delta$$m$ within designated limits and produces an initial estimate
of possible binary parameters for use in later fitting.  While computationally
intensive, the global minimum is found using a very fine grid spacing.  The 500 mas
limit of the GRIDFIT searches was chosen partly for a reasonable computational and
analysis effort and to avoid an extensive overlap with other techniques of courser
resolution such as speckle interferometry and adaptive optics (AO).  As was noted
in \S~\ref{intro}, previous well documented work has been published for speckle or
in some cases AO observations at separations of 500 mas and greater.

For those sources for which GRIDFIT demonstrated significant, global
$\chi_{\nu}$$^2$(min), we next proceeded to detailed modeling.  All binary modeling
was performed using C. A. Hummel's OYSTER software package
\textit{(www.eso.org/$\sim$chummel/oyster/ oyster.html)}, which has been the
standard for displaying, editing, averaging, calibrating, and modelling
interferometry data from the NPOI for many years \citep[e.g., ][]{hum03}.  OYSTER
was used to fit binary models on a night-by-night basis to the $V^2$ as a function
of ${\it u}$ and ${\it v}$ using a Levenberg-Marquardt nonlinear least-squares
technique.  Experience has shown that the most efficient path to fully optimized
models is a three-step iterative process: first, fitting the astrometric parameters
$\rho$ and $\theta$ while holding $\Delta$$m$ fixed at an estimated value, second,
fitting $\Delta$$m$ while holding $\rho$ and $\theta$ fixed at the values
determined in the first stage, and third, performing a final, simultaneous fit for
all three parameters.

To begin the modeling process, estimates of $\rho$ and $\theta$ for each night, the
$\Delta$$m$ of the system, and the angular diameters of the individual stars are
required.  In the case of the $\Delta$$m$ test binaries, initial estimates for
$\rho$ and $\theta$ were derived in most cases from published orbital elements, as
verified using the GRIDFIT results.  However, where the GRIDFIT results differed by
$>$ 2 mas from the position predicted from the orbital elements, or if no orbital
elements existed, the GRIDFIT values were used.

Initial estimates for $\Delta$$m$ were taken from the literature in most cases, or
from the GRIDFIT results where there were no published values in the visible.  For
the primary star angular diameters, we used values from the literature or estimates
from the NPOI data reduction software (\S~\ref{seqred}).  Secondary star angular
diameters were either taken from the literature, estimated from either the relative
colors, spectral types, or known masses of the primary and secondary stars, or set
to small values.  Unless otherwise noted, we subsequently held the component
angular diameters fixed in all three fitting stages outlined above since the
baselines we used were not long enough to reliably determine the component
diameters in the case of these stars.

The results of the fitting process for the $\Delta$$m$ test binaries are presented
in Tables~\ref{knownfit},~\ref{sigHerfits} \&~\ref{knowndelmag}.
Table~\ref{knownfit} lists the fitted angular separations and position angles for
each system (with the exception of $\sigma$ Her; Table~\ref{sigHerfits},
\S~\ref{sigher}), on each night, along with the parameters of the fit error
ellipse.  The uncertainty ellipses correspond to one-seventh of the
synthesized beam, which has been shown to give realistic estimates of the
astrometric accuracy of NPOI multi-baseline observations \citep{hum03,hum13}.  In
some cases, the fitted position angles for a system were adjusted by 180$\arcdeg$
for consistency with those listed in the $\mathit{Fourth\ catalog\ of\
interferometric\ measurements\ of\ binary\ stars}$ \citep[INT4,][]{hmw01b} at
similar epochs or with the predictions of the published orbits cited above.  The
lack of closure phase in our single-baseline data leads to a $\pm$ 180$\arcdeg$
ambiguity in the fitted position angles.  However, in the case of $\alpha$ Dra
and V1334 Cyg, where data from additional baselines were available, closure phase
data were used to independently determine the position angles.  For seven of the
sources in Table~\ref{knownfit}, where published orbits based on observations from
long-baseline optical interferometers exist (PTI: $\kappa$ UMa and $\beta$ CrB;
NPOI/Mark III/PTI: o Leo; NPOI: $\zeta$ UMa A and $\eta$ Peg; Mark III: 113 Her,
and CHARA: V1334 Cyg), the agreement of our nightly fits with the positions
calculated from the orbits is excellent, the typical observed minus calculated
(O-C) values being $\sim$ 0.5 mas.  For 59 Cyg, where the orbit of \citet{mas11}
was based mostly on speckle interferometry observations through 2008, the agreement
is $\lesssim$ 4 mas.  For $\zeta$ Ori A \citep{hum13} and $\phi$ Her \citep{zav07}
binary model fits to the observations listed in Table~\ref{binobs} have been
published, and so were not again fitted, the literature values being reproduced in
Table~\ref{knownfit}.  Columns 3 and 5 of Table~\ref{knowndelmag} contain the mean
of the nightly fitted $\rho$ values from Table~\ref{knownfit} and the log thereof,
with the exception of $\alpha$ Dra and $\sigma$ Her (Table~\ref{sigHerfits}), where
the values at the minimum and maximum angular separations are listed.

Table~\ref{knowndelmag}, Col. 6 contains the mean of the nightly fitted $\Delta$$m$
values at 700 nm ($\Delta$$m_{700}$, \S~\ref{bcfd}) for each star, while Col. 7
contains the standard deviation of the nightly fitted $\Delta$$m_{700}$ values.  We
conservatively quote standard deviations rather than the uncertainty of the mean
($\sigma$/\( \sqrt{N} \), where N is the number of nights) to compensate for the
fact that the systematic errors in the $V^2$ calibration (\S~\ref{seqred}) are not
explicitly accounted for in the modeling.  Once again, the modeling process for
$\zeta$ Ori A and $\phi$ Her was not repeated, so the mean $\Delta$$m_{700}$ values
($<$$\Delta$$m_{700}$$>$) and error estimates in Table~\ref{knowndelmag} for these
systems are taken from the literature.  In the case of $\zeta$ UMa A, where our
$\Delta$$m$ fit was of lesser quality, we have substituted the published values of
\citet{hum98}.

Figure~\ref{59cyg-uv-vis-fit} shows an example (for a binary of relatively large
$\rho$ and $\Delta$$m$) of the ${\it uv}$-plane sampling for a night's
observations, and model $V^2$ values resulting from the best-fit binary parameters
from Tables~\ref{knownfit} \&~\ref{knowndelmag} overlayed on the calibrated $V^2$
data.  Lastly, Figure~\ref{dmagsep} displays the $<$$\Delta$$m_{700}$$>$ value for
each $\Delta$$m$ test binary plotted against the log of its mean $\rho$, values
taken from Table~\ref{knowndelmag}.

Notes pertaining to the modeling of each $\Delta$$m$ test binary (except $\sigma$
Her; \S~\ref{sigher}, $\beta$ Sct; \S~\ref{betasct}, and V1334 Cyg;
\S~\ref{v1334cyg}) are as follows:

$\zeta$ Ori A (HR 1948, HD 37742, WDS J05407-0157Aa,Ab): Binary model fits to the
observations listed in Table~\ref{binobs} have been published by \citet{hum13},
and so were not again fitted.  The values of \citet{hum13} are reproduced in
Table~\ref{knownfit}.

$\kappa$ UMa (FK5 341, HR 3594, HIP 44471, HD 77327, WDS J09036+4709AB): An
estimated primary star angular diameter $\theta_{P}$ = 0.746 mas was provided by
the NPOI data reduction software (\S~\ref{seqred}).  A secondary star angular
diameter $\theta_{S}$ = 0.5 mas was assumed.  An initial $\Delta$$m_V$ = 0.38 was
taken from the $\mathit{Sixth\ catalog\ of\ orbits\ of\ visual\ binary\ stars}$
\citep[ORB6,][]{hmw01a}.  Initial estimates for $\rho$ and $\theta$ were derived
using the orbital elements of \citet{mut10a}.  These estimates agree with the
global minima from GRIDFIT to $\leq$ 0.4 mas on two of the three nights, the
GRIDFIT result on the third night being highly divergent and likely spurious.  The
O-C values of the fits listed in Table~\ref{knownfit} with respect to the
\citet{mut10a} orbit are all $\leq$ 0.5 mas.

o Leo (FK5 365, HR 3852, HIP 47508, HD 83809, WDS J09412+0954Aa,Ab): An estimated
primary star angular diameter $\theta_{P}$ = 1.273 mas was provided by the NPOI
data reduction software (\S~\ref{seqred}).  A secondary star angular diameter
$\theta_{S}$ = 0.5 mas and an initial $\Delta$$m_{700}$ = 1.05 were taken from
\citet{hum01}.  Initial estimates for $\rho$ and $\theta$ were derived using the
orbital elements of \citet{hum01}.  These estimates agree with the global minima
from GRIDFIT to $\leq$ 0.5 mas on all three nights.  The O-C values of the fits
listed in Table~\ref{knownfit} with respect to the \citet{hum01} orbit are all
$\leq$ 0.3 mas.

$\zeta$ UMa A (FK5 497, HR 5054, HD 116656, WDS J13239+5456Aa,Ab): Estimates for
the primary star angular diameter $\theta_{P}$ = 0.8 mas, the secondary star
angular diameter $\theta_{S}$ = 0.8 mas, and an initial $\Delta$$m_{800}$ = 0 were
taken from \citet{hum98}.  Initial estimates for $\rho$ and $\theta$ were derived
using the orbital elements of \citet{hum98}.  These estimates agree with the global
minima from GRIDFIT to $\leq$ 0.3 mas on all three nights.  The O-C values of the
fits listed in Table~\ref{knownfit} with respect to the \citet{hum98} orbit are all
$\leq$ 0.2 mas.

$\alpha$ Dra (FK5 521, HR 5291, HIP 68756, HD 123299): Estimates for the primary
star angular diameter $\theta_{P}$ = 0.6 mas, secondary star angular diameter
$\theta_{S}$ = 0.3 mas, an initial $\Delta$$m_V$ = 1.8, and initial $\rho$ and
$\theta$ values were all provided by Hummel (private communication).  For
consistency, only data from the AC-AE and AC-AW baselines were modeled.

$\beta$ CrB (FK5 572, HR 5747, HIP 75695, HD 137909, WDS J15278+2906AB): A primary
star angular diameter $\theta_{P}$ = 0.721 mas and a secondary star angular
diameter $\theta_{S}$ = 0.361 mas were estimated as per \S~\ref{seqred} using the
$m_V$ value from Table~\ref{known} (3.68), $m_K$ = 3.28 \citep{duc02}, and
assuming the same ${\it V - K}$ = 0.4 value for both components, along with an
initial $\Delta$$m_V$ = 1.5 from the ORB6.  Initial estimates for $\rho$ and
$\theta$ were derived using the orbital elements of \citet{mut10a}.  These
estimates agree with the global minima from GRIDFIT ($>$ 99\% confidence detections
on all three nights) to $\leq$ 0.7 mas on all three nights.  The O-C values of the
fits listed in Table~\ref{knownfit} with respect to the \citet{mut10a} orbit are
all $\leq$ 0.5 mas.

$\phi$ Her (FK5 601, HR 6023, HIP 79101, HD 145389, WDS J16088+4456AB): Binary
model fits to the observations listed in Table~\ref{binobs} have been published by
\citet{zav07}, and so were not again fitted.  The values of \citet{zav07} are
reproduced in Table~\ref{knownfit}.

113 Her (FK5 3508, HR 7133, HIP 92818, HD 175492, WDS J18547+2239Aa,Ab): An
estimated primary star angular diameter $\theta_{P}$ = 1.4 mas, an estimated
secondary star angular diameter $\theta_{S}$ = 0.2 mas, and an initial
$\Delta$$m_{800}$ = 3.00 were all taken from \citet{hum95}.  Initial estimates for
$\rho$ and $\theta$ were derived using the orbital elements of \citet{hum95}.
These estimates agree with the global minima from GRIDFIT ($\geq$ 99\% confidence
detections on two of three nights) to $\leq$ 1.2 mas on all three nights.  The O-C
values of the fits listed in Table~\ref{knownfit} with respect to the \citet{hum95}
orbit are all $\leq$ 0.9 mas.

32 Cyg (HR 7751, HIP 99848, HD 192910, WDS J20155+4743A): An estimated primary star
angular diameter $\theta_{P}$ = 5.361 mas was provided by the NPOI data reduction
software (\S~\ref{seqred}).  A secondary star angular diameter $\theta_{S}$ = 0.3
mas was assumed.  An initial $\Delta$$m_V$ = 3.8 was taken from \citet{paa98}.
Orbital elements from \citet{raf13}, and references therein, were used to
estimate $\rho$ and $\theta$.  Using these values in the OYSTER modelling
procedures produced better quality (lower $\chi_{\nu}$$^2$) models than the
alternative single star model on two nights, equal quality fits on two nights, and
a marginally lower quality fit compared to the single star model on one night.  The
application of GRIDFIT likewise produced no convincing evidence for a binary, with
a $\sim$ 99\% confidence detection on only one night, wildly divergent minimum in
position angle between the various nights, and separations smaller than the
estimated primary star angular diameter on three of the five nights.  We therefore
conclude that we have failed to detected the secondary star in this system.

59 Cyg (FK5 1551, HR 8047, HIP 103632, HD 200120, WDS J20598+4731Aa,Ab): An
estimated primary star angular diameter $\theta_{P}$ = 0.492 mas was provided by
the NPOI data reduction software (\S~\ref{seqred}).  A secondary star angular
diameter $\theta_{S}$ = 0.2 mas was assumed.  An initial $\Delta$$m_V$ = 2.8 was
taken from the ORB6.  Initial estimates for $\rho$ and $\theta$ were derived using
the orbital elements of \citet{mas11}.  These estimates, for this wide, slowly
moving binary (P = 161.5 yr), agree with the GRIDFIT results to $\approx$ 2.2 mas.
The O-C values of the fits listed in Table~\ref{knownfit} with respect to the
\citet{mas11} orbit are all $\leq$ 3.8 mas.

5 Lac (FK5 3799, HR 8572, HIP 111022, HD 213310, WDS J22295+4742AB): An estimated
primary star angular diameter $\theta_{P}$ = 5.290 mas was provided by the NPOI
data reduction software (\S~\ref{seqred}).  An secondary star angular diameter
$\theta_{S}$ = 0.5 mas was assumed.  An initial $\Delta$$m_V$ = 3.6 was taken from
\citet{paa98}.  The most closely contemporaneous speckle interferometry measurement
\citep[$\rho$ = 82 mas, $\theta$ = 43$\fdg$1][]{hmm97} was used for the initial
values of $\rho$ and $\theta$.  Using these values in the OYSTER modelling
procedures produced good quality binary fits on five of the six nights, but the
alternative single star models were essentially equal in quality
($\chi_{\nu}$$^2$) on these nights and significantly better on the sixth night
(2004 August 02 UT).  The application of GRIDFIT likewise produced no convincing
evidence for a binary, with wildly divergent minima in position angle and
separation between the various nights, and angular separations smaller than the
estimated primary star angular diameter on half the nights.  We therefore conclude
that we have failed to detected the secondary star in this system.

$\eta$ Peg (FK5 857, HR 8650, HIP 112158, HD 215182, WDS J22430+3013Aa,Ab): An
estimated primary star angular diameter $\theta_{P}$ = 3.045 mas was provided by
the NPOI data reduction software (\S~\ref{seqred}).  An estimated secondary star
angular diameter $\theta_{S}$ = 0.5 mas was assumed.  An initial $\Delta$$m_{800}$
= 3.61 was taken from \citet{hum98}.  Initial estimates for $\rho$ and $\theta$
were derived using the orbital elements of \citet{hum98}.  These estimates agree
with the global minima from GRIDFIT ($\gtrsim$ 99\% confidence detections on five
of six three nights) to $\leq$ 0.4 mas.  The O-C values of the fits listed in
Table~\ref{knownfit} with respect to the \citet{hum98} orbit are all $\leq$ 0.9
mas.

\subsubsection{$\sigma$ Her} \label{sigher}

For $\sigma$ Her (FK5 621, HR 6168, HIP 81126, HD 149630, WDS J16341+4226AB)
Tables~\ref{sigHercalibs} \&~\ref{sigHerobs} list the calibrator stars used and the
log of observations, respectively.  Multiple calibrator stars were used for the
calibration of the $\sigma$ Her visibility data on each night.  The calibrator star
data were time weighted as per \S~\ref{seqred}, but the calibrator stars were not
necessarily within 15$\arcdeg$ of $\sigma$ Her.

A primary star angular diameter $\theta_{P}$ = 0.52 mas and a secondary star
angular diameter $\theta_{S}$ = 0.16 mas were estimated as per \S~\ref{seqred}
using the $m_V$ value from Table~\ref{known} (4.20), $m_K$ = 4.05 \citep{cut03},
and assuming the same ${\it V - K}$ value for both components, along with an
initial $\Delta$$m_V$ = 2.6 estimated from the values listed in the INT4.  Initial
estimates for $\rho$ and $\theta$ were derived using the orbital elements of
\citet{bab88}.  A program within OYSTER, similar to GRIDFIT, was also used to
search a grid of 80 mas by 80 mas at 0.2 mas spacings, centered on the predicted
position of the secondary star, to verify these estimates.

Columns 1 through 8 of Table~\ref{sigHerfits} list the fitted angular separations
and position angles, and their errors, for this system on each night, in a format
similar to Table~\ref{knownfit}.  For the earlier 1997 through 2001 observations
listed in Table~\ref{sigHerobs}, data were obtained simultaneously on the three
baselines formed by the AC, AE, and AW (or E02) stations, so closure phase data
were available for inclusion in the fitting process.  The magnitude differences
were fitted at three wavelengths (550 nm, 700 nm, and 850 nm), the resulting mean
values being $\Delta$$m_{550}$ = 2.70 $\pm$ 0.05, $\Delta$$m_{700}$ = 2.34 $\pm$
0.05, and $\Delta$$m_{850}$ = 2.28 $\pm$ 0.05.  The minimum and maximum values of
$\rho$ from Table~\ref{sigHerfits} (14.06 mas, 115.13 mas), the log of these
values, and the mean $\Delta$$m_{700}$ are listed in Cols. 3, 5, and 6 of
Table~\ref{knowndelmag}.

\begin{figure}
\plottwo{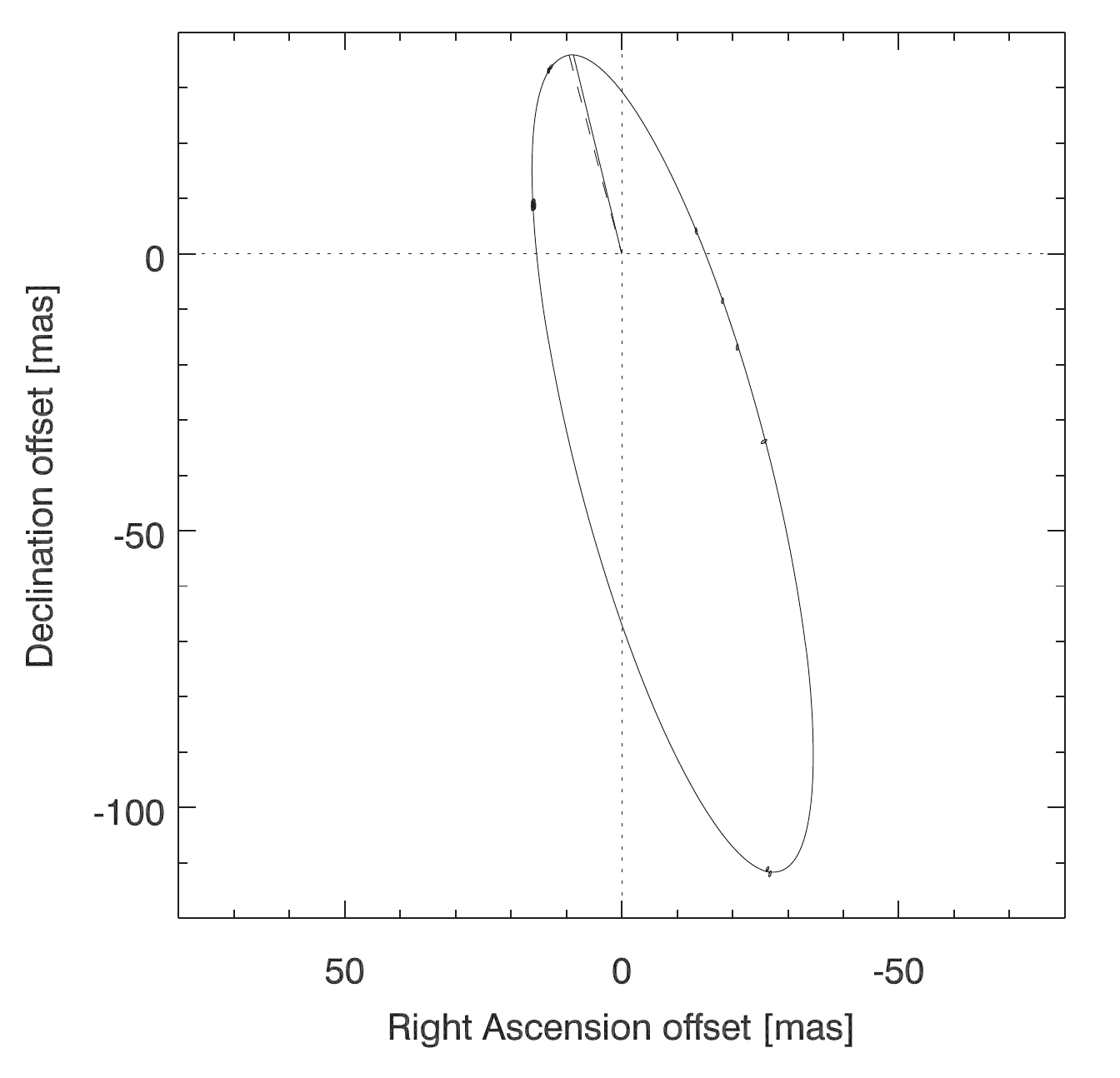}{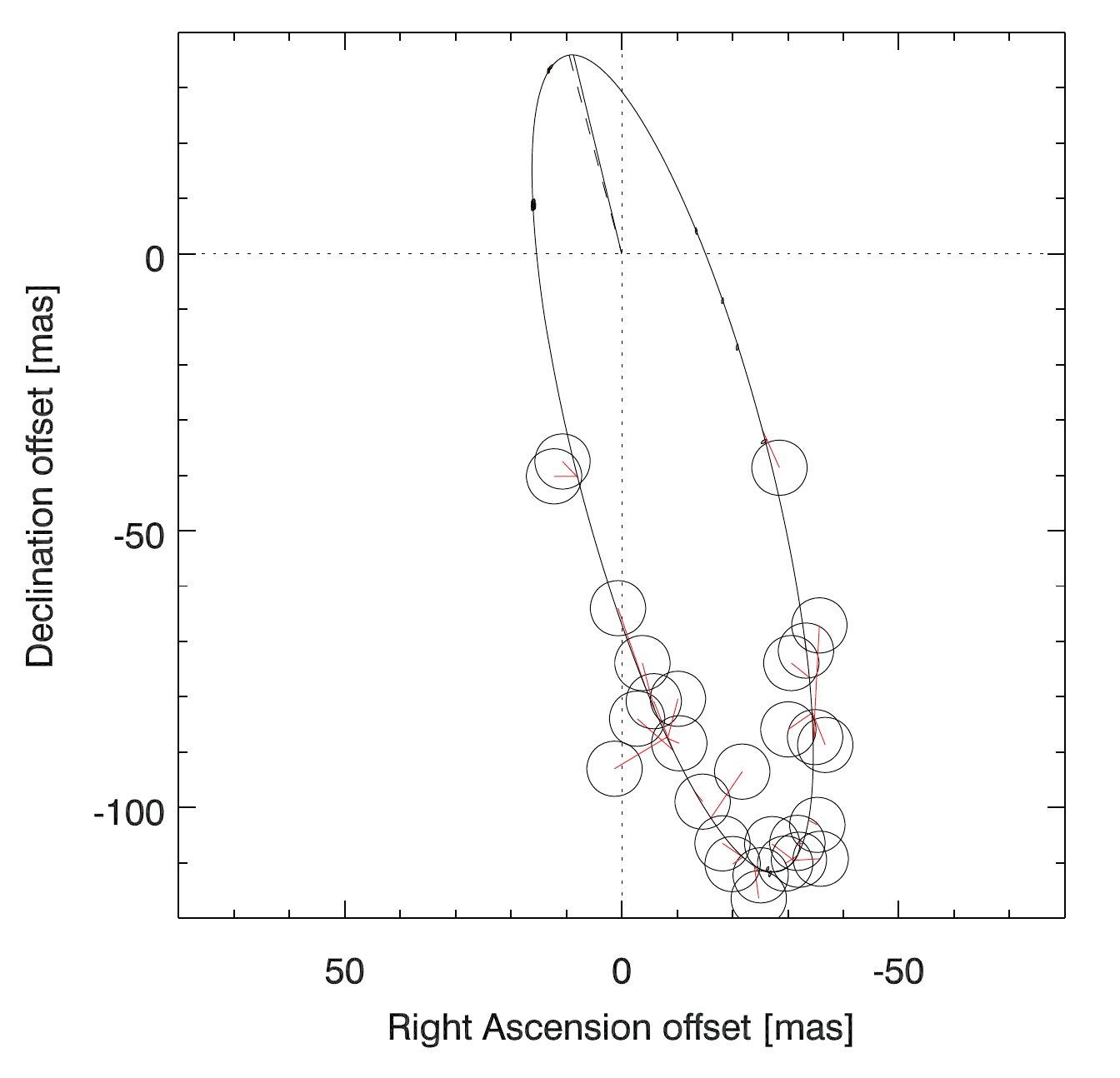}
\caption{Orbit of $\sigma$ Her AB: ${\it Left\ Panel:}$ Plot of the fitted relative
positions at each of the 16 epochs of observation (Table~\ref{sigHerfits}), with
error ellipses plotted to scale.  The apparent orbit corresponding to the best-fit
orbital solution to these points (Table~\ref{sig-Her-elements}) is overplotted.
${\it Right\ Panel:}$ Same as left panel, but with archival measurements from the
INT4 overplotted.  Red line segments indicate the offset of each INT4 measurement
with respect to the position predicted by the new NPOI orbit at the epoch of each
observation.  Assumed error circles of radius 5 mas are also plotted for INT4
entries that lack error estimates.}
\label{sig-Her-orbit}
\end{figure}

Given the more than seven-year timespan of the observations of $\sigma$ Her listed
in Table~\ref{sigHerobs}, it was also possible to obtain a high-quality orbit fit
using ${\it only}$ NPOI data.  The resulting orbital elements are listed in
Table~\ref{sig-Her-elements}, with the corresponding apparent orbit overplotted on
the nightly $\rho$, $\theta$ fits from Table~\ref{sigHerfits} in the left panel of
Figure~\ref{sig-Her-orbit}.  The ${\it O - C}$ values in $\rho$ and $\theta$ of the
nightly fits with respect to this orbit are listed in Cols. 9 and 10 of
Table~\ref{sigHerfits}, respectively.  For purposes of comparison, archival
observations from the INT4 ($\theta$ adjusted by 180$\arcdeg$ in some cases) are
added in the right panel of Figure~\ref{sig-Her-orbit}.

\subsubsection{$\beta$ Sct} \label{betasct}

For $\beta$ Sct (FK5 1489, HR 7063, HIP 92175, HD 173764) the application of
GRIDFIT to our eight nights of data (Table~\ref{binobs}) indicates marginal, $\leq$
68\% confidence interval \citep{avn76,wal96}, detections on two nights (2004 August
06 UT and 2004 August 27 UT) at $\Delta$m = 3.8 to 4.0.  For purposes of
comparison, we also applied the CANDID algorithm
\citep{gal15}\footnote{https://github.com/amerand/CANDID} to our 2004 August 06 UT
data.  CANDID reported a significant detection at a position (modulo 180$\arcdeg$)
within 2 mas of the GRIDFIT result.  We subsequently used the GRIDFIT position for
this night ($\rho$ = 15.57 mas, $\theta$ = 162$\fdg$28) as initial values in the
modelling of the data from all eight nights using OYSTER.  We also used an
estimated primary star angular diameter $\theta_{P}$ = 2.212 mas, as provided by
the NPOI data reduction software (\S~\ref{seqred}).  A secondary star angular
diameter $\theta_{S}$ = 0.1 mas was estimated from the ratio of the stellar radii,
derived using the adopted $\theta_{P}$ value and the parallax of \citet{paa05} to
estimate the radius of the primary, then using the spectral type and mass
\citep{paa05} and the tables of masses and radii of \citet{dal99} to estimate the
radius of the secondary.  An initial $\Delta$$m_V$ = 3.8 was estimated from the
GRIDFIT results.

In all cases, the binary models (Table~\ref{knownfit}) were a better, lower
$\chi^{2}$, fit than a single-star model.  However, the nightly fits for the
$\Delta$$m_{700}$ of this binary varied widely.  Given that our best nightly model
fit (2004 August 06 UT) reports an essentially identical result (3.59 $\pm$ 0.06)
as that found by CANDID, but in consideration of the other nightly OYSTER and
GRIDFIT results that report larger values, we adopt a tentative value of
$\Delta$$m_{700}$ = 3.6 +0.2/-0.1 for the binary.

The results reported here represent the first detection of the secondary star in
this system at precisely measured separations and position
angles\footnote{\citet{paa05} reported TRANS mode observations from the HST Fine
Guidance Sensor that demonstrated duplicity in the signal but failed to provide
reliable relative astrometry.}.  Given the limited time span of our observations
($\approx$ 21 days, Table~\ref{binobs}) relative to the orbital period of $\approx$
833 days \citep[e.g.,][]{paa05}, we did not attempt an orbital solution using the
combination of our astrometric data with those from spectroscopy
\citep[e.g.,][]{grf08}.  However, it is likely that our detection does correspond
to the secondary star detected by spectroscopy.  One can use the orbital elements
of \citet{grf08}, along with the largest likely value of the system semimajor axis
from \citet{paa05} ($\sim$ 18 mas) to predict separations and position angles on
the dates of our observations that are $\leq$ 4 mas in separation and $\leq$
30$\arcdeg$ in position angle of our measured positions (Table~\ref{knownfit}).

\subsubsection{V1334 Cyg} \label{v1334cyg}

For V1334 Cyg (HR 8157, HIP 105269, HD 203156, WDS J21194+3814Aa,Ab) an estimated
primary star angular diameter $\theta_{P}$ = 0.50 mas was provided by averaging the
values cited by \citet{gal13}, and references therein.  A secondary star angular
diameter $\theta_{S}$ = 0.2 mas was assumed.  An initial $\Delta$$m_{700}$ = 2.7
was derived using GRIDFIT (below).  Initial estimates for $\rho$ and $\theta$ were
derived using the orbital elements of \citet{gal13}.

We also performed a systematic search for the intermittently detected third (B)
component in this system \citep[][and references therein]{gal13}, using our three
nights of data (Table~\ref{binobs}).  GRIDFIT and the similar program within OYSTER
were each used to search a region of 500 mas radius around this system at a
resolution of 0.1 mas.  Neither program provided any evidence of a third component
in this system, but both recovered the Ab component at positions within 0.6 mas in
separation and 4$\arcdeg$ in position angle of those predicted by the \citet{gal13}
orbit in the data from our first two night nights of observation.  We also made use
of the CANDID algorithm \citep{gal15} to search a region within 50 mas of the
primary.  In this case, the Ab component was again recovered within 0.2 mas in
separation and 2$\arcdeg$ in position angle on the first two nights, but another
3$\sigma$ detection at $\rho$ = 20 mas and PA = 40$\arcdeg$ was also reported for
2009 December 01 UT.  Given that this result was not repeated on our other two
nights, we assume it to be spurious.

Both CANDID and GRIDFIT produced highly discrepant position angles for the Ab
component for our third night of observations (2009 December 18 UT), producing
values of approximately 45$\arcdeg$ and 136$\arcdeg$, respectively, from that
predicted by the \citep{gal13} orbit.  Given the lesser quantity and the relatively
poor qualitative appearance of the calibrated $V^2$ on this night, these results
may not be surprising.  Given all these indications, we chose to down-weight our
model fit for this night in determining a value of $\Delta$$m_{700}$ = 2.69 $\pm$
0.27 for this system (Table~\ref{knowndelmag}), weighting the model fits for the
individual nights by the number of $V^2$ measures on each night
(Table~\ref{binobs}).

\subsubsection{Nondetections} \label{nondet}

\begin{figure}
\epsscale{1.0} \plotone{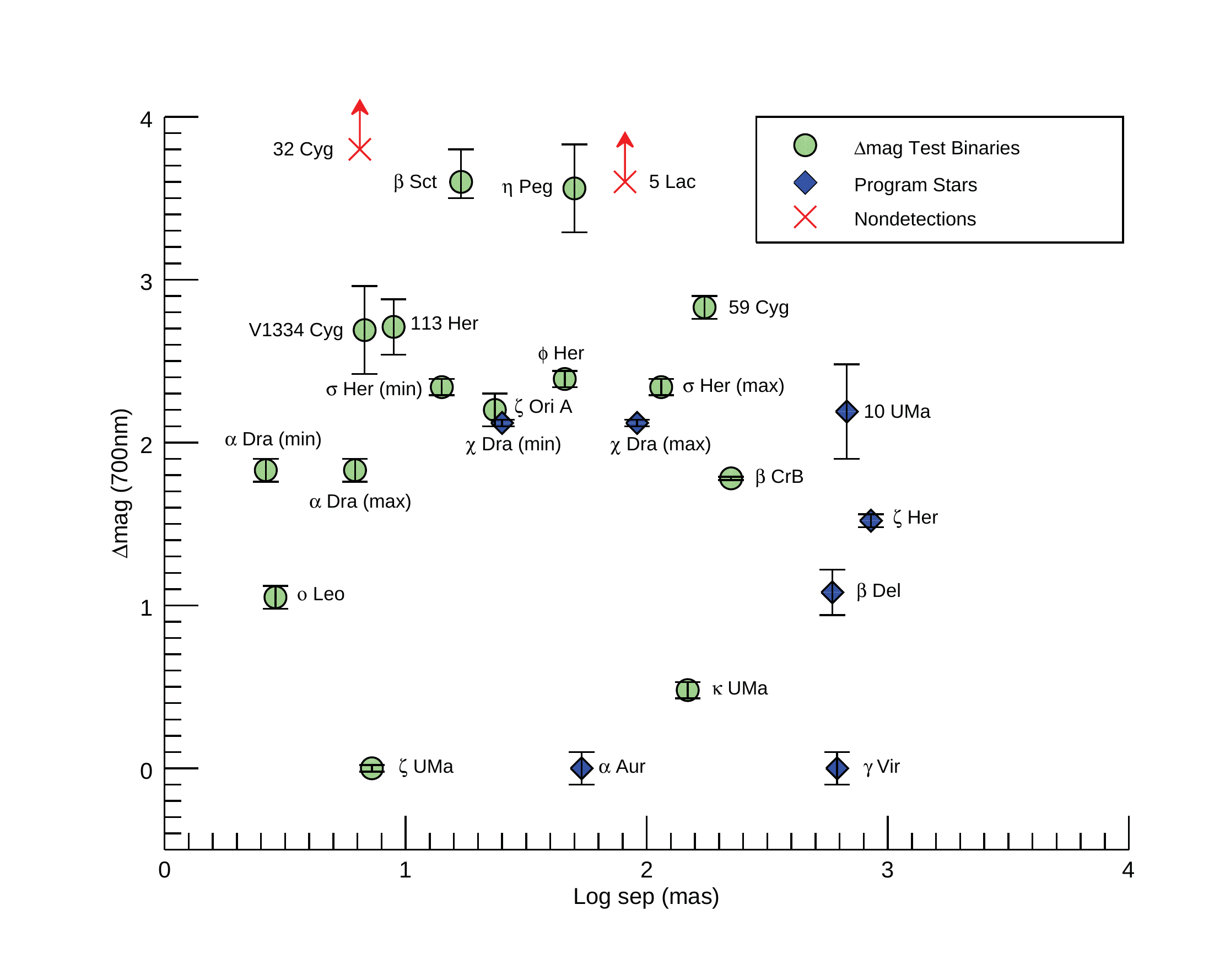}
\caption{Primary-secondary magnitude difference $<$$\Delta$$m_{700}$$>$ plotted
against the log of the mean component separation in milliarcseconds for binaries
observed by the NPOI.  $\Delta$$m$ test binaries
(Tables~\ref{knownfit},~\ref{sigHerfits} \&~\ref{knowndelmag}) are shown as filled
circles, program stars (Tables~\ref{programfit} \&~\ref{programdelmag}) are shown
as filled diamonds, and estimated lower limits for nondetections
(Table~\ref{knowndelmag}, with $\Delta$$m_{550}$ values plotted) are shown as
crosses with arrows.}
\label{dmagsep}
\end{figure}

As per \S~\ref{binknown}, two sources (HR 7751 = 32 Cyg and HR 8572 = 5 Lac) listed
in Tables~\ref{known} \&~\ref{binobs} do not appear in Table~\ref{knownfit} since
it was not possible to fit a statistically significant, low reduced $\chi$$^2$
binary model to the $V^2$ data.  For each of these sources, a single-star model of
angular diameter similar to the initial primary star diameter estimate provided an
equal or better model fit.  This leads to the conclusion that for these
large-$\Delta$$m$ stars, we have failed to detect the secondary component.
However, for purposes of comparison, we have added ${\it estimates}$ of likely
angular separations and $\Delta$$m$ values for these sources at the mean epoch of
our observations to Table~\ref{knowndelmag} and Figure~\ref{dmagsep}.

Examining Figure~\ref{dmagsep}, it initially appears curious that 32 Cyg and 5 Lac
were not detected by our observations given the close proximity of their plotted
$\Delta$$m$ ${\it vs}$ log$\rho$ values to those of $\beta$ Sct and $\eta$ Peg,
respectively.  However, on further examination, it may not have been surprising
that these systems were not detected as binaries by our observations.  They, along
with $\beta$ Sct were originally chosen for observation to test the limits of large
$\Delta$$m$ binary detection based on their very large ($\geq$ 3) ${\it estimated}$
$\Delta$$m$ values \citep{paa98}.  Each of these systems consists of a luminous
cool primary and a hot companion for which \citet{paa98} utilized IUE observations
and the total UV-optical energy distribution to determine the spectral class of the
hot secondary star along with an estimated $\Delta$$m_V$ of the binary.  Given that
the wavelength of the ``crossover point'' at which the dominance of the total flux
in the spectral fits transitions from the hot to cool components given by
\citet{paa98} is quite blue $\leq$ 380 nm), one would expect that the observed
$\Delta$$m$ would steadily increase as the effective wavelength of the observations
increases.  Such is demonstratively the case for 113 Her and $\eta$ Peg, where the
measured $\Delta$$m$ varies as 1.61 (500 nm), 2.01 (550 nm), and 3.00 (800 nm) for
113 Her \citep{hum95}, and 2.01 (450 nm), 2.76 (550 nm), and 3.61 (800 nm) for
$\eta$ Peg \citep{hum98}.  Given that the primary-secondary spectral type
differences are most extreme for 32 Cyg (K3Ib + B3V) and 5 Lac (M0II + B8V), large
for $\beta$ Sct (G4IIa + B9V), and relatively moderate for 113 Her (G4III + A6V)
and $\eta$ Peg (G2II-III + A5V), it is likely that the variation in $\Delta$$m$
from the ${\it V}$ band to the $\sim$ ${\it R}$ band of our observations is even
larger for 32 Cyg and 5 Lac, making their $\Delta$$m_{700}$ values considerably
larger than their estimated $\Delta$$m_{550}$ values plotted in
Figure~\ref{dmagsep}, thus accounting for their nondetection as binaries in the
current NPOI observations.  Likewise, the wavelength dependence of $\Delta$$m$ can
explain why, for 5 Lac, historical speckle interferometry observations (INT4) made
at effective wavelengths $<$ 500 nm consistently detect the secondary in this
system, while observations at $>$ 550 nm consistently fail to do
so\footnote{An alternative interpretation of this system is that of a binary in a
highly inclined orbit that occasionally closes to a separation less than the
angular resolution of our observations.  The INT4 records indicate an average
position angle $\approx$ 44$\arcdeg$, with a general trend toward decreasing
separation.  Hartkopf (private communication) notes that many of the observations
reported the INT4 were made using \textit{photographic} speckle interferometry
whose $\Delta$$m$ sensitivity is likely $<$ 3 magnitudes, implying the secondary
should be detectable by the NPOI when more widely separated.}.

\subsubsection{$\Delta$$m$ Limit} \label{delmaglimit}

Figure~\ref{dmagsep} and the discussion of \S~\ref{nondet} thus indicate that the
NPOI is sensitive to binary companions to $\Delta$$m_{700}$ $\approx$ 3 at angular
separations $\gtrsim$ 6 mas (e.g., V1334 Cyg) and to $\Delta$$m_{700}$ $\approx$ 2
down to 3 mas (e.g., $\alpha$ Dra).  Our $\Delta$$m_{700}$ limit is perhaps
somewhat larger ($\sim$ 3.5) for binaries where there is not more than a difference
of about two full spectral types between the components (e.g., $\eta$ Peg and
$\beta$ Sct).

\subsection{Detecting Binaries among the Program Stars} \label{bindet}

We next proceeded to systematically examine the calibrated visibility data for all
of the program sources observed to date (Table~\ref{progstarsobs}) for
statistically significant evidence (99\% confidence, \S~\ref{binknown}) of a
stellar companion using GRIDFIT.  Based on this examination, the observed sources
can be placed into several groups.

The first group ($\alpha$ Aur, 10 UMa, $\gamma$ Vir, $\zeta$ Her, $\chi$ Dra, and
$\beta$ Del) all display obvious, deep $\chi_{\nu}$$^2$ minima in the GRIDFIT plots
on all nights, similar to the example of Figure~\ref{grid-fit-chi-dra}.  These are
all well known binary systems for which we then proceed to detailed modeling of the
separation, position angle, and magnitude difference of the binary components in
\S~\ref{binBSS}.  For four of these six sources (10 UMa, $\gamma$ Vir, $\zeta$ Her,
and $\beta$ Del) the $\chi_{\nu}$$^2$(min) occur at angular separations at or very
close to the 500 mas outer edge of the grid search.  Subsequent detailed fitting in
\S~\ref{binBSS} shows the actual separations of the binary components to range up
to $\sim$ 860 mas and the implications of these results are discussed in
\S~\ref{limbindet}.

A second group of sources ($\alpha$ Ari, $\beta$ Gem, $\theta$ UMa, $\alpha$ Ser,
and $\eta$ Dra) all have significant $\chi_{\nu}$$^2$(min) on one or more nights
each, but these minima correspond to $\rho$ values $\lesssim$ (usually $\ll$) the
known or estimated (per the methods of \S~\ref{seqred}) angular diameters of the
primary stars.  We made no further attempt to model these sources as binaries,
rather proceeding to accurate single-star model fitting of all of these sources
(\S~\ref{diamBSS}), with the exception of $\theta$ UMa, where the estimated primary
diameter ($\sim$ 1.7 mas) is marginally too small to allow accurate
angular-diameter modeling from our data.  See \S~\ref{diamBSS} for a quantitative
assessment of our ability to fit angular diameters $<$ 2 mas. Conversely, in
\S~\ref{diamBSS} we fit single-star angular diameters to an additional six sources
($\beta$ Cas, 46 LMi, $\kappa$ Oph, $\beta$ Oph, $\eta$ Ser, and $\gamma$
Cep\footnote{\citet{nmf07} reported the detection of $\gamma$ Cep B at $\rho$
$\sim$ 890 mas and $\Delta$$m_K$ $\sim$ 6.3}) that had no significant
$\chi_{\nu}$$^2$(min) in any GRIDFIT output on any night of observation, but had
estimated angular diameters $\geq$ 2 mas.

A third group of two sources ($\eta$ Cep and $\iota$ Peg) may be considered
${\it marginal}$ detections in that GRIDFIT outputs for these stars show
significant $\chi_{\nu}$$^2$(min) on only one night each (2004 October 05 UT for
$\eta$ Cep, and 2004 September 30 UT for $\iota$ Peg, respectively).  For $\eta$
Cep, we attempt both binary- and single-star models in \S~\ref{etacep} and
\S~\ref{diamBSS}, respectively, while a binary model for $\iota$ Peg is attempted
in \S~\ref{iotpeg}.

\begin{figure}
\epsscale{.80}
\plotone{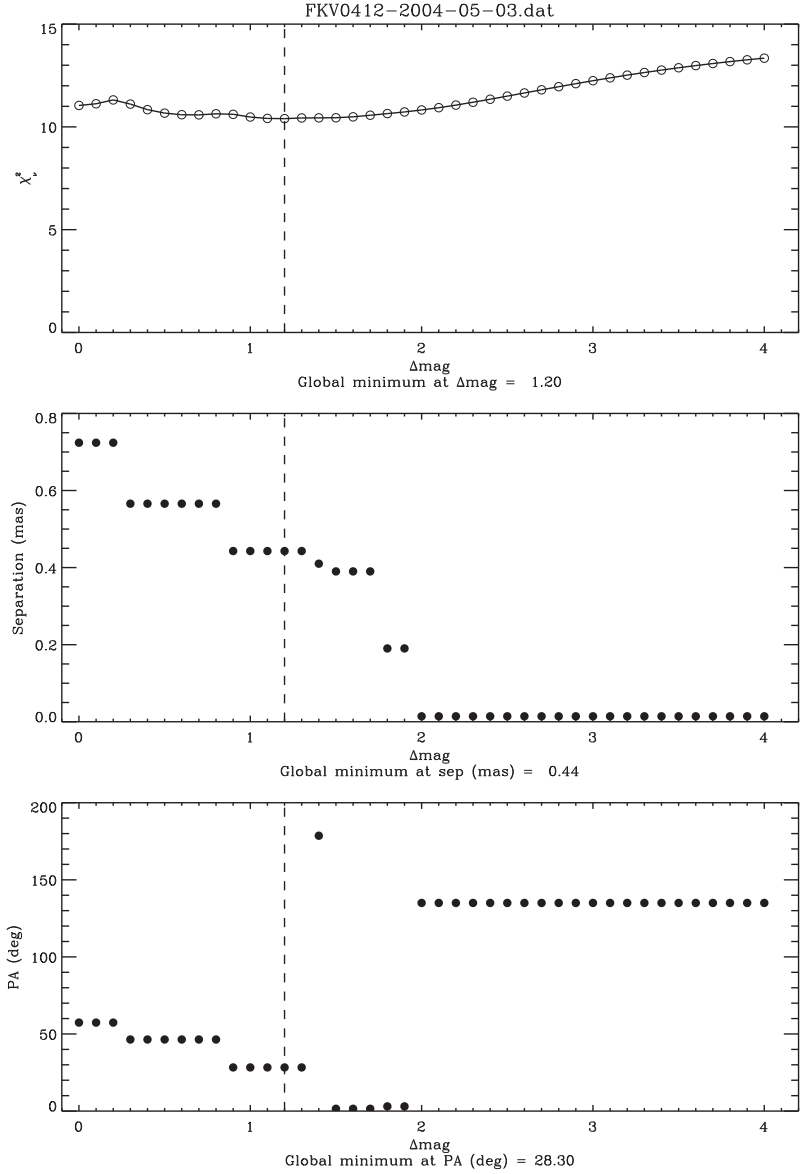}
\caption{Output of the GRIDFIT program for observations of 46 LMi on 2004 May 3 UT.
The format of the plot is the same as in Figure~\ref{grid-fit-chi-dra}.  The top
panel illustrates an example of a GRIDFIT search resulting in no statistically
significant $\chi_{\nu}$$^2$(min) for grid searches over separation and position
angle over the range 0 to 4 magnitudes.}
\label{grid-fit-46-LMi}
\end{figure}

None of the remaining sources observed show any indication of significant
$\chi_{\nu}$$^2$(min) in the GRIDFIT output plots.  An example of such a plot is
presented in Figure~\ref{grid-fit-46-LMi}.  However, to lend further confidence
as to the completeness of the GRIDFIT search results, we additionally conducted
an extensive search of the literature for references to multiplicity among our
observed sources (Table~\ref{progstarsobs}).  We searched the $\mathit{Hipparcos\
Catalogue\ Double\ and\ Multiple\ Systems\ Annex}$ \citep{esa97}, the ${\it
SIMBAD}$ database \citep{wen00}, the ${\it Seventh\ catalogue\ of\ the\ orbital\
elements\ of\ spectroscopic}$ ${\it binary\ systems}$ \citep{bat78}, ${\it The\
ninth\ catalogue\ of\ spectroscopic\ binary\ orbits}$ \citep{pou04}, the WDS, the
ORB6, the INT4, the ${\it Third\ photometric\ magnitude}$ ${\it difference\
catalog}$ \citep{mas08}, and papers 1 through 233 of Griffin
\citep[e.g.,][]{grf13} for references to binary companions, particularly at $\rho$
$\leq$ 1$\arcsec$ and $\Delta$$m$ $\leq$ 4.  This search revealed seven systems of
possible interest.  Five of these ($\beta$ Cas, $\eta$ Cas, $\alpha$ Tri, $\rho$
Gem, and $\theta$ Dra) contain known or suspected spectroscopic binaries.  However,
given the reported spectroscopic orbital solutions and the known parallaxes
(Table~\ref{targetlist}), the likely angular separations of these binaries are all
$\leq$ 1 mas, significantly below the resolution of the array configuration used
for our observations \citep[$\lambda$/(2*B) $\approx$ 2.6 mas, where $\lambda$ =
550 nm (\S~\ref{npoides}) and B = 22.2 m (\S~\ref{config});][]{trb00}\footnote{In
the case of $\beta$ Cas, the notes to the WDS list a P = 27 d spectroscopic binary
\citep{ait32}, but \citet{abt65} concluded there was no convincing evidence of
binary motion.}.  Therefore, it is not surprising that these systems produce no
evidence of binarity in the GRIDFIT searches.  The remaining two sources ($\beta$
CVn, and $\delta$ Aql) each have one measurement of a companion made by speckle
interferometry \citep{bnf80,bon80} but numerous nondetections as well (INT4).
Given that \citet{bnf80} found the $\Delta$$m$ for $\beta$ CVn to be large (``about
3 mag''), it could be that this companion, if it exists, lies beyond the
$\Delta$$m$ detection limit of our current observations (\S~\ref{delmaglimit}), but
this source will be observed again with the more capable current NPOI array.  For
$\delta$ Aql, \citet{kam89} concluded, based on an orbit derived from astrometric
and spectroscopic measurements, that the $\Delta$$m$ is also large
\citep[and the][observation likely spurious]{bon80}, so once again the lack of a
detection from the current data does not necessarily reduce confidence in the
completeness of the GRIDFIT search results and our claim of a $\Delta$$m$ = 3.0
detection limit.  However, given the \citet{kam89} and later ${\it Hipparcos}$
orbital solutions \citep{esa97}, we attempt a binary model of our $\delta$ Aql data
in \S~\ref{delaql}.

Thus, our search of the literature for references to multiplicity among our
observed sources does not immediately contradict the GRIDFIT search results.  In
the following two sections we then proceed to the detailed modeling of the six
detected binaries and the three possible binary detections ($\eta$ Cep, $\iota$
Peg, and $\delta$ Aql) discussed above.

\subsection{Binaries Among the Program Stars} \label{binBSS}

The detailed modeling of the six detected binaries among the program sources
observed to date (\S~\ref{bindet}) closely followed the procedures used for the
$\Delta$m test binaries described in \S~\ref{binknown}; fits for $\rho$ and
$\theta$ on individual nights were first performed, the component angular diameters
and $\Delta$$m$ held fixed, after which fits for $\Delta$$m$ were performed,
followed by a final, simultaneous fit for $\rho$, $\theta$, and $\Delta$$m$.  As in
\S~\ref{binknown}, the fitted position angles were adjusted by 180$\arcdeg$ as
needed for consistency with those listed in the INT4 and with the predictions of
published orbits.  The results are tabulated in Tables~\ref{programfit}
\&~\ref{programdelmag}, which follow the same format as Tables~\ref{knownfit}
\&~\ref{knowndelmag}, respectively.  Notes for individual systems follow.

\subsubsection{$\alpha$ Aur} \label{alpaur}

For $\alpha$ Aur (FK5 193, HR 1708, HIP 24608, HD 34029, WDS J05167+4600 Aa,Ab),
the orbital elements and component angular diameters (6.4 mas and 8.5 mas,
respectively) of \citet{hum94}, along with $\Delta$$m$ = 0, were used as inputs to
the binary models.  The grid fitting procedure within OYSTER was used to verify the
initial estimates of $\rho$ and $\theta$.  The grid search was performed in 0.5 mas
steps over a range of 50 mas in RA and DEC, centered on the position predicted by
the \citet{hum94} orbit.  The results of this search show excellent agreement ($<$
0.5 mas) with the orbit predictions on the first two nights, and for the position
angle on the third night, but with a discrepancy of $\sim$ 10 mas in separation,
evidently due to the poor quality of the $V^2$ calibration on this night.  The
final OYSTER model fitting produced nightly $\rho$, $\theta$ fits
(Table~\ref{programfit}) showing excellent agreement with the \citet{hum94} orbit
(nightly O-C values ranging from 0.2 mas to 0.3 mas.), albeit with a large error in
the fit for $\rho$ on the third night.  The fitted $\Delta$$m_{700}$ = 0.0 $\pm$
0.1 (Table~\ref{programdelmag}) is consistent with values of both
\citet[][$\Delta$$m_{800}$ = -0.05 $\pm$ 0.05]{hum94} and \citet{bld96}, determined
at similar effective wavelengths.

\subsubsection{10 UMa} \label{tenuma}

For 10 UMa (FK5 339, HR 3579, HIP 44248, HD 76943, WDS J09006+4147 AB), the orbital
elements of \citet{mut10a}, an estimated primary star angular diameter $\theta_{P}$
= 1.004 mas provided by the NPOI data reduction software \S~\ref{seqred}, an
assumed secondary star angular diameter $\theta_{S}$ = 0.1 mas, and $\Delta$$m$ =
2.3 (ORB6) were used as inputs to the binary models.  The initial estimates of
$\rho$ and $\theta$ were verified using the OYSTER `gridfit' procedure.

The O-C values of the final nightly $\rho$, $\theta$ fits from OYSTER with respect
to the \citet{mut10a} orbit show good agreement ($\leq$ 5 mas), while the fitted
$\Delta$$m_{700}$ = 2.19 $\pm$ 0.29 is consistent with several measurements at
similar effective wavelengths reported in the INT4.

\subsubsection{$\gamma$ Vir} \label{gamvir}

$\gamma$ Vir (HR 4825/6, HIP 61941, HD 110379, WDS J12417-0127 AB) is a visual
binary with over 200 years of relative astrometric measures listed in the INT4
catalog.  It is one of the first pairs of stars presented by Sir William Herschel
\citep{her03} that conclusively demonstrated orbital motion.  This was significant
as previously most of these double stars were presumed to be coincident optical
pairs as orbital motion had not been conclusively observed.  The recent periastron
passage in 2005 was key to a revision of some imprecisely known orbital elements
\citep{sca07}.  Most magnitude differences recorded in the INT4 for the pair are
$\sim$ 0.0.  A nearly equal brightness ratio is expected as the spectral types of
the two stars are the same \citep[][Table~\ref{targetlist}]{abt95}.  Interestingly,
some magnitude differences for measurements within the 21st century are decidedly
non-zero.  These include a $\Delta$$m$ = 0.69 with the RYTSI speckle camera
\citep{hor10} and as large as 0.9 at 745 nm with the Robo-AO system \citep{rid15}.
This motivates use of the NPOI visibility data to provide a precise $\Delta$$m$ to
investigate the significance of a non-zero brightness ratio for these two stars.

\begin{figure}
\epsscale{1.0}
\plotone{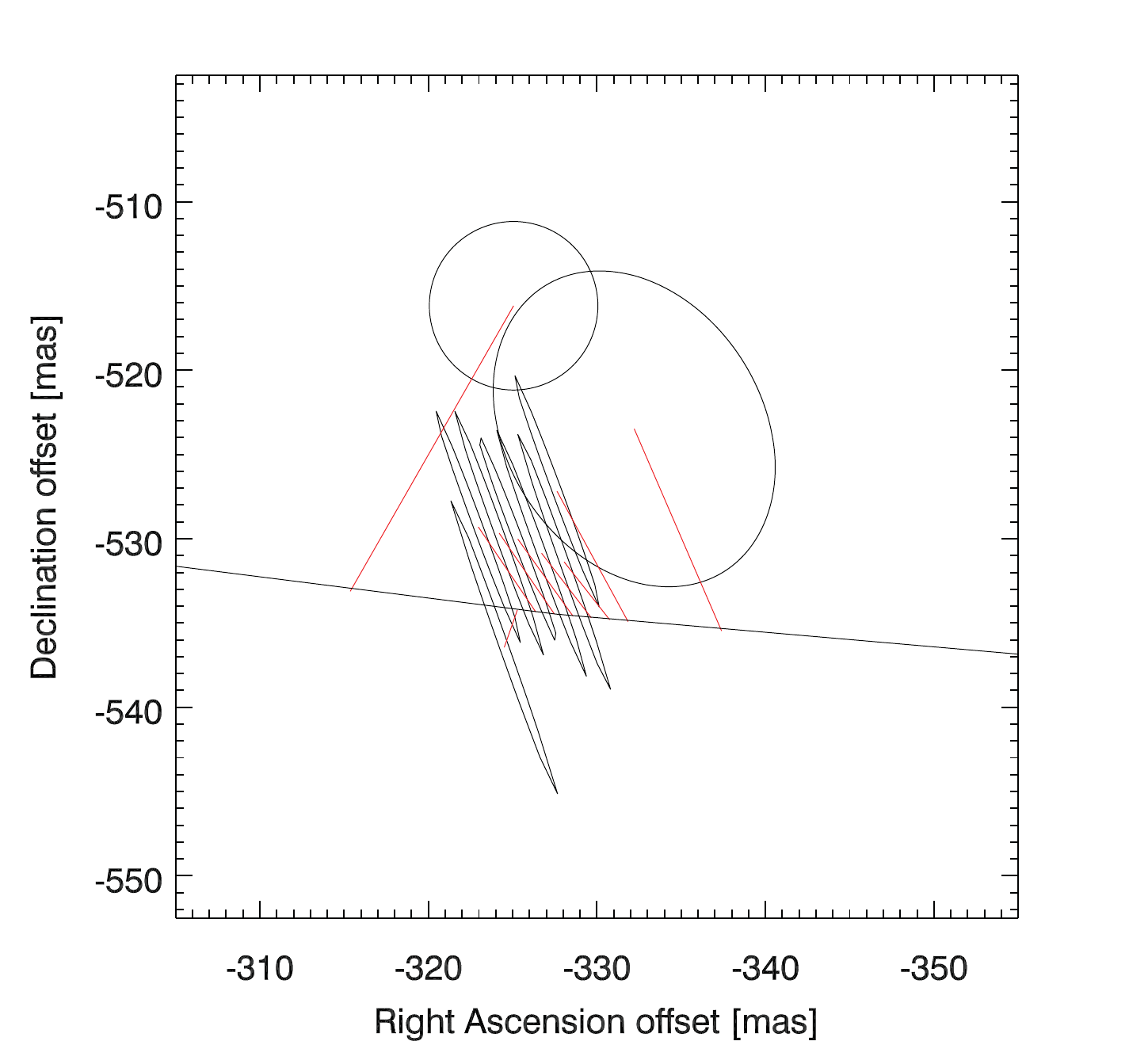}
\caption{Orbit of $\gamma$ Vir AB: Plot of the fitted relative positions at each
of the 7 epochs of NPOI observation (Table~\ref{programfit}; average position:
$\rho$ = 622.64 mas, $\theta$ = 211$\fdg$53), with (narrow) error ellipses
plotted to scale.  A small section of the apparent orbit corresponding to the
orbital elements of \citet{sca07} is overplotted in black.  Archival measurements
from the INT4 \citep{hmr08,sca05} are also overplotted, with error ellipses.  Red
line segments indicate the offset of each NPOI and INT4 measurement with respect
to the position predicted by the \citet{sca07} orbit at the epoch of each
observation.}
\label{gam-Vir-orbit}
\end{figure}

For $\gamma$ Vir, the orbital elements of \citet{sca07}, primary and secondary star
angular diameters of 1.59 mas \citep{ric05}, and a $\Delta$$m$ = 0.1 resulting from
the GRIDFIT search (\S~\ref{bindet}), were used as inputs to the binary model.  The
initial estimates of $\rho$ and $\theta$ were verified using the OYSTER `gridfit'
procedure.  As the raw $V^2$ on the AC-AW baseline are very low, varying from
$\sim$ 10$\%$ in the red to $\sim$ 3$\%$ in the blue, and because the predicted
period of the $V^2$ oscillations as a function of wavelength is \textit{much} less
than the widths of our spectral channels, calibrating and modelling these data
proved very difficult.  Therefore, these data were not used in the final modelling
process.  Our results (Table~\ref{programfit}) are nicely bracketed in position
angle by the contemporaneous observations of \citet{hmr08} and \citet{sca05}, and
the O-C values of the nightly $\rho$, $\theta$ fits with respect to the
\citet{sca07} orbit (Figure~\ref{gam-Vir-orbit}) show reasonably good agreement (2
mas to 9 mas) considering the relatively poor north-south resolution of the
single-baseline data used in our astrometric fits.  Our fitted $\Delta$$m_{700}$ =
0.0 $\pm$ 0.10 (Table~\ref{programdelmag}) is smaller than most INT4 values, but is
consistent with the identical spectral types for the components of $\gamma$ Vir
listed in Table~\ref{targetlist}.

\subsubsection{$\zeta$ Her} \label{zether}

For $\zeta$ Her (HR 6212, HIP 81693, HD 150680, WDS J16413+3136 AB), the orbital
elements of \citet{sod99}, a primary star angular diameter $\theta_{P}$ = 2.367 mas
\citep{moz03}, an estimated secondary star angular diameter $\theta_{S}$ = 0.77
mas, and $\Delta$$m$ = 1.54 (below) were initially used.  $\theta_{S}$ was
estimated from the spectral type (G7V) and parallax (92.63 mas) listed in
Table~\ref{targetlist} using the tables of $R/R_{\sun}$ of \citet{dal99}.

We next endeavored to improve our estimates for several of these parameters: First,
the OYSTER gridfit procedure was used to improve the initial estimates of $\rho$
and $\theta$.  This grid search was performed in 0.5 mas steps over a range of 150
mas in RA and DEC, centered on the position predicted by the \citet{sod99} orbit.
As there was no evidence ($V^2$ oscillations with wavelengh/channel) for the wide
companion in any of the observations on the AC-AW baseline, these data were not
used in the grid search.  However, such oscillations are often present for AC-AE
observations (Figure~\ref{V2-zeta-Her}).  Next, the improved $\rho$ and $\theta$
values, along with the initial $\theta_{S}$ and $\Delta$$m$ values (all held
fixed), were used to improve the $\theta_{P}$ value through a simultaneous fit of a
binary model to all the data from both baselines on the seven nights of observation
(Table~\ref{progstarsobs}).  The value thus obtained $\theta_{P}$ = 2.359 mas is
only slightly different from that of \citep{moz03}.

Lastly, final, nightly fits for $\rho$, $\theta$, and $\Delta$$m$, using only the
AC-AE baseline data were performed.  The results are presented in
Tables~\ref{programfit} \&~\ref{programdelmag}.  Our position results are all quite
similar and are bracketed in position angle by contemporaneous results at 2004.3031
\citep{hmr08} and 2004.659 \citep{sca06}.  The average separation is about 1.8 mas
less than the \citet{sod99} orbit, while the average position angle is $\approx$
1$\fdg$6 more, corresponding to $\approx$ 25 mas in a northwesterly direction.  The
average of the nightly fits for $\Delta$$m$ ($\Delta$$m_{700}$ = 1.52 $\pm$ 0.04)
is much smaller than the values typically quoted in the INT4 for observations made
at similar effective wavelengths \citep[$\sim$2.6; e.g.,][]{hor04,hor08,dru14}, but
is clearly consistent with our data for observations where the period of the
observed $V^2$ oscillations with wavelength is large with respect to the width of
the NPOI spectral channels (e.g., panel ``a'' of Figure~\ref{V2-zeta-Her}).

\subsubsection{$\chi$ Dra} \label{chidra}

For $\chi$ Dra (FK5 695, HR 6927, HIP 89937, HD 170153, WDS J18211+7244 Aa,Ab), the
orbital elements of \citet{far10}, an estimated primary star angular diameter
$\theta_{P}$ = 1.531 mas provided by the NPOI data reduction software
(\S~\ref{seqred}), an assumed secondary star angular diameter $\theta_{S}$ = 0.1
mas, and $\Delta$$m$ = 2.13 (ORB6) were used as inputs to the binary models.
OYSTER's gridfit procedure was used to improve the initial estimates of $\rho$ and
$\theta$.  This grid search was performed in 0.5 mas steps over a range of 50 mas
in RA and DEC, centered on the position predicted by the \citet{far10} orbit.  The
O-C values of the nightly $\rho$, $\theta$ fits from OYSTER
(Table~\ref{programfit}) with respect to the \citet{far10} orbit range from 2 - 11
mas.  The fitted $\Delta$$m_{700}$ = 2.12 $\pm$ 0.02 (Table~\ref{programdelmag}) is
consistent with the values of both \citet{sco98} [$\Delta$$m_{656}$ = 2.10 $\pm$
0.15; $\Delta$$m_{755}$ = 2.08 $\pm$ 0.15] and \citet{hor08} [$\Delta$$m_{698}$ =
2.03 $\pm$ 0.10; $\Delta$$m_{754}$ = 2.09 $\pm$ 0.10], determined at similar
effective wavelengths.

\subsubsection{$\beta$ Del} \label{betdel}

For $\beta$ Del (HR 7882, HIP 101769, HD 196524, WDS J20375+1436 AB), the orbital
elements of \citet{mut10b}, an estimated primary star angular diameter $\theta_{P}$
= 1.188 mas provided by the NPOI data reduction software (\S~\ref{seqred}), an
assumed secondary star angular diameter $\theta_{S}$ = 0.5 mas, and $\Delta$$m$ =
0.91 (ORB6) were used as inputs to the binary models.  OYSTER's gridfit procedure
was used to verify the initial estimates of $\rho$ and $\theta$ calculated from the
\citet{mut10b} orbit.  This grid search was performed in 0.5 mas steps over a range
of 50 mas in RA and DEC, centered on the positions predicted by the \citet{mut10b}
orbit.  The O-C values of the nightly $\rho$, $\theta$ model fits from OYSTER
(Table~\ref{programfit}) with respect to the \citet{mut10b} orbit are all $\leq$
1.2 mas.  The fitted $\Delta$$m_{700}$ = 1.08 $\pm$ 0.14
(Table~\ref{programdelmag}) is consistent with the mean ($\approx$1.1 $\pm$ 0.1) of
the numerous measurements at $\approx$ 700 nm reported in the INT4.

\subsection{Marginal Detections} \label{mardet}

Lastly, we attempted to model the three binaries ($\eta$ Cep, $\iota$ Peg, and
$\delta$ Aql) possibly resolved by our observations (\S~\ref{bindet}).

\subsubsection{$\eta$ Cep} \label{etacep}

For $\eta$ Cep (FK5 783, HR 7957, HIP 102422, HD 198149, WDS J20453+6150 A), we
used the position and $\Delta$$m$ values from the GRIDFIT output for 2004 October
05 UT (99\% confidence detection, at $\rho$ = 7.51 mas, $\theta$ = 25$\fdg$30, and
$\Delta$$m$ = 3.70), an estimated primary star angular diameter $\theta_{P}$ = 2.85
mas provided by the NPOI data reduction software (\S~\ref{seqred}), and an assumed
secondary star angular diameter $\theta_{S}$ = 0.5 mas as initial values in binary
model fits to the $V^2$ data from the three nights of observations listed in
Table~\ref{progstarsobs}.

We also applied CANDID to the identical data set from 2004 October 05 UT, producing
very similar results ($\rho$ = 7.50 mas, $\theta$ = 26$\fdg$70, $\Delta$$m$ = 3.65,
and $\theta_{P}$ = 2.58).  These values were likewise used as inputs to models of
the data from all three nights.

We also fit an alternative model of a single star with an initial diameter of 2.85
mas or 2.58 mas to the data from all three nights, producing identical results for
each night.

The binary models, starting with either the GRIDFIT or CANDID values, proved to be
the best (much lower $\chi_{\nu}$$^2$) fits to the data of 2004 October 05 UT, and
marginally better fits for 2004 October 01 UT, while the single star model was a
slightly better fit for 2004 September 30 UT.  Thus, we are left with a rather
inconclusive situation where the detection of a new stellar companion remains a
possibility.  (The literature search of \S~\ref{bindet} produced no references to
any likely companions in the detection range of our observations.)  New NPOI
observations are clearly indicated.

\subsubsection{$\iota$ Peg} \label{iotpeg}

For the very well studied binary $\iota$ Peg \citep[FK5 831, HR 8430, HIP 109176,
HD 210027, WDS J22070+2521A;][]{bod99,mor00,kon10}, we fit both binary and
single-star models to the calibrated $V^2$ data resulting from the two nights of
observations listed in Table~\ref{progstarsobs}.

We fit two different binary models using the orbital parameters of the ``Primary
Data Set'' of \citet{bod99}, then those of \citet{kon10}.  In the first model, we
also used an estimated primary star angular diameter $\theta_{P}$ = 1.273 mas
provided by the NPOI data reduction software (\S~\ref{seqred}), a secondary star
angular diameter $\theta_{S}$ = 0.71 \citep{bod99}, and a component magnitude
difference $\Delta$$m$ = 2 based on values at various bandpasses, cited in the
previous references.  In the second model, we used the primary and secondary star
angular diameters of \citet{kon10}, $\theta_{P}$ = 1.06 mas and $\theta_{S}$ = 0.6,
respectively, along with a component magnitude difference $\Delta$$m$ = 2.  For the
single-star model, an estimated angular diameter 1.06 mas was used as the initial
input.

The resulting optimized models on each night show those based on the \citet{bod99}
orbit to be marginally the best fit, followed by those based on the \citet{kon10}
orbit, then the single star model, but the $\chi_{\nu}$$^2$ values do not differ
widely and are all relatively large (2.9 to 5.9).  Thus, we conclude that our
observations have failed to resolve the secondary component in this system.  This
result is not surprising given that the angular separations of the secondary
predicted by the \citet{bod99} and \citet{kon10} orbits for these dates ($\rho$
$\approx$ 1.1 mas to 1.3 mas) are well below the angular resolution of the
observations (\S~\ref{bindet}), the secondary star lying on portions of the highly
inclined orbit \citep[i = 95$\fdg$83,][]{bod99} close to the points of minimum
projected separation.

We also independently examined the calibrated $V^2$ data from both nights using
GRIDFIT, and those from 2004 October 05 UT using CANDID.  GRIDFIT indicated a 99\%
confidence detection at $\rho$ = 9.39 mas, $\theta$ = 26$\fdg$65, and $\Delta$$m$ =
3.20 on 2004 September 30 UT, and a $\sim$ 68\% confidence detection at $\rho$ =
24.80 mas, $\theta$ = 7$\fdg$44, and $\Delta$$m$ = 4.00 on 2004 October 05 UT.
CANDID also produced a weak detection at very similar values ($\rho$ = 24.33 mas,
$\theta$ = 7$\fdg$34, and $\Delta$$m$ = 3.76) on 2004 October 05 UT.  Given that
these $\rho$ and $\theta$ values are widely divergent from the predictions of the
\citet{bod99} and \citet{kon10} orbits, and that a binary model based on the
GRIFIT/CANDID results for the second night produces a fit no better than those
cited above, we again conclude we have not detected the secondary star with our
current observations.  The NPOI currently has operational baselines to $\approx$
100 m, which should make future observations that resolve this binary eminently
practical.

\subsubsection{$\delta$ Aql} \label{delaql}

For $\delta$ Aql (FK5 730, HR 7377, HIP 95501, HD 182640, WDS J19255+0307 A), we
fit four different models to the calibrated $V^2$ data resulting from the four
nights of observations listed in Table~\ref{progstarsobs}:

1) First, using the parameters of the ${\it Hipparcos}$ orbital solution
\citep{esa97}, with an estimated primary star angular diameter of 1.288 mas from
the NPOI data reduction software (\S~\ref{seqred}), an assumed secondary star
angular diameter of 0.5 mas, and an assumed component magnitude difference of
$\Delta$$m$ = 2.

2) Second, using the parameters $\rho$ = 16.8 mas, $\theta$ = 147$\fdg$6, and
$\Delta$$m$ = 3.9 corresponding to the very marginal ($<$ 1 $\sigma$)
$\chi_{\nu}$$^2$(min) seen in the GRIDFIT output for 2004 September 23 UT as an
initial model guess, along with the same component angular diameters.  (CANDID was
also applied to the data from this night, producing very similar results: $\rho$ =
17.2 mas, $\theta$ = 148$\fdg$3, and $\Delta$$m$ = 4.1).

3) Third, using the epoch 1979.4615 speckle observation of \citet{bon80} ($\rho$ =
132 mas, $\theta$ = 130$\arcdeg$) as an initial model guess, with $\Delta$$m$ = 3.9
and the same assumed component angular diameters.

4) Lastly, a model of a single star of an estimated angular diameter of 1.288 mas
as an initial model guess.

The results of these efforts were that the best-fit optimized model for the first
night and marginally best fit to the data from the third night were derived using
the second starting model, while the single-star model was clearly the best fit to
the data for the second and fourth nights.  These results, along with the
significantly different final $\rho$ and $\theta$ values for the binary models for
the first and third nights (16.9 mas and 147$\fdg$8 ${\it vs}$ 10.8 mas and
125$\fdg$6, respectively), lead to little confidence of a definite detection of a
second star.  Once again, additional observations are needed.

\subsection{Limits of Binary Detection} \label{limbindet}

The generally excellent agreement in the results of applying GRIDFIT and CANDID to
our data on five stars ($\beta$ Sct: \S~\ref{betasct}, V1334 Cyg:
\S~\ref{v1334cyg}, $\eta$ Cep: \S~\ref{etacep}, $\iota$ Peg: \S~\ref{iotpeg}, and
$\delta$ Aql: \S~\ref{delaql}) and our successful recovery of 13 of the 15
$\Delta$$m$ Test Binaries (\S~\ref{binknown}) using GRIDFIT lend confidence to our
claim of a limiting $\Delta$$m_{700}$ sensitivity of $\approx$ 3.0, and to perhaps
$\Delta$$m_{700}$ $\sim$ 3.5 for binaries where there is not more than a difference
of two full spectral types between the components (\S~\ref{delmaglimit},
Figure~\ref{dmagsep}, and Table~\ref{known}).  These results, along with the fact
that the effects of bandwidth smearing are explicitly accounted for in GRIDFIT
(\S~\ref{ifov}), also lend confidence in the completeness of our survey out to it's
originally intended 500 mas angular separation limit
(\S~\ref{intro},~\ref{binknown}).

Having completed the detailed modeling of the six detected binaries among the
program sources observed to date, we added the log of the mean of the nightly
$\rho$ values, and the mean $\Delta$$m_{700}$ values for each star from
Tables~\ref{programfit} \&~\ref{programdelmag} to Figure~\ref{dmagsep}.  The
inclusion of the four very wide binaries (10 UMa, $\gamma$ Vir, $\zeta$ Her, and
$\beta$ Del) in this group greatly extends the range of angular separations over
which we have fitted $\Delta$$m_{700}$ values to $\sim$ 860 mas ($\zeta$ Her), and
appears likely consistent with the $\Delta$$m_{700}$ $\approx$ 3.0 limit at smaller
separations.  However, can we claim completeness in our search for new binaries in
the region $>$ 500 mas in Figure~\ref{dmagsep} without performing finely sampled
GRIDFIT searches (\S~\ref{bindet}) out to such large angular separations?  As an
alternative to extending the computationally intensive GRIDFIT process, we
generated simulated $V^2$ ${\it vs}$ wavelength plots using the actual
${\it uv}$-plane coverage on typical nights for $\gamma$ Vir and $\zeta$ Her
($\rho$ $\approx$ 620 mas and $\approx$ 860 mas, respectively) at various binary
position angles and $\Delta$$m$ values.  These simulations show that for the full
range of position angles up to $\pm$ 90$\arcdeg$ from the observed $\theta$ values
for these stars, one or more observations would have shown $\pm$ 5$\%$ - 10$\%$
$V^2$ oscillations so long as $\Delta$$m$ $\leq$ 2.6.  While, as noted in
\S~\ref{zether}, the model fitting process for $\zeta$ Her was not straightforward,
the facts that good-quality fits were still possible and that the simulations
predict similar $V^2$ oscillations regardless of binary position angle indicate
that similarly sampled NPOI observations would have detected (via the GRIDFIT
process) all binaries out to the same range of angular separation and $\Delta$$m$.
We also direct the reader to our plots of typical \textit{uv} coverage in
Figures~\ref{uv} \&~\ref{59cyg-uv-vis-fit}.  Taken together, these demonstrate how
difficult it is for a binary to remain undetected within our limits.  Just a few
scans are needed to show a binary oscillation and deviation from a single star
assumption.

Conversely, it can be shown that near the narrow-separation end of our search range
($\approx$ 3 mas) our search is reasonably complete with respect to the limited
time span of the observations (average of 8 nights per source) occurring at times
when companions in eccentric apparent orbits might lie at angular separations below
our resolution limit.  For the example of the Main Sequence primary stars in our
sample, a limiting detectable magnitude difference of $\Delta$$m_R$ $\approx$ 3
implies primary/secondary spectral type pairings of $\sim$ F0V/K0V at the blue
limit of our sample and $\sim$ K2V/M2V at the red limit, respectively
\citep{dal99}.  These in turn imply mass sums $M_{P}$ + $M_{S}$ of $\approx$ 2.4
$M_{\sun}$ and 1.1 $M_{\sun}$, respectively.  Examples of such systems with
apparent semi-major axes near the 3 mas resolution limit and at the distance limit
of our sample (30 pc) would have physical semi-major axes of $\approx$ 0.1 AU and
orbital periods of $\sim$ 7 - 11 days.  Since it is known that almost all binaries
with periods this short have circular orbits as a result of tidal dissipation
\citep{rag10,dak13}, it is likely that they would have been detectable unless
viewed at inclinations ${\it i}$ $\gtrsim$ 60$\arcdeg$.  A similar analysis
pertains to the systems in our sample with giant primaries or of smaller
$\Delta$$m_R$ (greater mass sum), or at distances $<$ 30 pc (smaller physical
semi-major axes at our angular resolution limit), where the resulting orbital
periods would be even shorter.

\subsection{Angular Diameters of Program Stars} \label{diamBSS}

For those sources where the results of \S~\ref{bindet} and \S~\ref{binBSS}
indicated the lack of detectable stellar companions, we next fit a circular,
uniform intensity disk model \citep[Eq. 9,][]{hbd67a} to all the calibrated,
squared visibility measurements on the nights listed in Table~\ref{progstarsobs}.
The resulting uniform disk angular diameters $\theta_{UD}$, with assigned errors
$\sigma_{UD}$, for 11 sources with $\theta_{UD}$ $\geq$ 2.0 mas are listed in
Table~\ref{progdiams}.  Two contributing factors were combined in quadrature to
determine the errors.

First, a bootstrap Monte Carlo technique \citep{tyc10} was used to estimate the
likely effects on the fitted diameters of quasi-random atmospheric variations on
timescales shorter than the cadence between the program star - calibrator pairs,
as well as the systematic effects of the choice of a standard calibration
weighting interval (\S~\ref{seqred}).  In this technique, synthetic data sets
(here 5,000 per source) are created using the actual observed data points picked
at random, with the total number of points in each set being the same.  The
distribution of the best-fit model diameter fitted to the bootstrapped data sets
was then used to estimate the uncertainty associated with the diameter fitted to
the actual data.  The resulting errors range from 0.2\% to 2.3\% of the fitted
diameters for the stars in Table~\ref{progdiams}.  (Note that the star with the
largest quoted error, $\eta$ Cep, was also examined as a possible binary in
\S~\ref{etacep}.)  Fits to other program sources with angular diameters even
marginally smaller than 2 mas had rapidly increasing uncertainties with decreasing
angular diameter; therefore we decided to limit the stars included in
Table~\ref{progdiams} to $\theta_{UD}$ $\geq$ 2.0 mas, where the errors of
$\lesssim$ 2\% meet the generally recognized minimal standard of accuracy for
astrophysically useful stellar angular diameter measurements \citep{bth97, hlm09}.

The second factor used in estimating the total error in $\theta_{UD}$ for each star
is that due to the estimated 10\% uncertainty in the angular diameters of our
calibrations stars (\S~\ref{seqred}).  The effect of the uncertainty of the
calibrator diameter on the fitted program star diameter was estimated for each
source by applying a $\pm$ 10\% offset to the estimated calibrator angular
diameter, recalculating its expected visibility at the time observation, then
recalibrating and refitting the source visibilities.  The resulting changes in the
fitted $\theta_{UD}$ for the sources in Table~\ref{progdiams} ranged from 0.1\% to
2.4\%, depending on the relative size of the source and its calibrator.  This
result is not surprising given the small angular diameters of the calibrator stars
observed for each of these 11 sources ($\leq$ 1.0 mas,
Tables~\ref{calibs} \&~\ref{progstarsobs}) and the relatively short baselines used
\citep[see e.g.,][]{cru10}.

\begin{figure}
\epsscale{1.0}
\plotone{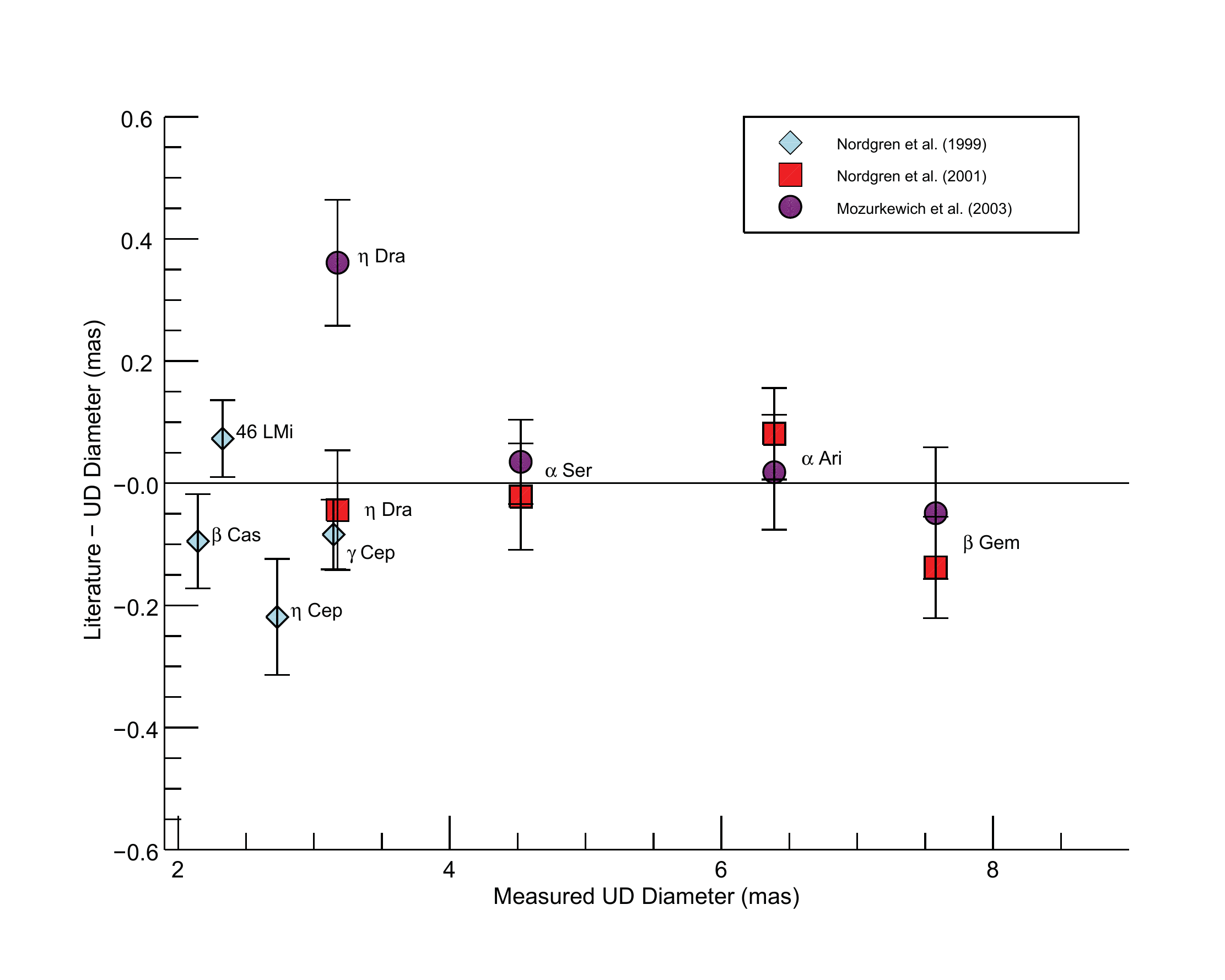}
\caption{{\it Direct Measures}: Comparison of measured uniform disk angular
diameters from this study (``UD Diameter''; Table~\ref{progdiams}, Col. 3) with
previous measurements made using the NPOI \citep{nor99,nor01} and the Mark III
interferometer \citep{moz03}.  The difference (literature - UD Diameter) in
milliarcseconds is plotted against the measured UD Diameter.  Plotted errors are
the quadratic combination of the Table~\ref{progdiams} and literature values.}
\label{UDdiffs}
\end{figure}

\begin{figure}
\epsscale{1.0}
\plotone{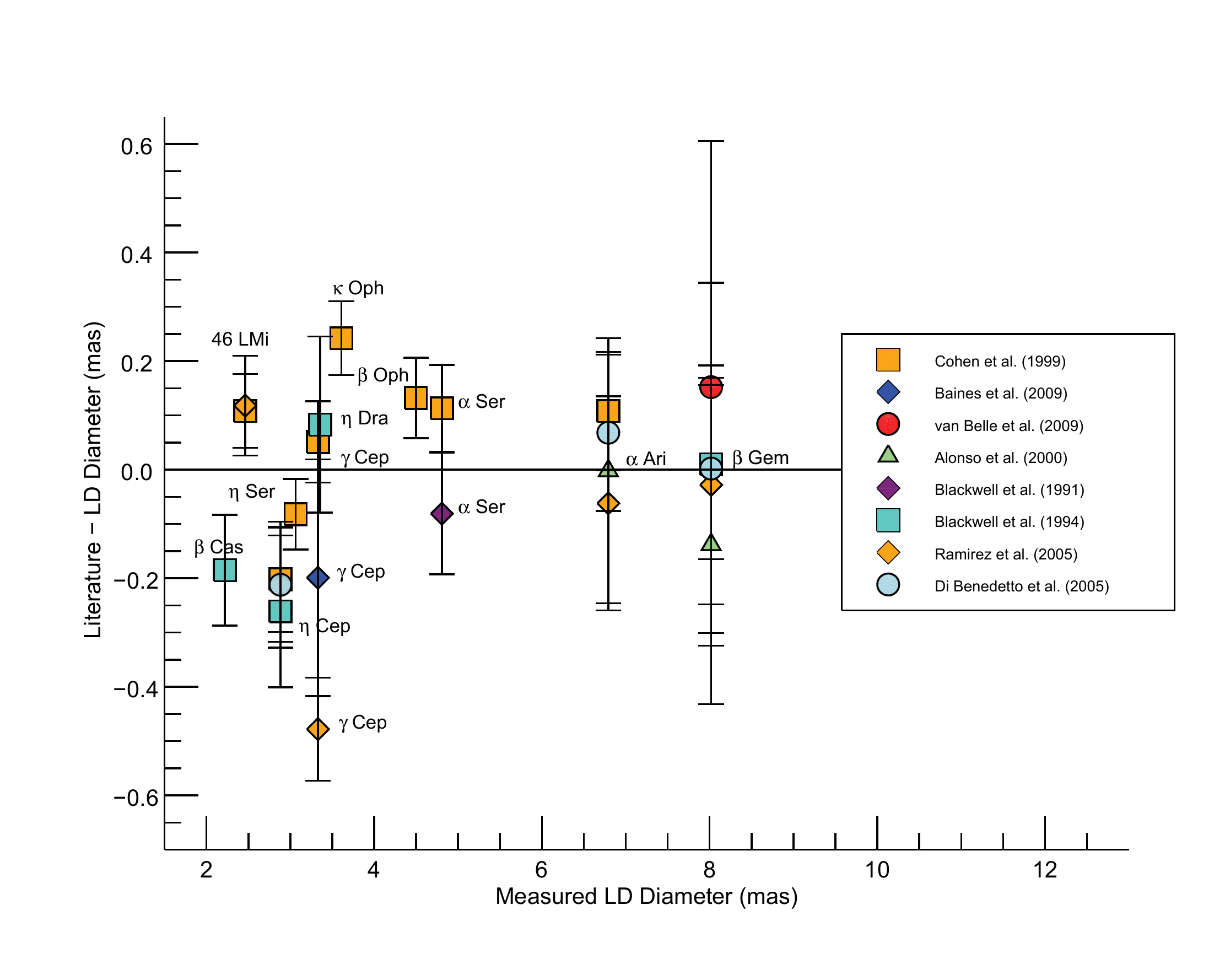}
\caption{{\it Indirect Measures}: Comparison of limb darkened angular diameters
from this study (``LD Diameter''; Table~\ref{progdiams}, Col. 6) with {\it
estimates} derived from radiometric methods \citep{coh99}, SED fits
\citep{bai09,gvb09}, the infrared flux method \citep{aln00,blk91,blk94,ram05}, and
the infrared surface brightness method \citep{dib05}.  The difference (literature -
LD Diameter) in milliarcseconds is plotted against the measured LD Diameter.  The
plotted errors were calculated as per Figure~\ref{UDdiffs}.}
\label{LDdiffsestimated}
\end{figure}

After combining in quadrature the above error estimates, the final estimated errors
for our fitted diameters in Table~\ref{progdiams} range from 0.2\% to 3\%.  As a
further check on their accuracy, we compared these results with other high
precision measurements only available in publications since ${\it c.}$ 1999 for
these sources.  In Figure~\ref{UDdiffs} we plot the diameter difference
(literature minus the Table~\ref{progdiams} $\theta_{UD}$ value) ${\it vs}$ our
$\theta_{UD}$ values for the eight of the 11 stars for which published results
exist \citep[all from the NPOI or Mark III interferometers;][]{nor99,nor01,moz03}.
The overall agreement with these previous results is good, the weighted mean of
the differences being only -0.023 $\pm$ 0.016 mas.  The most discordant point in
Figure~\ref{UDdiffs} is a Mark III measurement for $\eta$ Draconis \citep{moz03}.
We can find no obvious explanation for this difference, but note that the previous
NPOI measurement of this star plotted in Figure~\ref{UDdiffs} \citep{nor01}, and
results from the infrared flux method (see discussion of
Figure~\ref{LDdiffsestimated}, below) show good agreement.

A second comparison can be made between our results, converted to equivalent limb
darkened angular diameters ($\theta_{LD}$), and those from the literature.  We
converted our $\theta_{UD}$ using the limb darkening correction (LDC) factors of
\citet{nor99} for the same stars, or stars of the same spectral types
(Table~\ref{progdiams}, Col. 5).  The resulting equivalent $\theta_{LD}$ values are
listed in Table~\ref{progdiams}, Col. 6, with the estimated errors listed in Col.
7.  These latter errors were derived by scaling our $\theta_{UD}$ errors by the LDC
factor for each star, then combining the result with the estimated error in the LDC
\citep[0.5\% of the derived $\theta_{LD}$;][]{nor99}.  In
Figure~\ref{LDdiffsestimated} we plot the differences between our equivalent
$\theta_{LD}$ values and those from the literature ${\it vs}$ our $\theta_{LD}$
values, for all 11 stars from Table~\ref{progdiams}.  The $\theta_{LD}$ from the
literature were derived either by radiometric methods \citep{coh99}, SED fits
\citep{bai09,gvb09}, the infrared flux method \citep{aln00,blk91,blk94,ram05}, or
the infrared surface brightness method \citep{dib05}.  The overall agreement with
the literature is again excellent, the weighted mean of the differences being
0.007 $\pm$ 0.019 mas.  The largest outlier in Figure~\ref{LDdiffsestimated} is one
of three diameter estimates for $\gamma$ Cephei \citep{ram05}.

\section{Discussion} \label{resdisc}

Our survey efforts to date have yielded a significant number of separation and
position angle measurements for six binary systems from our program sample
(\S~\ref{binBSS}), as well as for 13 additional systems observed to assist in
establishing the NPOI's range of sensitivity to binary detection in $\Delta$$m$
and angular separation (\S~\ref{binknown}).  We have also obtained astrophysically
useful angular diameter measurements of 11 stars (\S~\ref{diamBSS}).  However,
while we have obtained the first visual detection of the secondary star in the
$\beta$ Sct system, one of our test binaries, we have failed to unambiguously
detect any new binary systems among our program stars (\S~\ref{bindet}), with the
possible exception of $\eta$ Cep (\S~\ref{etacep}).  This null result might be
tentatively construed as consistent with the results of \citet{rag12} in support of
the conclusion of \citet{rag10} that the once presumed gap between the angular
sensitivity ranges of spectroscopic and visual techniques for the detection of
stellar companions to nearby solar-type stars has been largely filled by systematic
surveys utilizing radial-velocity and interferometric techniques.  However, while
our survey has an order-of-magnitude greater range of angular separation
sensitivity for binary detection (3 - 860 mas) as compared to that of
\citet[8 - 80 mas]{rag12}, and perhaps greater $\Delta$$m$ sensitivity
($\Delta$$m_{700}$ = 3.0 ${\it vs}$ $\Delta$$m_K$ $\lesssim$ 1.5), the size of our
sample to date [41 stars, only 27 of which are Main Sequence stars by our common
definition (\S~\ref{targsel})] is small compared to that of
\citet[186 stars]{rag12}.

On the other hand, many more stars are potentially available for observation with
the NPOI.  As Figure~\ref{color-mag-observed-and-not} shows, there are still a
significant number (13) of Main Sequence stars from our original sample yet to be
observed as compared to only five evolved stars not observed.  Additionally, the
sensitivity of the NPOI has been significantly improved over the past decade.  As
noted in \S~\ref{npoides} and \S~\ref{targsel}, respectively, the practical limit
for NPOI observations was $m_V$ $\approx$ 4.3 ${\it c.}$ 2004, but is now (2016)
routinely $m_V$ = 5.5, and $m_V$ = 6.0 under good observing conditions.  This
opens the opportunity to greatly expand observational coverage of Main Sequence
stars.  For example, in the range 0.30 $\leq$ ${\bv}$ $\leq$ 0.70 (spectral types
F0 through G5), to a limiting magnitude $m_V$ = 6.0, it is possible to define a
complete, volume-limited sample of $\sim$ 50 stars within $\approx$ 17 pc, with
parallax errors $\leq$ 5\% \citep{vlw07} and $\delta$ $\geq$ -20$\arcdeg$.  The
combination of a greater $m_V$ limit with a smaller distance limit for this sample
results in proportionately many more intrinsically fainter stars in the range 0.57
$\lesssim$ ${\bv}$ $\leq$ 0.70 (spectral types G0 - G5) relative to our current
program sample (Figure~\ref{color-mag-bss-with-Main-Seq}).  Observations of this
list would provide a more robust comparison with the results of other surveys of
solar-type stars \citep[e.g.,][]{rag12} using a complementary wavelength bandpass,
coupled with a capability for binary detection over a wide range of angular
separations.  Observing a sample of $\sim$ 50 stars in a reasonable length of time
with the NPOI is also eminently practical.  Based on the experience of the work
reported here, approximately five nights of multiple observations of each star
under reasonable observing conditions should be adequate to detect binary
companion(s) in the $\Delta$$m$ - angular separation range demonstrated in this
paper.  Given the known weather and seeing statistics at the NPOI, and a
reasonably-sized nightly observing list (5 - 7 sources, plus calibrators), the
survey of a 50-star sample could be completed within $\sim$ 1 year utilizing
$\approx$ 20\% - 25\% of all scheduled NPOI observing time.  Also, the substitution
of siderostat stations more recently commissioned than those used in the present
study could increase the angular resolution of future NPOI multiplicity surveys by
up to a factor of three while maintaining the same simple fringe detection
configuration described in \S~\ref{config}.  This increase in angular resolution
would provide obvious benefits in terms of resolving far more of the known or
suspected spectroscopic binaries (\S~\ref{bindet}), and possible, as yet unknown,
stellar companions at similar angular separations.  The bandpass of NPOI
observations is also expected to be extended towards the blue (to 450 - 860 nm in
32 channels) in the near future, making the detection of binaries with components
of widely differing spectral types (\S~\ref{nondet}) more likely.

\section{Conclusions} \label{conclus}

We have presented the first results from an ongoing survey for multiplicity among
the bright stars using the NPOI.  The initial source sample emphasized bright stars
of spectral types F0 through K2.  We report observations of 41 stars ($m_V$ $\leq$
4.30), all having been observed on multiple nights.  Observations of known binary
systems among the program star sample, combined with additional observations of
other known binaries, including the first ``visual'' detection of the secondary
star in the $\beta$ Sct system at precisely measured separations and position
angles, demonstrate the NPOI's sensitivity for binary detection over a wide range
of angular separations (3 - 860 mas) at component magnitude differences
$\Delta$$m_{700}$ $\lessapprox$ 3, and to perhaps $\Delta$$m_{700}$ $\sim$ 3.5 for
binaries where the component spectral types differ by less than two.  Fitted
angular separations, position angles, and component magnitude differences for six
previously known binaries from the program sample are presented, along with angular
diameters for 11 resolved stars from the sample that have no detected stellar
companions.  In the light of the significant improvements made to the limiting
sensitivity of NPOI observations ($m_V$ $\approx$ 6.0) in recent years, plans are
being drafted to extend the initial survey to a complete, volume-limited sample of
stars of spectral types F0 through G5.

\acknowledgments

We gratefully acknowledge NPOI observers D. Allen, L. Bright, B. Burress, C.
Denison, L. Foley, B. O'Neill, C. Sachs, S. Strosahl, D. Theiling, W. Wack, and R.
Winner for their careful and efficient operation of the NPOI over the many epochs
of data collection required for this paper.  We also gratefully acknowledge the
skillful instrument makers at the USNO and Lowell Observatory instrument shops, and
thank the staff of the USNO Library for assistance with our many reference requests.
Lastly, we thank Dr. Paul Shankland, Director of USNO Flagstaff Station, for his
long-standing and continuous support of the NPOI Program.

The NPOI project is funded by the Oceanographer of the Navy and the Office of Naval
Research.  This research has made use of the SIMBAD literature database and the
VizieR catalog access tool, both operated at CDS, Strasbourg, France, the
Washington Double Star Catalog and associated catalogs, maintained at the U.S.
Naval Observatory, Washington, DC, the NASA/IPAC Infrared Science Archive, operated
by the Jet Propulsion Laboratory, California Institute of Technology, under
contract with the National Aeronautics and Space Administration, and the JSTOR
digital archive.


\clearpage



\clearpage


\clearpage


\clearpage


\clearpage


\clearpage


\clearpage


\clearpage


\clearpage


\clearpage


\clearpage


\clearpage


\clearpage


\clearpage


\clearpage


\clearpage


\clearpage



\begin{deluxetable}{rcccccrccc}
\tabletypesize{\scriptsize}
\tablecaption{Program Stars \label{targetlist}}
\tablewidth{0pt}
\tablehead{
\colhead{} & \colhead{} & \colhead{} & \colhead{Other} & \colhead{$m_V$} & \colhead{$\bv$} & \colhead{$\pi$} &
\colhead{Distance} & \colhead{$M_V$} & \colhead{Spectral} \\
\colhead{HIP} & \colhead{HR} & \colhead{FK5} & \colhead{Name} & \colhead{(mag)} & \colhead{(mag)} & \colhead{(mas)} &
\colhead{(pc)} & \colhead{(mag)} & \colhead{Type} \\
\colhead{(1)} & \colhead{(2)} & \colhead{(3)} & \colhead{(4)} & \colhead{(5)} & \colhead{(6)} & \colhead{(7)} & \colhead{(8)} &
\colhead{(9)} & \colhead{(10)} \\
}
\startdata
   746 &      21   &    2    & $\beta$ Cas    & 2.27 & 0.37 &  59.89 & 16.70 &  1.16 & F2III-IV     \\
  3419 &     188   &   22    & $\beta$ Cet    & 2.04 & 1.04 &  34.04 & 29.38 & -0.30 & G9.5IIICH-1  \\
  3821 &     219   & \nodata & $\eta$ Cas     & 3.45 & 0.58 & 167.99 &  5.95 &  4.58 & F9V+dM0      \\
  7513 &     458   & 1045    & $\upsilon$ And & 4.10 & 0.54 &  74.25 & 13.47 &  3.45 & F8V          \\
  8102 &     509   &   59    & $\tau$ Cet     & 3.49 & 0.73 & 274.17 &  3.65 &  5.68 & G8V          \\
  8796 &     544   &   64    & $\alpha$ Tri   & 3.41 & 0.48 &  50.87 & 19.66 &  1.94 & F6IV         \\
  9884 &     617   &   74    & $\alpha$ Ari   & 2.01 & 1.17 &  49.48 & 20.21 &  0.48 & K2-IIICa-1   \\
 12777 &     799   &   93    & $\theta$ Per   & 4.10 & 0.50 &  89.03 & 11.23 &  3.85 & F8V          \\
\enddata
\tablecomments{Cols. (1), (7), and (8): ${\it Hipparcos}$ number, parallax, and distance from the
$\mathit{Hipparcos\ Catalogue}$ \citep{esa97}.  Cols. (2), (3), and (4): HR number \citep{hof91}, FK5
number \citep{frk88}, and other name (Bayer or Flamsteed designation) from the $\mathit{SIMBAD}$
database \citep{wen00}, and references therein.  Cols. (5), (6), and (9): Apparent $m_V$ magnitude,
$\bv$ color index, and absolute $M_V$ magnitude from TPF-I SWG spreadsheet \citep{law07}, and
references therein.  Col. (10): Spectral type from the $\mathit{The\ Bright\ Star\ Catalogue}$, 5th
Rev. Ed. \citep{hof91}, unless otherwise noted.}
\tablenotetext{a}{\citet{edw76}}
\end{deluxetable}

\clearpage


\begin{deluxetable}{rcccccrc}
\tabletypesize{\scriptsize}
\tablecaption{Calibrator Stars \label{calibs}}
\tablewidth{0pt}
\tablehead{
\colhead{} & \colhead{} & \colhead{} & \colhead{} & \colhead{} & \colhead{} & \colhead{} & \colhead{Adopted} \\
\colhead{} & \colhead{} & \colhead{} & \colhead{Other} & \colhead{$m_V$} & \colhead{${\it V - K}$} & \colhead{Spectral} &
\colhead{$\theta_{LD}$} \\
\colhead{HIP} & \colhead{HR} & \colhead{FK5} & \colhead{Name} & \colhead{(mag)} & \colhead{(mag)} & \colhead{Type} &
\colhead{(mas)} \\
\colhead{(1)} & \colhead{(2)} & \colhead{(3)} & \colhead{(4)} & \colhead{(5)} & \colhead{(6)} & \colhead{(7)} & \colhead{(8)} \\
}
\startdata

  2920 &  153 &      17 & $\zeta$ Cas      & 3.66 & -0.59 & B2IV       & 0.432     \\
  8886 &  542 &      63 & $\epsilon$ Cas   & 3.38 & -0.58 & B3III      & 0.493     \\
  9598 &  580 &      70 & 50 Cas           & 3.98 &  0.06 & A2V        & 0.565     \\
 10670 &  664 &      79 & $\gamma$ Tri     & 4.01 &  0.05 & A1Vnn      & 0.555     \\
 23767 & 1641 &     185 & $\eta$ Aur       & 3.17 & -0.69 & B3V        & 0.507     \\
 26311 & 1903 &     210 & $\epsilon$ Ori   & 1.70 & -0.57 & B0Ia       & 0.876$^a$ \\
 33018 & 2540 &     261 & $\theta$ Gem     & 3.60 &  0.44 & A3III      & 0.797$^b$ \\
 35350 & 2763 &     277 & $\lambda$ Gem    & 3.58 &  0.04 & A3V        & 0.673     \\
\enddata
\tablecomments{Cols.(1) through (4) per Table~\ref{targetlist}.  Cols. (5) and (7):
$m_V$ magnitude and spectral type from $\mathit{The\ Bright\ Star\ Catalogue,}$ 5th
Rev. Ed. \citep{hof91}.  Col. (6): ${\it V - K}$ color index calculated from Col. (5)
value and ${\it K}$ magnitude from the $\mathit{2MASS}$ catalog \citep{cut03}.  Col.
(8): Adopted limb darkened angular diameter (\S~\ref{seqred}).}
\tablenotetext{a}{\citet{moz91}, corrected to limb darkened diameter
(\S~\ref{seqred}).}
\tablenotetext{b}{\citet{blk91}}
\end{deluxetable}

\clearpage


\begin{deluxetable}{rccccr}
\tabletypesize{\scriptsize}
\tablecaption{Program Stars - Observation Log \label{progstarsobs}}
\tablewidth{0pt}
\tablehead{
\colhead{Object} & \colhead{Other} & \colhead{UT Date} & \colhead{No. of} & \colhead{No. of} & \colhead{Calibrator} \\
\colhead{FK5/HR} & \colhead{Name} & \colhead{(yyyy-mm-dd)} & \colhead{Observations} & \colhead{Visibilities} & \colhead{FK5/HR} \\
\colhead{(1)} & \colhead{(2)} & \colhead{(3)} & \colhead{(4)} & \colhead{(5)} & \colhead{(6)} \\
}
\startdata
FK5 2        & $\beta$ Cas    & 2004-09-23 &  9 & 279 & FK5 63   \\
             &                & 2004-09-24 &  8 & 248 & FK5 63   \\
             &                & 2004-09-27 &  9 & 279 & FK5 63   \\
             &                &            &    &     &          \\
HR 219       & $\eta$ Cas     & 2004-09-30 &  1 &  31 & FK5 17   \\
             &                & 2004-10-01 &  7 & 217 & FK5 17   \\
             &                & 2004-10-05 &  4 & 108 & FK5 17   \\
             &                &            &    &     &          \\
\enddata
\tablecomments{Cols. (1) and (6): FK5 \citep{frk88} or HR number \citep{hof91},
as per NPOI scheduling and archiving software.  Col. (2): Per Table~\ref{targetlist}.
Col. (3): UT date of NPOI observations.  Col. (4): Total number of interferometric
observations using AC-AE and AC-AW baselines.  Col. (5): Number of $V^2$ measurements
made over available spectral channels, over all recorded beam combiner outputs (less
edits).}
\tablenotetext{a}{HR 4826 = $\gamma$ Vir B, but both A and B components of
$\gamma$ Vir = HIP 61941 observed.}
\end{deluxetable}

\clearpage


\begin{deluxetable}{rcclcrc}
\tabletypesize{\scriptsize}
\tablecaption{$\Delta$$m$ Test Binaries \label{known}}
\tablewidth{0pt}
\tablehead{
\colhead{} & \colhead{} & \colhead{} & \colhead{} & \colhead{Other} & \colhead{$m_V$} & \colhead{Spectral} \\
\colhead{HR} & \colhead{FK5} & \colhead{HIP} & \colhead{WDS} & \colhead{Name} & \colhead{Magnitude} & \colhead{Type} \\
\colhead{(1)} & \colhead{(2)} & \colhead{(3)} & \colhead{(4)} & \colhead{(5)} & \colhead{(6)} & \colhead{(7)} \\
}
\startdata
1948 & \nodata & \nodata & J05407$-$0157Aa,Ab   & $\zeta$ Ori A & 2.05 & O9.5Iab+B1IV$^a$  \\
3594 &     341 &   44471 & J09036+4709AB        & $\kappa$ UMa  & 3.60 & A1Vn              \\
3852 &     365 &   47508 & J09412+0954Aa,Ab     & o Leo         & 3.52 & F9III+A5m$^b$     \\
5054 &     497 & \nodata & J13239+5456Aa,Ab     & $\zeta$ UMa A & 2.27 & A2V+A2V$^c$       \\
5291 &     521 &   68756 & \nodata              & $\alpha$ Dra  & 3.65 & A0III             \\
5747 &     572 &   75695 & J15278+2906AB        & $\beta$ CrB   & 3.68 & F0p               \\
6023 &     601 &   79101 & J16088+4456AB        & $\phi$ Her    & 4.26 & B9VspHgMn+A8V$^d$ \\
6168 &     621 &   81126 & J16341+4226AB        & $\sigma$ Her  & 4.20 & B9V               \\
\enddata
\tablecomments{Cols. (1) and (2): HR \citep{hof91} and FK5 \citep{frk88} numbers, as per the
NPOI scheduling and archiving software, or from the $\mathit{SIMBAD}$ database \citep{wen00}.
Cols. (3), (4), and (5): ${\it Hipparcos}$ number, WDS number, and other name (Bayer or
Flamsteed designation), unless otherwise noted, are from the $\mathit{SIMBAD}$ database, and
references therein.  Cols. (6) and (7): $m_V$ magnitude and spectral type from
$\mathit{The\ Bright\ Star\ Catalogue,}$ 5th Rev. Ed. \citep{hof91}, unless otherwise noted.}
\tablenotetext{a}{\citet{hum13}}
\tablenotetext{b}{\citet{hum01}}
\tablenotetext{c}{\citet{hum98}}
\tablenotetext{d}{Spectral type of secondary: \citet{zav07}}
\tablenotetext{e}{\citet{paa05}}
\tablenotetext{f}{\citet{lsh68}}
\tablenotetext{g}{\citet{mas01}}
\tablenotetext{h}{\citet{ksc15}}
\end{deluxetable}

\clearpage


\begin{deluxetable}{rccccr}
\tabletypesize{\scriptsize}
\tablecaption{$\Delta$$m$ Test Binaries - Observation Log \label{binobs}}
\tablewidth{0pt}
\tablehead{
\colhead{Object} & \colhead{Other} & \colhead{UT Date} & \colhead{No. of} & \colhead{No. of} & \colhead{Calibrator} \\
\colhead{FK5/HR} & \colhead{Name} & \colhead{(yyyy-mm-dd)} & \colhead{Observations} & \colhead{Visibilities} & \colhead{FK5/HR} \\
\colhead{(1)} & \colhead{(2)} & \colhead{(3)} & \colhead{(4)} & \colhead{(5)} & \colhead{(6)} \\
}
\startdata
HR 1948      & $\zeta$ Ori A  & 2004-03-11 &  4 & 124 & FK5 210  \\
             &                & 2004-03-12 &  3 &  93 & FK5 210  \\
             &                &            &    &     &          \\
FK5 341      & $\kappa$ UMa   & 2004-03-11 &  3 &  93 & FK5 368  \\
             &                & 2004-03-14 &  8 & 248 & FK5 368  \\
             &                & 2004-03-15 & 10 & 310 & FK5 368  \\
             &                &            &    &     &          \\
FK5 365      & o Leo          & 2004-03-17 & 10 & 310 & FK5 379  \\
\enddata
\tablecomments{Cols. (1) and (6): FK5 \citep{frk88} or HR number
\citep{hof91}, as per the NPOI scheduling and archiving software.
Col. (2): Per Table~\ref{known}.  Col. (3): UT date of NPOI
observations.  Col. (4): Total number of interferometric observations
using AC-AE and AC-AW baselines, unless otherwise noted.  Col. (5):
Number of $V^2$ measurements made over available spectral channels,
over all recorded beam combiner outputs (less edits).}
\tablenotetext{a}{See Table~\ref{sigHerobs}}
\tablenotetext{b}{AC-AE and AE-AN baselines used.  AE-AN baseline =
34.9 m, with AN station at 330$\arcdeg$ with respect to AE.}
\tablenotetext{c}{AC-AE, AE-AN, AN-AC, and AC-AW baselines used.
AN-AC baseline = 22.8 m, with AC station at 180$\arcdeg$ with respect
to AN.  Closure data available between the first three baselines.}

\end{deluxetable}

\clearpage


\begin{deluxetable}{lcccrrrrr}
\tabletypesize{\scriptsize}
\tablecaption{$\Delta$$m$ Test Binaries - Binary Model Fits \label{knownfit}}
\tablewidth{0pt}
\tablehead{
\colhead{Object} & \colhead{Other} & \colhead{UT Date} & \colhead{} & \colhead{$\rho$} & \colhead{$\theta$} &
\colhead{$\sigma_{maj}$} & \colhead{$\sigma_{min}$} & \colhead{$\phi$} \\
\colhead{FK5/HR} & \colhead{Name} & \colhead{(yyyy-mm-dd)} & \colhead{MJD} & \colhead{(mas)} & \colhead{(deg)} &
\colhead{(mas)} & \colhead{(mas)} & \colhead{(deg)} \\
\colhead{(1)} & \colhead{(2)} & \colhead{(3)} & \colhead{(4)} & \colhead{(5)} & \colhead{(6)} & \colhead{(7)} & \colhead{(8)} &
\colhead{(9)} \\
}
\startdata
HR 1948$^a$ & $\zeta$ Ori A & 2004-03-11 & 53075.32 &  23.5  & 332.2  & 0.49 & 0.28 & 178   \\
            &               & 2004-03-12 & 53076.30 &  23.6  & 332.5  & 0.48 & 0.29 &   6   \\
            &               &            &          &        &        &      &      &       \\
FK5 341     & $\kappa$ UMa  & 2004-03-11 & 53075.32 & 145.92 & 306.58 & 0.58 & 0.27 & 130.6 \\
            &               & 2004-03-14 & 53078.31 & 146.25 & 306.51 & 0.54 & 0.27 & 145.4 \\
            &               & 2004-03-15 & 53079.33 & 146.37 & 306.53 & 0.48 & 0.29 & 137.1 \\
            &               &            &          &        &        &      &      &       \\
FK5 365     & $o$ Leo       & 2004-03-17 & 53081.34 &   3.50 & 219.72 & 0.65 & 0.27 & 163.4 \\
\enddata
\tablecomments{Col. (1): FK5 \citep{frk88} or HR number \citep{hof91}, as per NPOI scheduling and
archiving software.  Col. (2): per Table~\ref{known}.  Col. (3): UT date of NPOI observations.
Col. (4): Modified Julian Date (MJD) of NPOI observations.  Col. (5): Fitted binary separation.
Col. (6): Fitted binary position angle.  Col. (7): Semimajor axis of error ellipse.  Col. (8):
Semiminor axis of error ellipse.  Col. (9): Position angle of error ellipse.}
\tablenotetext{a}{Fitted values from \citet{hum13}}
\tablenotetext{b}{Models fitted only to data from AC-AE and AC-AW baselines}
\tablenotetext{c}{Fitted values from \citet{zav07}}
\tablenotetext{d}{See Table~\ref{sigHerfits}}
\end{deluxetable}

\clearpage


\begin{deluxetable}{rcccccrc}
\tabletypesize{\scriptsize}
\tablecaption{$\sigma$ Her - Calibrator Stars \label{sigHercalibs}}
\tablewidth{0pt}
\tablehead{
\colhead{} & \colhead{} & \colhead{} & \colhead{} & \colhead{} & \colhead{} & \colhead{} & \colhead{Adopted} \\
\colhead{} & \colhead{} & \colhead{} & \colhead{Other} & \colhead{$m_V$} & \colhead{${\it V - K}$} & \colhead{Spectral} &
\colhead{$\theta_{LD}$} \\
\colhead{HIP} & \colhead{HR} & \colhead{FK5} & \colhead{Name} & \colhead{(mag)} & \colhead{(mag)} & \colhead{Type} &
\colhead{(mas)} \\
\colhead{(1)} & \colhead{(2)} & \colhead{(3)} & \colhead{(4)} & \colhead{(5)} & \colhead{(6)} & \colhead{(7)} & \colhead{(8)} \\
}
\startdata
  48319 & 3888 &     368 & $\upsilon$ UMa   & 3.80 &  0.65 & F2IV      & 0.89  \\
  49583 & 3975 &     379 & $\eta$ Leo       & 3.49 &  0.19 & A0Ib      & 0.75  \\
  50372 & 4033 &     383 & $\lambda$ UMa    & 3.40 & -0.02 & A2IV      & 0.71  \\
  54879 & 4359 &     423 & $\theta$ Leo     & 3.34 &  0.26 & A2V       & 0.86  \\
  55434 & 4386 &     427 & $\sigma$ Leo     & 4.06 & -0.08 & B9.5Vs    & 0.49  \\
  61281 & 4787 &     472 & $\kappa$ Dra     & 3.87 &  0.05 & B6IIIpe   & 0.59  \\
  63125 & 4915 &     485 & $\alpha$$^2$ CVn & 2.90 & -0.24 & A0pSiEuHg & 0.76  \\
  64844 & 5017 &     494 & 20 CVn           & 4.73 &  0.72 & F3III     & 0.60  \\
\enddata
\tablecomments{Cols. (1) through (4): per Table~\ref{targetlist}.  Col. (5): From
the $\mathit{Hipparcos\ Catalogue}$ \citep{esa97}, or the $\mathit{HD\ Catalogue}$
\citep{hde18}.  Col. (6): ${\it V - K}$ color index calculated from Col. (5) value
and ${\it K}$ magnitude from the $\mathit{2MASS}$ catalog \citep{cut03}.  Col. (7):
Spectral type from $\mathit{The\ Bright\ Star\ Catalogue,}$ 5th Rev. Ed.
\citep{hof91}.  Col. (8): Adopted limb darkened angular diameter
(\S~\ref{seqred}).}
\end{deluxetable}

\clearpage


\begin{deluxetable}{lccl}
\tabletypesize{\scriptsize}
\tablecaption{$\sigma$ Her - Observation Log \label{sigHerobs}}
\tablewidth{0pt}
\tablehead{
\colhead{UT Date} & \colhead{No. of} & \colhead{No. of} & \colhead{Calibrators} \\
\colhead{(yyyy-mm-dd)} & \colhead{Observations} & \colhead{Visibilities} & \colhead{FK5/HR} \\
\colhead{(1)} & \colhead{(2)} & \colhead{(3)} & \colhead{(4)} \\
}
\startdata
1997-05-29$^a$ &  8 &  674 & FK5 423, FK5 494, FK5 609, FK5 641 \\
               &    &      & FK5 668, FK5 768 \\
1997-05-30$^a$ &  8 &  680 & FK5 423, FK5 494, FK5 609, FK5 641 \\
               &    &      & FK5 668, FK5 768 \\
1997-05-31$^a$ &  3 &  255 & FK5 501, FK5 641, FK5 668, FK5 768 \\
1998-03-20$^a$ &  6 &  570 & FK5 472, FK5 641, FK5 681 \\
1998-05-23$^a$ & 16 & 1520 & FK5 383, FK5 641, FK5 716, FK5 1558 \\
1998-07-04$^a$ &  9 &  852 & FK5 485, FK5 641, FK5 717, FK5 733 \\
\enddata
\tablecomments{Col. (1): UT date of NPOI observations.  Col. (2): Total
number of interferometric observations on AC-AE and AC-AW baselines, unless
otherwise noted.  Col. (3): Number of $V^2$ measurements made over available
spectral channels, over all recorded beam combiner outputs (less edits).
Col. (4): per Table~\ref{sigHercalibs}.}
\tablenotetext{a}{AC-AE, AC-AW, and AE-AW baselines used.  AE-AW baseline =
37.5 m, with AW station at 266$\arcdeg$ with respect to AE.  Data collected
using earlier three-station beam combiner, producing 32 spectral channel data,
covering 450 nm - 860 nm, on each of three spectrometer outputs (less one
defective channel, for a total of 95 $V^2$ measurements per observation).}
\tablenotetext{b}{AC-AE, AC-E02, and AE-E02 baselines used.  AC-E02 baseline =
9.6 m, with E02 station at 39$\arcdeg$ with respect to AC.  AE-E02 baseline =
15.9 m, with E02 station at 314$\arcdeg$ with respect to AE.  Data collected
using earlier three-station beam combiner, producing 32 spectral channel data,
covering 450 nm - 860 nm, on each of three spectrometer outputs (less one
defective channel, for a total of 95 $V^2$ measurements per observation).}
\end{deluxetable}

\clearpage


\begin{deluxetable}{ccrrrrrrr}
\tabletypesize{\scriptsize}
\tablecaption{$\sigma$ Her - Binary Model Fits \label{sigHerfits}}
\tablewidth{0pt}
\tablehead{
\colhead{UT Date} & \colhead{} & \colhead{$\rho$} & \colhead{$\theta$} & \colhead{$\sigma_{\rm maj}$} &
\colhead{$\sigma_{\rm min}$} & \colhead{$\phi$} & \colhead{O--C$_\rho$} & \colhead{O--C$_\theta$} \\
\colhead{(yyyy-mm-dd)} & \colhead{MJD} & \colhead{(mas)} & \colhead{(deg)} & \colhead{(mas)} & \colhead{(mas)} &
\colhead{(deg)} & \colhead{(mas)} & \colhead{(deg)} \\
\colhead{(1)} & \colhead{(2)} & \colhead{(3)} & \colhead{(4)} & \colhead{(5)} & \colhead{(6)} & \colhead{(7)} & \colhead{(8)} &
\colhead{(9)} \\
}
\startdata
1997-05-29 & 50597.36 &  35.67 &  21.71 &  0.45 &  0.17 & 171.3 &  0.45 & -0.7 \\
1997-05-30 & 50598.34 &  35.57 &  21.75 &  0.46 &  0.17 & 173.0 &  0.28 & -0.5 \\
1997-05-31 & 50599.41 &  36.02 &  20.85 &  0.58 &  0.18 & 145.9 &  0.65 & -1.3 \\
1998-03-20 & 50892.47 &  14.06 & 286.96 &  0.60 &  0.16 &   7.5 &  0.02 & -2.0 \\
1998-05-23 & 50956.32 &  20.07 & 244.99 &  0.50 &  0.16 & 171.7 &  0.08 & -0.5 \\
1998-07-04 & 50998.19 &  26.83 & 231.05 &  0.57 &  0.17 & 175.1 &  0.12 & -0.6 \\
1998-10-07 & 51093.11 &  42.53 & 217.17 &  0.60 &  0.21 & 124.3 & -0.49 & -0.1 \\
2001-05-10 & 52039.41 & 115.13 & 193.45 &  0.56 &  0.16 & 165.5 &  0.40 &  0.1 \\
\enddata

\tablecomments{Col. (1): UT date of NPOI observations.  Col. (2): MJD of NPOI observations.
Col. (3): Fitted binary separation.  Col. (4): Fitted binary position angle.  Col. (5):
Semimajor axis of error ellipse.  Col. (6): Semiminor axis of error ellipse.  Col. (7):
Position angle of error ellipse.  Col.(8): Difference between fitted binary separation and
prediction from new NPOI orbit (\S~\ref{sigher}, Figure~\ref{sig-Her-orbit}).  Col.(9):
Difference between fitted binary position angle and prediction from new NPOI orbit
(\S~\ref{sigher}, Figure~\ref{sig-Her-orbit}).}
\end{deluxetable}

\clearpage


\begin{deluxetable}{lc}
\tabletypesize{\scriptsize}
\tablecaption{$\sigma$ Her - Orbital Elements \label{sig-Her-elements}}
\tablewidth{0pt}
\tablehead{
\colhead{Parameter} & \colhead{Value} \\
}
\startdata
${\it a}$ (mas)           &    76.21    $\pm$  0.27$^a$ \\
${\it e}$                 &     0.5135  $\pm$  0.0028   \\
${\it i}$ (deg)           &   105.25    $\pm$  0.51     \\
$\omega$ (deg)            &   184.97    $\pm$  0.40     \\
$\Omega$ (deg) (J2000.0)  &    14.95    $\pm$  0.47     \\
${\it P}$ (days)          &  2706.19    $\pm$  4.89     \\
${\it T}$ (JD - 244E4)    & 10665.4     $\pm$  2.68     \\
$\chi_{\nu}$$^2$          &     1.34                    \\
\enddata
\tablenotetext{a}{Possible systematic error 0.5 mas (\S~\ref{channels})}
\end{deluxetable}

\clearpage


\begin{deluxetable}{lcrcccc}
\tabletypesize{\scriptsize}
\tablecaption{$\Delta$$m$ Test Binaries - Mean Separations and $\Delta$$m$ Fits \label{knowndelmag}}
\tablewidth{0pt}
\tablehead{
\colhead{Object} & \colhead{Other} & \colhead{$<$$\rho$$>$} & \colhead{$\sigma_{\rho}$} & \colhead{log$<$$\rho$$>$} &
\colhead{$<$$\Delta$$m_{700}$$>$} & \colhead{$\sigma_{\Delta m}$} \\
\colhead{FK5/HR} & \colhead{Name} & \colhead{(mas)} & \colhead{(mas)} & \colhead{mas} & \colhead{(mag)} & \colhead{(mag)} \\
\colhead{(1)} & \colhead{(2)} & \colhead{(3)} & \colhead{(4)} & \colhead{(5)} & \colhead{(6)} & \colhead{(7)} \\
}
\startdata
HR 1948     & $\zeta$ Ori A      &  23.55     & 0.07    & 1.37 & 2.20$^a$ & 0.1$^a$  \\
FK5 341     & $\kappa$ UMa       & 146.18     & 0.23    & 2.17 & 0.48     & 0.05     \\
FK5 365     & $o$ Leo            &   2.86     & 0.59    & 0.46 & 1.05     & 0.07     \\
FK5 497     & $\zeta$ UMa A      &   7.32     & 0.32    & 0.86 & 0.00$^b$ & 0.02$^b$ \\
FK5 521     & $\alpha$ Dra (min) &   2.61     & \nodata & 0.42 & 1.83     & 0.07     \\
            & $\alpha$ Dra (max) &   6.23     & \nodata & 0.79 & 1.83     & 0.07     \\
FK5 572     & $\beta$ CrB        & 224.95     & 0.36    & 2.35 & 1.78     & 0.01     \\
FK5 601     & $\phi$ Her         &  45.74     & 2.41    & 1.66 & 2.39$^c$ & 0.05$^c$ \\
\enddata
\tablecomments{Col. (1): FK5 \citep{frk88} or HR number \citep{hof91}, as per NPOI
scheduling and archiving software.  Col. (2): per Table~\ref{known}.  Col. (3): Mean of
nightly separation measurements (except for $\alpha$ Dra and $\sigma$ Her).  Col. (4):
Standard deviation of the nightly separation measurements.  Col. (5): log of Col. (3)
value.  Col. (6): Average of the nightly fitted component magnitude difference at 700
nm, unless otherwise noted.  Col. (7): Standard deviation of the nightly
$\Delta$$m_{700}$ fits.}
\tablenotetext{a}{Values from \citet{hum13}}
\tablenotetext{b}{Values from \citet{hum98}}
\tablenotetext{c}{Values from \citet{zav07}}
\tablenotetext{d}{Not detected; estimated using orbital elements of \citet{raf13}
and references therein, see text}
\tablenotetext{e}{Not detected; $\Delta$$m_V$ value from \citet{paa98}, see text}
\tablenotetext{f}{Not detected; value from \citet{hmm97}, see text}
\end{deluxetable}

\clearpage


\begin{deluxetable}{lcccrrrrr}
\tabletypesize{\scriptsize}
\tablecaption{Binary Program Stars - Binary Model Fits \label{programfit}}
\tablewidth{0pt}
\tablehead{
\colhead{Object} & \colhead{Other} & \colhead{UT Date} & \colhead{} & \colhead{$\rho$} & \colhead{$\theta$} &
\colhead{$\sigma_{maj}$} & \colhead{$\sigma_{min}$} & \colhead{$\phi$} \\
\colhead{FK5/HR} & \colhead{Name} & \colhead{(yyyy-mm-dd)} & \colhead{MJD} & \colhead{(mas)} & \colhead{(deg)} &
\colhead{(mas)} & \colhead{(mas)} & \colhead{(deg)} \\
\colhead{(1)} & \colhead{(2)} & \colhead{(3)} & \colhead{(4)} & \colhead{(5)} & \colhead{(6)} & \colhead{(7)} & \colhead{(8)} &
\colhead{(9)} \\
}
\startdata
FK5 193       & $\alpha$ Aur  & 2004-03-20 & 53084.28 &  52.81 & 244.16 &  0.85 & 0.28 &  0.0 \\
              &               & 2004-03-21 & 53085.31 &  53.46 & 241.20 &  0.88 & 0.29 &  0.0 \\
              &               & 2004-03-23 & 53087.27 &  52.72 & 235.94 & 15.57 & 0.20 &  0.0 \\
              &               &            &          &        &        &      &      &       \\
FK5 339       & 10 UMa        & 2004-03-11 & 53075.32 & 687.37 &   1.91 & 0.72 & 0.28 &  13.9 \\
              &               & 2004-03-14 & 53078.31 & 681.96 &   2.08 & 0.69 & 0.29 &   9.2 \\
              &               & 2004-03-15 & 53079.33 & 686.93 &   2.03 & 0.71 & 0.28 &  12.6 \\
              &               & 2004-05-28 & 53153.30 & 677.94 & 359.86 & 3.40 & 0.16 &  25.7 \\
\enddata
\tablecomments{Col. (1): FK5 \citep{frk88} or HR number \citep{hof91}, as per NPOI scheduling and
archiving software.  Col. (2): per Table~\ref{targetlist}.  Col. (3): UT date of NPOI observations.
Col. (4): MJD of NPOI observations.  Col. (5): Fitted binary separation.  Col. (6): Fitted binary
position angle.  Col. (7): Semimajor axis of error ellipse.  Col. (8): Semiminor axis of error ellipse.
Col. (9): Position angle of error ellipse.}
\tablenotetext{a}{$\gamma$ Vir AB fitted.}
\end{deluxetable}

\clearpage


\begin{deluxetable}{lcrcccc}
\tabletypesize{\scriptsize}
\tablecaption{Binary Program Stars - Mean Separations and $\Delta$$m$ Fits \label{programdelmag}}
\tablewidth{0pt}
\tablehead{
\colhead{Object} & \colhead{Other} & \colhead{$<$$\rho$$>$} & \colhead{$\sigma_{\rho}$} & \colhead{log$<$$\rho$$>$} &
\colhead{$<$$\Delta$$m_{700}$$>$} & \colhead{$\sigma_{\Delta m}$} \\
\colhead{FK5/HR} & \colhead{Name} & \colhead{(mas)} & \colhead{(mas)} & \colhead{mas} & \colhead{(mag)} & \colhead{(mag)} \\
\colhead{(1)} & \colhead{(2)} & \colhead{(3)} & \colhead{(4)} & \colhead{(5)} & \colhead{(6)} & \colhead{(7)} \\
}
\startdata
FK5 193       & $\alpha$ Aur     &  53.66 & 0.97    & 1.73 & 0.00 & 0.1  \\
FK5 339       & 10 UMa           & 679.76 & 6.83    & 2.83 & 2.19 & 0.29 \\
HR 4825/6$^a$ & $\gamma$ Vir     & 622.64 & 2.46    & 2.79 & 0.00 & 0.10 \\
HR 6212       & $\zeta$ Her      & 860.03 & 1.45    & 2.93 & 1.52 & 0.04 \\
FK5 695       & $\chi$ Dra (min) &  25.14 & \nodata & 1.40 & 2.12 & 0.02 \\
\nodata       & $\chi$ Dra (max) &  90.47 & \nodata & 1.96 & 2.12 & 0.02 \\
HR 7882       & $\beta$ Del      & 582.52 & 0.38    & 2.77 & 1.08 & 0.14 \\
\enddata
\tablecomments{Col. (1): FK5 \citep{frk88} or HR number \citep{hof91}, as per
NPOI scheduling and archiving software.  Col. (2): per Table~\ref{targetlist}.
Col. (3): Mean of nightly separation measurements from Table~\ref{programfit}
(except for $\chi$ Dra).  Col. (4): Standard deviation of the nightly
separation measurements.  Col. (5): log of Col. (3) value.  Col. (6): Average
of the nightly fitted component magnitude difference at 700 nm.  Col. (7):
Standard deviation of the nightly $\Delta$$m_{700}$ fits.}
\tablenotetext{a}{$\gamma$ Vir AB fitted.}
\end{deluxetable}

\clearpage


\begin{deluxetable}{lcccccc}
\tabletypesize{\scriptsize}
\tablecaption{Single Program Stars - Angular Diameter Fits \label{progdiams}}
\tablewidth{0pt}
\tablehead{
\colhead{Object} & \colhead{Other} & \colhead{$\theta_{UD}$} & \colhead{$\sigma_{UD}$} & \colhead{} & \colhead{$\theta_{LD}$} &
\colhead{$\sigma_{LD}$} \\
\colhead{FK5/HR} & \colhead{Name} & \colhead{(mas)} & \colhead{(mas)} & \colhead{LDC} & \colhead{(mas)} & \colhead{(mas)} \\
\colhead{(1)} & \colhead{(2)} & \colhead{(3)} & \colhead{(4)} & \colhead{(5)} & \colhead{(6)} & \colhead{(7)} \\
}
\startdata
FK5 2   & $\beta$ Cas  & 2.145 & 0.055 & 1.034 & 2.218 & 0.058 \\
FK5 74  & $\alpha$ Ari & 6.389 & 0.024 & 1.063 & 6.792 & 0.043 \\
FK5 295 & $\beta$ Gem  & 7.578 & 0.015 & 1.058 & 8.018 & 0.043 \\
FK5 412 & 46 LMi       & 2.327 & 0.050 & 1.058 & 2.462 & 0.054 \\
FK5 582 & $\alpha$ Ser & 4.522 & 0.024 & 1.063 & 4.807 & 0.035 \\
HR 6132 & $\eta$ Dra   & 3.174 & 0.071 & 1.057 & 3.355 & 0.077 \\
FK5 633 & $\kappa$ Oph & 3.394 & 0.035 & 1.063 & 3.608 & 0.041 \\
FK5 665 & $\beta$ Oph  & 4.231 & 0.021 & 1.063 & 4.498 & 0.032 \\
FK5 688 & $\eta$ Ser   & 2.894 & 0.043 & 1.058 & 3.062 & 0.048 \\
FK5 783 & $\eta$ Cep   & 2.729 & 0.082 & 1.056 & 2.882 & 0.088 \\
FK5 893 & $\gamma$ Cep & 3.144 & 0.037 & 1.059 & 3.329 & 0.042 \\
\enddata
\tablecomments{Col. (1): FK5 \citep{frk88} or HR number
\citep{hof91}, as per NPOI scheduling and archiving software.
Col. (2): per Table~\ref{targetlist}.  Col. (3): Uniform disk
angular diameter.  Col. (4): Error in uniform disk angular
diameter.  Col. (5): Limb darkening correction factor.  Col. 6):
Limb darkened angular diameter.  Col. (7): Error in limb darkened
angular diameter.}
\end{deluxetable}

\clearpage

\end{document}